\documentclass[a4paper,12pt]{article}
\usepackage{jheppub,graphicx,slashed,xcolor,multirow,bbold,mathtools,sidecap,tikz,bm,ulem,enumitem,booktabs,array,amsmath,braket,import,hyperref,setspace}
\usepackage[T1]{fontenc}
\usepackage[compat=1.1.0]{tikz-feynman}
\graphicspath{ {figures/} }
\pdfoutput=1  
\allowdisplaybreaks

\let\emph\textit

\newcolumntype{M}[1]{>{\centering\arraybackslash}m{#1}}

\title{Anatomy of the Real Higgs Triplet Model}

\author[a,b,c,d]{Saiyad Ashanujjaman}
\emailAdd{saiyad.ashanujjaman@kit.edu}
\author[e,f]{Sumit Banik}
\emailAdd{sumit.banik@psi.ch}
\author[e,f]{Guglielmo Coloretti}
\emailAdd{guglielmo.coloretti@physik.uzh.ch}
\author[e,f]{Andreas Crivellin}
\emailAdd{andreas.crivellin@psi.ch}
\author[g,h]{Siddharth P.~Maharathy}
\emailAdd{siddharth.m@students.iiserpune.ac.in}
\author[g,i]{Bruce Mellado}
\emailAdd{bmellado@mail.cern.ch}
\affiliation[a]{Institut f\"ur Theoretische Teilchenphysik, Karlsruhe Institute of Technology, Engesserstra\ss e 7, D-76128 Karlsruhe, Germany}
\affiliation[b]{Institut f\"ur Astroteilchenphysik, Karlsruhe Institute of Technology, Hermann-von-Helmholtz-Platz 1, D-76344 Eggenstein-Leopoldshafen, Germany}
\affiliation[c]{Institute of High Energy Physics, Chinese Academy of Sciences, Beijing 100049, China}
\affiliation[d]{Kaiping Neutrino Research Center, Jiangmen 529399, China}
\affiliation[e]{Laboratory for Particle Physics, PSI Center for Neutron and Muon Sciences,
Forschungsstrasse 111, 5232 Villigen PSI, Switzerland}
\affiliation[f]{Physik-Institut, Universitat Zurich, Winterthurerstrasse 190, CH–8057 Zurich}
\affiliation[g]{School of Physics and Institute for Collider Particle Physics, University of the Witwatersrand, Johannesburg, Wits 2050, South Africa}

\affiliation[h]{Indian Institute of Science Education and Research Pune,
Dr. Homi Bhabha Road, Pune 411008, India}
\affiliation[i]{iThemba LABS, National Research Foundation, PO Box 722, Somerset West 7129, South Africa}

\abstract{In this article, we examine the Standard Model extended by a $Y=0$ real Higgs triplet, the $\Delta$SM. It contains a $CP$-even neutral Higgs ($\Delta^0$) and two charged Higgs bosons ($\Delta^\pm$), which are quasi-degenerate in mass. We first study the theoretical constraints from vacuum stability and perturbative unitarity and then calculate the Higgs decays, including the loop-induced modes such as di-photons ($\gamma\gamma$) and $Z\gamma$. In the limit of a small mixing between the SM Higgs and $\Delta^0$, the latter decays dominantly to $WW$ and can have a sizable branching ratio to di-photon. The model predicts a positive definite shift in the $W$ mass, which agrees with the current global electroweak fit. At the Large Hadron Collider, it leads to a $(i)$ stau-like signature from $pp\to \Delta^+\Delta^-\to \tau^+\tau^-\nu\bar\nu$, $(ii)$ multi-lepton final states from $pp\to \gamma^*\to \Delta^+\Delta^-\to W^+W^-ZZ$ and $pp\to W^{*} \to \Delta^\pm\Delta^0\to W^\pm Z W^+W^-$ as well as $(iii)$ associated di-photon production from $pp\to W^{*} \to \Delta^\pm(\Delta^0\to\gamma\gamma)$. Concerning $(i)$, the reinterpretation of the recent supersymmetric tau partner search by ATLAS and CMS excludes $m_{\Delta^\pm}<110$\,GeV at 95\% CL. From $(ii)$, some of the signal regions of multi-lepton searches lead to bounds close to the predicted cross-section, but electroweak scale masses are still allowed. For $(iii)$, the recast of the associated di-photon searches by ATLAS and a combined log-likelihood fit of signal and background to data find that out of the 25 signal regions, 10 provide relevant limits on Br$(\Delta^0\to\gamma\gamma)$ at the per cent level. Interestingly, 6 signal regions show excesses at around 152\,GeV, leading to a preference for a non-zero di-photon branching ratio of about 0.7\% with the corresponding significance amounting to about $4\sigma$. While the minimalistic $\Delta$SM does not fully describe the discrepancies between the data and the SM, this study indicates that the Drell-Yan production mechanisms can contribute to the explanation of the narrow excesses at 152\,GeV at the LHC. Furthermore, extended models involving the triplet are capable of explaining the multi-lepton anomalies. In the appendix, we provide the Feynman rules for the model along with an analysis of charge-breaking minima.}

\preprint{TTP24-044, P3H-24-089, PSI-PR-24-25, ZU-TH 59/24, ICPP-88}
\begin{document} 
\compress
\maketitle
\flushbottom  

\newpage \section{Introduction}
\label{sec:intro}
The Standard Model (SM) of particle physics is the currently accepted theory describing the fundamental constituents of matter and their interactions. It has been tested with high accuracy~\cite{ParticleDataGroup:2022pth,HFLAV:2019otj,ALEPH:2005ab}, and the discovery of the 125 GeV Brout-Englert-Higgs boson~\cite{Higgs:1964ia,Englert:1964et,Higgs:1964pj,Guralnik:1964eu} at the Large Hadron Collider (LHC)~\cite{Aad:2012tfa,Chatrchyan:2012ufa} provided its last missing piece. Therefore, any observation of new (fundamental) particles would prove the existence of physics beyond the SM (BSM). In fact, the SM cannot be the ultimate fundamental theory of Nature as it fails to account for several experimental observations, such as neutrino masses and mixing or the existence of Dark Matter. Therefore, the SM must be extended by new particles and new interactions. 

While one can account for Dark Matter and neutrino masses in many ways and at very different energies, anomalies, i.e.~deviations from the SM predictions, point towards new physics at or below the TeV scale (see Ref.~\cite{Crivellin:2023zui} for a review). In fact, many of these anomalies can be explained by extensions of the SM Higgs sector, whose minimality, {\it i.e.} the presence of a single $SU(2)_L$ doublet scalar that simultaneously gives mass to the electroweak (EW) gauge bosons and all fermions, is not guaranteed by any guiding principle or symmetry. Furthermore, while the measured properties of the $125\,$GeV Higgs are consistent with the SM expectations~\cite{Chatrchyan:2012jja,Aad:2013xqa,ATLAS:2016neq,Langford:2021osp,ATLAS:2021vrm}, this does not exclude the existence of additional scalar bosons as long as their role in electroweak symmetry breaking is minute. A plethora of models going beyond the SM Higgs sector have been proposed in the literature, including the addition of $SU(2)_L$ singlets~\cite{Silveira:1985rk,Pietroni:1992in,McDonald:1993ex}, doublets~\cite{Lee:1973iz,Haber:1984rc,Kim:1986ax,Peccei:1977hh,Turok:1990zg} and triplets~\cite{Konetschny:1977bn,Cheng:1980qt,Lazarides:1980nt,Schechter:1980gr,Magg:1980ut,Mohapatra:1980yp}, etc. While, in the past, the main focus has been on singlet and doublet extensions that preserve the custodial symmetry at tree level (i.e.~$\rho=1$), the larger-than-expected $W$ mass measured by the CDF-II collaboration~\cite{CDF:2022hxs} has led to a renaissance of scalar multiplet models~\cite{Butterworth:2022dkt,Heeck:2022fvl,Strumia:2022qkt,Dorsner:2007fy,FileviezPerez:2022lxp,Cheng:2022hbo,Rizzo:2022jti,Wang:2022dte,Chabab:2018ert,Shimizu:2023rvi,Crivellin:2023xbu,Chowdhury:2022moc,Dcruz:2022dao,Babu:2022pdn,Arcadi:2022dmt,Kim:2022hvh,Kim:2022xuo,Chakrabarty:2022voz,Chowdhury:2023uyd,Chen:2022ocr,Kanemura:2022ahw,Ashanujjaman:2022ofg}. In particular, the Higgs triplet with hypercharge $Y=0$ is the most minimal extension of the SM that leads to a positive shift in the $W$ mass at tree-level~\cite{Ross:1975fq,Gunion:1989ci,Blank:1997qa,Forshaw:2003kh,Chankowski:2006hs,Chen:2006pb,Chivukula:2007koj,Dorsner:2007fy,Chabab:2018ert,Bandyopadhyay:2020otm,Strumia:2022qkt,FileviezPerez:2022lxp,Cheng:2022hbo,Rizzo:2022jti,Wang:2022dte,Chen:2022ocr,Lazarides:2022spe,Shimizu:2023rvi,Crivellin:2023xbu,Butterworth:2023rnw,Senjanovic:2022zwy,Crivellin:2023gtf,Chen:2023ins,Ashanujjaman:2023etj,Degrassi:2024qsf}.

Though the intensified LHC searches for new particles, particularly new Higgses, have not led to any discovery yet, several ``multi-lepton anomalies''---statistically significant deviations from the SM predictions in final states with multiple leptons, missing energy and possibly ($b$-)jets ~\cite{vonBuddenbrock:2016rmr,vonBuddenbrock:2017gvy,vonBuddenbrock:2019ajh,vonBuddenbrock:2020ter,Hernandez:2019geu,Coloretti:2023wng,Banik:2023vxa,Coloretti:2023yyq}---have emerged; see Refs.~\cite{Fischer:2021sqw,Crivellin:2023zui} for a review. They are compatible with the production of a $\approx$270\,GeV Higgs decaying into a pair of lighter Higgses ($S$ and $S^\prime$), which dominantly decay to $WW$ and $b\bar b$, respectively. The opposite-sign di-lepton invariant mass from the leptonic $W$ decays is sensitive to the mass $m_S$ and fitting the invariant mass spectra, Ref.~\cite{vonBuddenbrock:2017gvy} found $m_S=150 \pm 5$\,GeV. In this context, it is important to note that the neutral component of the $Y=0$ triplet naturally has a dominant branching ratio to $W$ bosons.

Refs.~\cite{Crivellin:2021ubm,Bhattacharya:2023lmu}, analyzing the sidebands of the SM Higgs boson searches by the ATLAS and CMS collaborations~\cite{Sirunyan:2021ybb,ATLAS:2020pvn,Aad:2020ivc,Sirunyan:2020sum,Aad:2021qks,CMS:2018nlv, Sirunyan:2018tbk,ATLAS:2020fcp}, find that the data suggest the presence of a new narrow resonance in the di-photon and $Z\gamma$ spectra with a mass around 152$\,$GeV produced in association with leptons, ($b$-)jets and missing energy. The overall global significance of this resonant excess, using the Fischer method, has surpassed the $5\sigma$ mark within a simplified model.\footnote{This is obtained by adding the ATLAS excess in $\gamma\gamma+\tau$ $(\approx3\sigma)$~\cite{ATLAS:2024lhu} to the combination of Ref.~\cite{Bhattacharya:2023lmu}.} Since there is no excess in $ZZ$ final states, it is therefore interesting to explore the phenomenology of the Real Higgs Triplet model in this context. 
As such, Refs.~\cite{Ashanujjaman:2024pky,Crivellin:2024uhc,Banik:2024ftv} showed that the latest ATLAS analyses~\cite{ATLAS:2023omk,ATLAS:2024lhu} targeting associated productions of the SM Higgs in various di-photon channels, are consistent with the Drell-Yan production of a $\approx 152\,$GeV Higgs triplet with a corresponding significance of around $4\sigma$.\footnote{Hints for the existence of a neutral scalar with a mass around 95$\,$GeV were presented by the CMS collaboration~\cite{CMS:2018cyk,CMS:2023yay,CMS:2022goy}, not excluded by the ATLAS experiment~\cite{ATLAS:2018xad,ATLAS:2022yrq} and consistent with a mild excess reported by the LEP experiments~\cite{LEPWorkingGroupforHiggsbosonsearches:2003ing}. As a possible explanation for these excesses, the scalar triplet with $Y=0$ was proposed in Ref.~\cite{Ashanujjaman:2023etj,Chen:2023bqr}. However, we will find here that the updated ATLAS stau search with full run 2 luminosity excludes this possibility.}

We take the above phenomenological motivations to perform a comprehensive study of the $Y=0$ Higgs triplet model~\cite{Ross:1975fq,Gunion:1989ci,Chankowski:2006hs,Blank:1997qa,Forshaw:2003kh, Chen:2006pb,Chivukula:2007koj,Bandyopadhyay:2020otm,Butterworth:2022dkt}, which after spontaneous symmetry breaking contains, in addition to the SM Higgs, a $CP$-even Higgs ($\Delta^0$) and two charged Higgs bosons ($\Delta^\pm$). The outline of this article is as follows. We present in Sec.~\ref{sec:model} details of the model, followed by the vacuum stability and perturbative unitarity constraints. The decay rates of the triplet Higgses are discussed in Sec.~\ref{sec:prodDecay} and the phenomenology in Sec.~\ref{sec:pheno}. Finally, we summarise our findings in Sec.~\ref{sec:summary}. Further, we collect the relevant Feynman rules in Appendix~\ref{app:Feynman}, and in Appendix~\ref{app:globality}, we discuss possible vacua configurations and the stability of the neutral ones against the charge-breaking ones.
\section{The $\Delta$SM}
\label{sec:model}

The Higgs sector of the real Higgs triplet model, called the $\Delta$SM, is composed of the SM Higgs $SU(2)_L$ doublet ($\Phi$) with $Y=1/2$ (in our convention)
\begin{equation}
\label{eq:doublet}
\Phi = \begin{pmatrix} h_\Phi^+ \\ \frac{1}{\sqrt{2}} (v_\Phi + h_\Phi^0 + iG^0) \end{pmatrix},
\end{equation}
and the Higgs triplet ($\Delta$) with $Y=0$ transforming in the adjoint representation of $SU(2)_L$
\begin{equation}
\label{eq:triplet}
\Delta = \frac{1}{2} \begin{pmatrix}  v_\Delta + h_\Delta^0 & \sqrt{2}h_\Delta^+ \\ \sqrt{2}h_\Delta^- & -(v_\Delta + h_\Delta^0) \end{pmatrix},
\end{equation}
where $h_{\Phi,\Delta}^0$ are real scalar fields, $h_{\Phi,\Delta}^- = (h_{\Phi,\Delta}^+)^*$, and $v_\Phi$ and $v_\Delta$ are the respective vacuum expectation values (VEVs). With these conventions, the canonically normalised gauge-kinetic part of the Lagrangian involving the triplet is
\begin{equation}
\mathcal{L} \supset {\rm Tr}[(D_{\mu} \Delta)^\dagger (D^{\mu} \Delta)],
\end{equation}
where $D_\mu \Delta= \partial_\mu \Delta + ig \left[\frac{1}{2}\sigma^k W^k_\mu, \Delta\right]$, with $g$ being the $SU(2)_L$ gauge coupling, $\sigma^k$ the Pauli matrices, and the square bracket denotes the commutator.

After EW symmetry breaking, the $SU(2)_L$ doublet Higgs contributes to both $W$ and $Z$ boson masses at the tree level, preserving the so-called custodial symmetry, i.e.~$\rho = 1$. However, the Higgs triplet only contributes to the $W$-boson mass (at the tree level), thereby breaking the custodial symmetry maximally such that
\begin{align}
\rho_{\rm tree} = \frac{m_W^2}{m_Z^2\cos^2\theta_w} = \frac{\frac{1}{4} g^2 (v_\Phi^2 + 4 v_\Delta^2)}{\frac{1}{4} (g^2 + g^{\prime 2}) v_\Phi^2 \cos^2\theta_w} \approx 1 + \frac{4v_\Delta^2}{v^2},
\label{eq:rho}
\end{align}
where $g^\prime$ is the $U(1)_Y$ gauge coupling, $\theta_w$ is the Weinberg angle, and $v\approx v_\Phi \approx 246$\,GeV. Note that the electroweak precision data require $\rho$ to be close to unity at the sub-percent level, allowing us to expand in $v_\Delta/v_\Phi$ in the last step.

At the phenomenological level, the $W$-boson mass is usually calculated in the so-called ``$G_F$ scheme'' input scheme using the $Z$-boson mass, the fine–structure constant $\alpha$, and the Fermi constant $G_F$ from muon decay, resulting in
\begin{align}
m_W^2 \left(1-\frac{m_W^2}{m_Z^2}\right) = \frac{\pi \alpha}{\sqrt{2} G_F}(1+\Delta r).
\end{align}
Here, $m_W$ and $m_Z$ are renormalised masses in the on-shell scheme, and $\Delta r = \Delta \alpha -\frac{c_W^2}{s_W^2} \Delta \rho + \Delta r_{\rm rem}$ represents radiative corrections to $\alpha$, the $\rho$ parameter, and the remaining part of the gauge boson two-point functions as well as the vertex and box diagram corrections to the light fermion scattering process~\cite{Lopez-Val:2014jva}. Combining the expressions above, we have%\footnote{Note that $\Delta r$ is a function of $m_W^2$, such that this equation is solved iteratively to obtain the SM prediction for $m_W$.}
\begin{align}
m_W^2 = \frac{m_Z^2}{2}\left[ 1 + \sqrt{1- \frac{4 \pi \alpha_{\rm em}}{\sqrt{2} G_F m_Z^2}(1+\Delta r(m_W^2))} \right].
\label{eq:Wmass}
\end{align}
Taking into account that the triplet Higgs contribution to $\Delta r$ at one-loop is suppressed on account of its small mixing with the SM Higgs and its degenerate masses (see later), the only relevant contribution to $\Delta r$ appears through $\rho$ at tree-level. Thus, to first–order in $v_\Delta$, Eq.~\eqref{eq:Wmass} implies a shift in $\rho_{\rm tree}$: $\delta \rho = \rho_{\rm tree}-1$, which translates to a shift in $m_W$~\cite{Lopez-Val:2014jva,Degrassi:2024qsf}
\begin{align}
\Delta m_W \approx \frac{m_W}{2} \frac{\cos^2\theta_w}{\cos^2\theta_w-\sin^2\theta_w} \delta \rho\,.
\end{align}
%Furthermore, the shift in the $W$ mass is positive definite, as preferred by the global fit EW fit~\cite{Bagnaschi:2022whn,deBlas:2022hdk,ParticleDataGroup:2022pth}.

\subsection{Masses and couplings}

The most general scalar potential of the $\Delta$SM model is
\begin{align}
\label{eq:pot}
V = & -\mu_\Phi^2 \Phi^\dag \Phi + \frac{\lambda_\Phi}{4} \left(\Phi^\dag \Phi\right)^2 - \mu_\Delta^2 {\rm Tr}\left(\Delta^\dag \Delta\right) + \frac{\lambda_\Delta}{4} \left[ {\rm Tr}\left(\Delta^\dag \Delta\right) \right]^2 \nonumber
\\
& + A \Phi^\dag \Delta \Phi + \lambda_{\Phi \Delta} \Phi^\dag \Phi {\rm Tr}\left(\Delta^\dag \Delta\right).
\end{align}
Without loss of generality, all couplings can be assumed to be real. Consequently, the potential is \textit{CP}-conserving. %\footnote{Though the potential is \textit{CP}-conserving, there still remains, in principle, the possibility for a spontaneous breakdown of this \textit{CP} symmetry. Although this is an interesting possibility to address, it is beyond the scope of this work.} The dimensionful coupling $A$ can be chosen to be real and positive without loss of generality. 
Note that in the limit $A\to 0$, the potential possesses a global $O(4)_\Phi \times O(3)_\Delta$ symmetry and the discrete $Z_{2,\Delta}$ ($\Delta \to -\Delta$) symmetry. Therefore, a non-zero $A$ leads to a soft breaking of this symmetry such that small values of it are natural in the sense ’t Hooft defined it~\cite{tHooft:1979rat}.

Minimizing the potential in Eq.~\eqref{eq:pot}, we find the conditions
\begin{align}
\mu_\Phi^2 &= -\frac{A v_\Delta}{2}+\frac{1}{4} v_\Phi^2 \lambda _{\Phi }+\frac{1}{2} \lambda _{\Phi \Delta } v_\Delta^2,
\label{eq:minimization-1}
\\
\mu_{\Delta}^2 &=  -\frac{A v_\Phi^2}{4 v_\Delta}+\frac{1}{2} v_\Phi^2 \lambda _{\Phi \Delta }+\frac{1}{4} \lambda _{\Delta } v_\Delta^2,
\label{eq:minimization-2}
\end{align}
which can be used to replace $\mu_\Phi^2$ and $\mu_\Delta^2$ such that the mass matrices for the (\textit{CP}-even) neutral and charged Higgses read
\begin{align}
M^2_0 &= \begin{pmatrix}
\frac{\lambda_\Phi v_\Phi^2}{2} & \left( \lambda _{\Phi \Delta } v_\Delta-\frac{A}{2} \right) v_\Phi \\
\left( \lambda _{\Phi \Delta } v_\Delta-\frac{A}{2} \right) v_\Phi & \frac{\lambda_\Delta v_\Delta^2}{2} + \frac{Av_\Phi^2}{4v_\Delta} \\
\end{pmatrix},\\
M^2_\pm &= \begin{pmatrix}
A v_\Delta & \frac{A v_\Phi}{2} \\
\frac{A v_\Phi}{2} & \frac{A v_\Phi^2}{4 v_\Delta}
\end{pmatrix},
\end{align}
in the interaction basis $(h^{0,\pm}_\phi,h^{0,\pm}_\Delta)$. We diagonalize these mass matrices by rotating the interaction states to the physical basis where the mass matrices are diagonal. These mass eigenstates are given by
\begin{align}
\begin{pmatrix} h \\ \Delta^0 \end{pmatrix} &= \begin{pmatrix} \cos\alpha & \sin\alpha \\ -\sin\alpha & \cos\alpha \end{pmatrix} \begin{pmatrix} h_\Phi^0 \\ h_\Delta^0 \end{pmatrix},
\\
\begin{pmatrix} G^\pm \\ \Delta^\pm \end{pmatrix} &= \begin{pmatrix} \cos\beta & \sin\beta \\ -\sin\beta & \cos\beta \end{pmatrix} \begin{pmatrix} h_\Phi^\pm \\ h_\Delta^\pm \end{pmatrix},
\label{Eq:gague_to_physical}
\end{align}
with the eigenvalues
\begin{align}
m_{h}^2 &= \frac{\lambda_\Phi v_\Phi^2}{2} + \tan\alpha \left(\lambda_{\Phi\Delta} v_\Delta -\frac{A}{2}\right) v_\Phi,
\\
m_{\Delta^0}^2 &= \frac{\lambda_\Delta v_\Delta^2}{2} + \frac{Av_\Phi^2}{4v_\Delta} - \tan\alpha \left(\lambda_{\Phi\Delta} v_\Delta -\frac{A}{2}\right) v_\Phi, \label{eq:mH0}
\\
m_{\Delta^\pm}^2 &= A \frac{v_\Phi^2+4v_\Delta^2}{4v_\Delta}, \label{eq:mHp}
\\
m_{G^\pm}^2 &= 0,
\end{align}
and the mixing angles
\begin{align}
& \tan 2\alpha = \dfrac{4v_\Phi v_\Delta \left(2\lambda_{\Phi\Delta} v_\Delta -A\right)}{2\lambda_\Phi v_\Phi^2 v_\Delta -2\lambda_\Delta v_\Delta^3 -A v_\Phi^2}\,,
\\
& \tan 2\beta = -\dfrac{4v_\Phi v_\Delta}{v_\Phi^2-4v_\Delta^2} \quad \left( {\rm equivalently,}~ \tan\beta = -\frac{2v_\Delta}{v_\Phi}\right).
\end{align}
Note that the mass eigenstate $h$ is identified as the 125\,GeV Higgs observed at the LHC, and the zero eigenvalue of the charged Higgs mass matrix corresponds to the {\it would-be} Goldstone boson. Note that we did not explicitly write down the neutral Goldstone $G^0$, which comes purely from the doublet Higgs and thus has the same properties as in the SM. 

From Eq.~\eqref{eq:mH0} and Eq.~\eqref{eq:mHp}, we have
\begin{align}
m_{\Delta^\pm}^2 - m_{\Delta^0}^2 = Av_\Delta - \frac{\lambda_\Delta v_\Delta^2}{2} + \tan\alpha \left(\lambda_{\Phi\Delta} v_\Delta -\frac{A}{2}\right) v_\Phi\,.
\end{align}
Therefore, $\Delta^0$ and $\Delta^\pm$ are nearly mass-degenerate for $\alpha\approx 0$ and $v_\Delta\approx0$. Even though $\alpha$ and $v_\Delta$ break this mass-degeneracy and sufficiently large values could {induce} sizable mass-splitting among them, the requirement of vacuum stability and perturbative unitarity (see Sec.~\ref{sec:stability_unitarity}) together with the experimental limit on the $\rho$ parameter restrict the mass-splitting to at most a few GeV.\footnote{Note that EW radiative corrections, driven by the EW gauge bosons, induce a mass-splitting $m_{\Delta^\pm} - m_{\Delta^0}$ of 160--170\,MeV~\cite{Cirelli:2005uq}.} However, such a small mass-splitting is of little consequence as far as the LHC phenomenology {or} electroweak precision observables---in particular, the oblique parameters---are concerned~\cite{FileviezPerez:2008bj,Kanemura:2012rs,Cheng:2022hbo}. Therefore, we take $m_{\Delta^0} \approx m_{\Delta^\pm} \approx m_\Delta$ further in this work unless stated otherwise.

We can trade all trilinear and quartic couplings of the Lagrangian in Eq.~\eqref{eq:pot} for the physical Higgs masses ($m_h \approx 125$\,GeV, $m_{\Delta^0}$, $m_{\Delta^\pm}$), the VEVs ($v_\Phi \approx 246$\,GeV, $v_\Delta$) and the mixing angle ($\alpha$):
\begin{align}
& \lambda_\Phi = \frac{2}{v_\Phi^2} \left[ \cos^2\alpha ~m_{h}^2 + \sin^2\alpha ~m_{\Delta^0}^2 \right]\,,
\\
&\lambda_\Delta = \frac{2}{v_\Delta^2} \left[ \sin^2\alpha ~m_{h}^2 + \cos^2\alpha ~m_{\Delta^0}^2 - \frac{v_\Phi^2}{v_\Phi^2+4v_\Delta^2} m_{\Delta^\pm}^2 \right]\,,
\\
&\lambda_{\Phi\Delta} = \frac{1}{2v_\Phi v_\Delta} \left[ \sin 2\alpha \left(m_{h}^2-m_{\Delta^0}^2\right) + \frac{4v_\Phi v_\Delta}{v_\Phi^2+4v_\Delta^2} m_{\Delta^\pm}^2 \right]\,,
\\
&A = \frac{4v_\Delta}{v_\Phi^2+4v_\Delta^2} m_{\Delta^\pm}^2\,.
\end{align}
Therefore, the Higgs sector has only four free parameters: $m_{\Delta^0}, m_{\Delta^\pm}, \alpha$ and $v_\Delta$. To get a better analytic understanding of these equations, we expand them in $v_\Delta/v_\Phi\approx v_\Delta/v$ to obtain
\begin{align}
& \lambda_\Phi = \frac{2}{v^2} \left[ \cos^2\alpha ~m_{h}^2 + \sin^2\alpha ~m_{\Delta^\pm}^2 \right]\,,
\\
& \lambda_\Delta = \frac{2 }{v_\Delta^2} \left[ \sin^2\alpha~m_h^2 +\cos^2\alpha~m_{\Delta^0}^2 - ~m_{\Delta^\pm}^2 \right]\,, 
\label{eq:lamDelta}
\\
& \lambda_{\Phi\Delta} = \frac{1}{2v v_\Delta} \sin 2\alpha \left(m_{\Delta^0}^2-m_{\Delta^\pm}^2\right) + \frac{2}{v^2} m_{\Delta^\pm}^2\,, 
\label{eq:lamPhiDelta}
\\
& A = \frac{4v_\Delta}{v^2} m_{\Delta^\pm}^2\,.
\end{align}
Further expanding in the mixing angle $\alpha$, one finds
\begin{align}
& \lambda_\Phi = \frac{2m_{h}^2}{v^2}\,,
\\
& \lambda_\Delta = \frac{2}{v_\Delta^2} \left[m_{\Delta^0}^2 - m_{\Delta^\pm}^2 \right]\,, 
\\
& \lambda_{\Phi\Delta} = \frac{\alpha}{v v_\Delta} \left(m_{\Delta^0}^2-m_{\Delta^\pm}^2\right) + \frac{2}{v^2} m_{\Delta^\pm}^2\,,
\\
& A = \frac{4v_\Delta}{v^2} m_{\Delta^\pm}^2\,.
\end{align}

Note that the Higgs triplet field ($\Delta$) does not couple to the SM fermions at the Lagrangian level. Therefore, the Yukawa Lagrangian is the same as that of the SM (before EW symmetry breaking),
\begin{equation}
    \label{eq:Yuk}
    \mathcal{L}_\mathrm{Yuk} = - \bar{Q}_L Y_d \, \Phi d_R - \bar{Q}_L Y_u \, \Phi^{c} u_R- \bar{L}_L Y_l \, \Phi e_R
\end{equation}
with the only difference arising from the mixing of the scalar states. The resulting Feynman rules concerning their gauge, Yukawa, and self-interactions are given in Appendix~\ref{app:Feynman}.

\subsection{Vacuum stability and perturbative unitarity}
\label{sec:stability_unitarity}

While writing Eq.~\eqref{eq:doublet} and Eq.~\eqref{eq:triplet}, we have implicitly assumed that the EW symmetry is spontaneously broken at some electrically neutral point in the field space and that the corresponding vacuum is the global minimum of the potential. Although the conditions in Eq.~\eqref{eq:minimization-1} and Eq.~\eqref{eq:minimization-2} ensure that the desired EW vacuum corresponds to an extremum of the potential in Eq.\eqref{eq:pot}, one still needs to check that this extremum is indeed stable, {\it i.e.}~not a saddle-point or a local maximum. As we show in Appendix~\ref{app:globality}, the absence of tachyonic modes in the Higgs sector, {\it i.e.}~$m_h^2 > 0$, $m_{\Delta^0}^2 > 0$ and $m_{\Delta^\pm}^2 > 0$, suffices to ensure that the desired EW vacuum corresponds to the global minimum of the potential. Further discussions on the possible vacua configurations and stability of the neutral ones against the charge-breaking ones are deferred till Appendix~\ref{app:globality}.

A necessary condition for the stability of the vacuum is that the potential is bounded from below in all directions in field space. At large field values, the potential in Eq.~\eqref{eq:pot} is dominated by the quadratic terms
\begin{align}
V_4(\Phi,\Delta) = & \frac{\lambda_\Phi}{4} \left(\Phi^\dag \Phi\right)^2 + \frac{\lambda_\Delta}{4} \left[ {\rm Tr}\left(\Delta^\dag \Delta\right) \right]^2 + \lambda_{\Phi \Delta} \Phi^\dag \Phi {\rm Tr}\left(\Delta^\dag \Delta \right)
\\
= & \begin{pmatrix} |\Phi|^2 & |\Delta|^2 \end{pmatrix} \begin{pmatrix} \frac{\lambda_\Phi}{4} & \frac{\lambda_{\Phi\Delta}}{4} \\ \frac{\lambda_{\Phi\Delta}}{4} & \frac{\lambda_\Delta}{8} \end{pmatrix} \begin{pmatrix} |\Phi|^2 \\ |\Delta|^2 \end{pmatrix}, \nonumber
\\
= & x^T \Lambda x,
\end{align}
where $x = \begin{pmatrix} |\Phi|^2 \\ |\Delta|^2 \end{pmatrix} \geq 0$ and $\Lambda = \begin{pmatrix} \frac{\lambda_\Phi}{4} & \frac{\lambda_{\Phi\Delta}}{4} \\ \frac{\lambda_{\Phi\Delta}}{4} & \frac{\lambda_\Delta}{8} \end{pmatrix}$. The requirement $V_4(\Phi,\Delta) \geq 0$, thus, implies that $\Lambda$ has to be a copositive matrix. Applying the copositivity conditions to $\Lambda$, we find
\begin{align}
& \lambda_\Phi > 0, \quad \lambda_\Delta > 0, \quad \sqrt{2} \lambda_{\Phi\Delta} + \sqrt{\lambda_\Phi \lambda_\Delta} > 0.
\end{align}
These conditions are sufficient and necessary to ensure that the potential is bounded from below in all directions in field space at the tree level.\footnote{In this work, we do not attempt to find the possible quantum modifications to these constraints.} The constraint $\lambda_\Delta > 0$ implies that $m_{\Delta^0}^2 - m_{\Delta^\pm}^2 > \sin^2\alpha \left(m_{\Delta^0}^2 - m_h^2 \right)$, see Eq.~\eqref{eq:lamDelta}. This, therefore, fixes the hierarchy of the triplet-like Higgs spectra for $m_{\Delta^0} > m_h$: $m_{\Delta^0} > m_{\Delta^\pm}$.

The model parameter space can also be constrained by requiring perturbative unitarity in scattering processes. This sets limits on interactions in $2\to 2$ scalar scattering processes.\footnote{On account of the equivalence theorem~\cite{Pal:1994jk,Horejsi:1995jj}, we can use unphysical scalar states instead of longitudinal components of the gauge bosons in the high energy limit. Compared to $2\to 2$ scattering processes, $2\to 3$ partial-wave amplitudes can be neglected as the latter scales as the inverse of the energy scale. Further, the amplitudes containing trilinear vertices are generally suppressed by factors accruing from the intermediate propagators.} The partial-wave decomposition of the scattering amplitude $\mathcal{M}_{i\to f}$ reads as 
\begin{equation}
 \mathcal{M}_{fi} = iT^{fi} = 16i\pi \sum_j (2j+1) a^{fi}_j(s) P_j(\cos\theta),   
\end{equation}
where $a_j$ denotes the $j$-th partial-wave amplitude, $\theta$ is the polar angle between
the $i$ and $f$ directions, and $P_j$ is the Legendre polynomial of degree $j$. In the high energy (massless) limit, the most dominant contribution comes from the $j=0$ partial-wave ($s$-wave) at tree-level\footnote{While the loop corrections to $2\to 2$ scattering amplitudes might affect the allowed parameter space, the difference is expected to be marginal. We checked numerically using {\tt Vevacious}~\cite{Camargo-Molina:2014pwa,Camargo-Molina:2013qva}, {\tt SPheno}~\cite{Porod:2003um,Porod:2011nf} and {\tt BSMArt}~\cite{Goodsell:2023iac} that the inclusion of the one-loop effective potential and meta stability, indeed, has only a marginal effect on vacuum stability and perturbative unitarity.}
\begin{equation}
a_0^{fi} = -\frac{i}{16\pi} \mathcal{M}_{fi}.
\end{equation}
The $S$-matrix unitarity for the scattering processes requires $|(a_0)| \leq 1$, $|{\rm Re}(a_0)| \leq \frac{1}{2}$, and $0 \leq {\rm Im}(a_0) \leq 1$. However, in practice, it suffices to require $|(a_0)| \leq 1$ or $|{\rm Re}(a_0)| \leq \frac{1}{2}$, which, in turn, implies that the eigenvalues $x_i$ of the scattering submatrices: $|x_i| \leq \kappa \pi$, where $\kappa = 16$ or 8  depending on whether we demand the former or the latter. This is largely a matter of choice \cite{Logan:2022uus}, and we choose the former.

In the following, we present the resulting submatrices structured in terms of net electric charge in the initial/final states, with their entries corresponding to the quartic couplings that mediate the scalar-scalar scattering processes
\begin{align}
\label{eq:unitarity1}
& \mathcal{M}^{2} = {\rm diag} \left( \frac{\lambda_\Phi}{2}, \frac{\lambda_\Delta}{2}, \lambda_{\Phi\Delta} \right),
\\
& 
\mathcal{M}^{1} = {\rm diag} \left( \frac{\lambda_\Phi}{2}, \lambda_{\Phi\Delta}, \frac{\lambda_\Phi}{2}, \lambda_{\Phi\Delta}, \frac{\lambda_\Delta}{2}, \lambda_{\Phi\Delta} \right),
\\
& \mathcal{M}^{0}_{(a)} = {\rm diag} \left( \lambda_{\Phi\Delta}, \lambda_{\Phi\Delta}, \lambda_{\Phi\Delta}, \frac{\lambda_\Phi}{2}, \lambda_{\Phi\Delta} \right),
\\
& \mathcal{M}^{0}_{(b)} = \begin{pmatrix}
\lambda_\Phi & \lambda_{\Phi\Delta} & \frac{\lambda_\Phi}{2\sqrt{2}} & \frac{\lambda_{\Phi\Delta}}{\sqrt{2}} & \frac{\lambda_\Phi}{2\sqrt{2}}
\\
\lambda_{\Phi\Delta} & \lambda_\Delta & \frac{\lambda_{\Phi\Delta}}{\sqrt{2}} & \frac{\lambda_\Delta}{2\sqrt{2}} & \frac{\lambda_{\Phi\Delta}}{\sqrt{2}}
\\
\frac{\lambda_\Phi}{2\sqrt{2}} & \frac{\lambda_{\Phi\Delta}}{\sqrt{2}} & \frac{3\lambda_\Phi}{4} & \frac{\lambda_{\Phi\Delta}}{2} & \frac{\lambda_\Phi}{4}
\\
\frac{\lambda_{\Phi\Delta}}{\sqrt{2}} & \frac{\lambda_\Delta}{2\sqrt{2}} & \frac{\lambda_{\Phi\Delta}}{2} & \frac{3\lambda_\Delta}{4} & \frac{\lambda_{\Phi\Delta}}{2}
\\
\frac{\lambda_\Phi}{2\sqrt{2}} & \frac{\lambda_{\Phi\Delta}}{\sqrt{2}} & \frac{\lambda_\Phi}{4} & \frac{\lambda_{\Phi\Delta}}{2} & \frac{3\lambda_\Phi}{4}
\end{pmatrix},
\label{eq:unitarity2}
\end{align}
where the submatrices, respectively, correspond to scattering processes with initial and final states 
$\left( \frac{1}{\sqrt{2}} h_\Phi^+ h_\Phi^+, \frac{1}{\sqrt{2}} h_\Delta^+ h_\Delta^+, h_\Phi^+ h_\Delta^+ \right)$, 
$\left( h_\Phi^0 h_\Phi^+, h_\Delta^0 h_\Phi^+, G^0 h_\Phi^+, h_\Phi^0 h_\Delta^+, h_\Delta^0 h_\Delta^+, G^0 h_\Delta^+ \right)$,\\
$\left( h_\Phi^+ h_\Delta^-, h_\Delta^+ h_\Phi^-, h_\Phi^0 h_\Delta^0, h_\Phi^0 G^0, h_\Delta^0 G^0 \right)$
and $\left( h_\Phi^+ h_\Phi^-, h_\Delta^+ h_\Delta^-, \frac{1}{\sqrt{2}} h_\Phi^0 h_\Phi^0, \frac{1}{\sqrt{2}} h_\Delta^0 h_\Delta^0, \frac{1}{\sqrt{2}} G^0 G^0 \right)$; $\sqrt{2}$ accounts for identical particle statistics. Now, requiring the moduli of the eigenvalues of the submatrices in Eq.~\eqref{eq:unitarity1}--Eq.~\eqref{eq:unitarity2} to be $\leq \kappa \pi$, we get
\begin{align}
& |\lambda_\Phi| \leq 2\kappa \pi, \quad |\lambda_\Delta| \leq 2\kappa \pi, \quad \\
&|\lambda_{\Phi\Delta}| \leq \kappa \pi, \quad |6\lambda_\Phi + 5\lambda_\Delta \pm \sqrt{(6\lambda_\Phi - 5\lambda_\Delta)^2 + 192\lambda_{\Phi\Delta}^2}| \leq 8\kappa \pi.
\end{align}
These conditions ensure that perturbative unitarity is respected in all $2\to 2$ scalar scattering processes and put non-trivial constraints on the parameter space. In particular, the conditions $|\lambda_\Delta| \leq 2\kappa \pi$ and $|\lambda_{\Phi\Delta}| \leq \kappa \pi$ restricts the mass-splitting $m_{\Delta^\pm}-m_{\Delta^0}$ to a few GeV, see Eq.~\eqref{eq:lamDelta} and Eq.~\eqref{eq:lamPhiDelta}.
\section{Higgs Production and Decays}
\label{sec:prodDecay}

\subsection{Production}
The SM-like Higgs $h$ is dominantly produced via gluon-gluon fusion (ggF) and vector-boson fusion (VBF) processes, and the cross-sections are obtained in the $\Delta$SM by multiplying the corresponding SM cross-sections (48.58 pb~\cite{Anastasiou:2016cez} and 3.78 pb~\cite{LHCHiggsCrossSectionWorkingGroup:2013rie}) by $\cos^2\alpha$ and 
\begin{align}
&g_{WW}^h = \frac{v_\Phi \cos\alpha + 4v_\Delta\sin\alpha}{v} \approx \cos\alpha + \frac{4v_\Delta}{v}\sin\alpha,
\end{align}
respectively.\footnote{As occasioned by the EW precision data, $v_\Delta \ll v$; therefore, $v_\Phi \approx v$. In what follows, we neglect the corrections in $v_\Phi/v$ while writing analytical expressions for the decay widths.} The triplet-like neutral Higgs $\Delta^0$ is also produced via ggF and VBF processes by mixing with the SM Higgs. The cross-sections are obtained from SM Higgs cross-sections by a rescaling with $\sin^2\alpha$ and 
\begin{align}
&g_{WW}^{\Delta^0} = \frac{-v_\Phi \sin\alpha + 4v_\Delta\cos\alpha}{v} \approx -\sin\alpha + \frac{4v_\Delta}{v}\cos\alpha,
\end{align}
respectively.\footnote{The triplet-like Higgs states are also pair-produced via VBF processes. Likewise, the charged Higgs states are also pair-produced via the $t/u$-channel photon-photon fusion processes. However, such processes are sub-dominant and can be safely neglected.}

\begin{figure}[htb!]
\resizebox{0.99\columnwidth}{!}{
\begin{tikzpicture}
\begin{feynman}
\vertex(a);
\vertex[right=of a] (b);
\vertex[above left=of a] (i1){\(q\)};
\vertex[below left=of a] (i2){\(\bar{q}^\prime\)};
\vertex[above right=of b] (f1){\(\Delta^\pm\)};
\vertex[below right=of b] (f2){\(\Delta^0\)};
\diagram* {
(i1)-- [fermion] (a),
(a)-- [fermion] (i2),
(a)-- [boson,edge label'=\(W^{\pm *}\)] (b),
(f1)-- [scalar] (b),
(b)-- [scalar] (f2),
};
\end{feynman}
\end{tikzpicture}
\quad
\begin{tikzpicture}
\begin{feynman}
\vertex(a);
\vertex[right=of a] (b);
\vertex[above left=of a] (i1){\(q\)};
\vertex[below left=of a] (i2){\(\bar{q}\)};
\vertex[above right=of b] (f1){\(\Delta^+\)};
\vertex[below right=of b] (f2){\(\Delta^-\)};
\diagram* {
(i1)-- [fermion] (a),
(a)-- [fermion] (i2),
(a)-- [boson,edge label'=\(\gamma^*/Z^*\)] (b),
(f1)-- [scalar] (b),
(b)-- [scalar] (f2),
};
\end{feynman}
\end{tikzpicture}
\quad
\begin{tikzpicture}
\begin{feynman}
\vertex(a);
\vertex[above left=of a] (b);
\vertex[below left=of a] (c);
\vertex[left=of b] (i1){\(g\)};
\vertex[left=of c] (i2){\(g\)};
\vertex[right=of a] (f){\(\Delta^0\)};
\diagram* {
(i1)-- [gluon] (b),
(i2)-- [gluon] (c),
(a)-- [fermion] (b),
(b)-- [fermion, edge label'=\(t/b\)] (c),
(c)-- [fermion] (a),
(a)-- [scalar] (f),
};
\end{feynman}
\end{tikzpicture}
}
\caption{Example Feynman diagrams showing the production of the triplet-like Higgs at the LHC: Drell-Yan (left and middle) and ggF (right).}
\label{fig:FeynProd}
\end{figure}
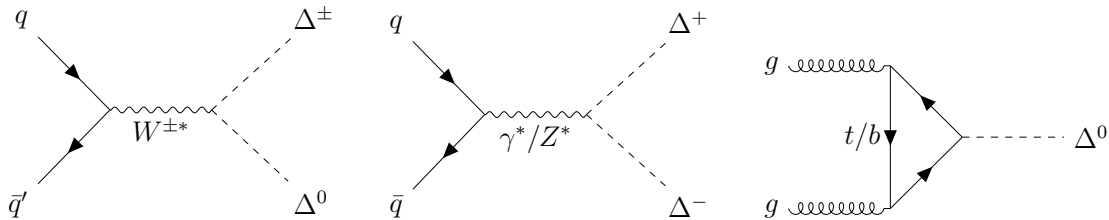

\begin{figure}[htb!]
\centering
\includegraphics[width=0.65\columnwidth]{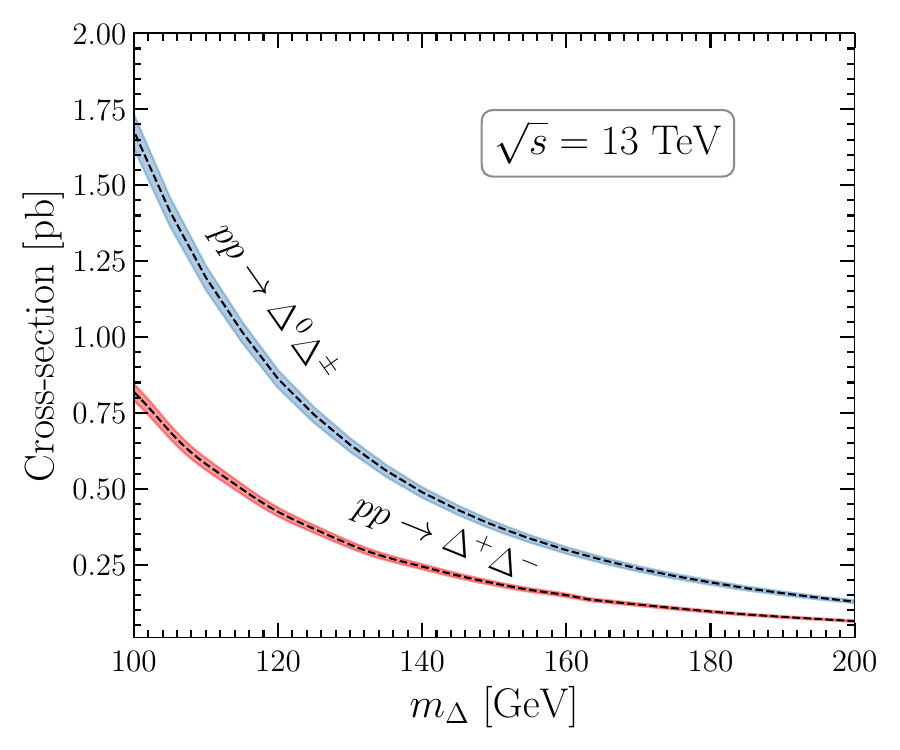}
\caption{Drell-Yan production cross-sections for the triplet-like Higgses, including the NLO+NNLL QCD corrections and uncertainties of Refs.~\cite{Ruiz:2015zca,Ajjath:2023ugn}, at the 13\,TeV LHC. Here, we disregarded small corrections due to the mixing angles $\alpha$ and $\beta$.}
\label{fig:xsec}
\end{figure}

In the $\Delta$SM, an additional production mechanism is relevant: the Drell-Yan (DY) production of Higgs pairs via $\gamma^*/Z^*$ and $W^{\pm *}$
\[
q\bar{q^\prime} \to W^{\pm *} \to \Delta^0 \Delta^\pm, \quad q\bar{q} \to \gamma^*/Z^* \to \Delta^+ \Delta^-.
\]
Fig.~\ref{fig:FeynProd} shows the dominant Feynman diagrams for the production of the triplet-like Higgs states.\footnote{We obtain the leading-order Drell-Yan production cross-sections by using the UFO modules generated from {\tt SARAH}~\cite{Staub:2013tta,Staub:2015kfa} and {\tt Feynrules}~\cite{Degrande:2011ua,Alloul:2013bka, Degrande:2014vpa} in {\tt MadGraph5\_aMC\_v3.5.3}~\cite{Alwall:2011uj,Alwall:2014hca} with the {\tt NNPDF23\_nlo\_as\_0118\_qed} parton distribution function~\cite{Ball:2013hta}. The higher-order corrections can be included via a $k$ factor, which at the next-to-leading order (NLO) in QCD matched with next-to-next-to-leading-logarithmic (NNLL) threshold resummation is $\approx$1.15 for the mass range of our interest~\cite{Ruiz:2015zca,Ajjath:2023ugn}. Further, Refs.~\cite{Ruiz:2015zca,Ajjath:2023ugn} showed that the NLO+NNLL differential $k$-factors for heavy lepton kinematics are substantially flat, suggesting that naive scaling by an overall $k$-factor is a very good approximation.} Fig.~\ref{fig:xsec} shows the Drell-Yan production cross-sections for the triplet-like Higgs states at the 13 TeV LHC, including the higher-order corrections via $k$ factor, as a function of their common mass. The SM-like Higgs $h$ is also DY-produced in the $\Delta$SM. However, the corresponding cross-sections are suppressed by the square of the small mixing angles $\alpha$ and $\beta$.

\subsection{Decays of the SM-like Higgs}
\label{sec:hDecays}
The properties of the SM-like Higgs boson $h$ have been studied in detail at the LHC~\cite{Chatrchyan:2012jja,Aad:2013xqa,ATLAS:2016neq,Langford:2021osp,ATLAS:2021vrm,ATLAS:2022vkf,CMS:2022dwd}. These data, particularly on $h \to \gamma\gamma$ and $h \to ZZ^*$, provide nontrivial constraints on New Physics (NP) impacting the SM Higgs. It is customary to define the corresponding signal strengths for a decay mode $h \to XY$, normalised to the SM value
\begin{align}
\mu_{XY} = \frac{\sigma_h}{\sigma_h^{\rm SM}} \times \frac{\Gamma_{h \to XY}}{\Gamma_{h \to XY}^{\rm SM}} \times \frac{\Gamma_{h, {\rm tot}}^{\rm SM}}{\Gamma_{h, {\rm tot}}}
\end{align}
where $\sigma_h/\sigma_h^{\rm SM} \approx \cos^2\alpha$ is the production cross-section of $h$ normalised to its SM value and $\Gamma_{h, {\rm tot}}^{\rm SM} = 4.07$\,MeV is the total decay width in the SM. For $\gamma\gamma$ and $Z\gamma$, the partial widths are given by~\cite{Chen:2013vi}
\begin{align}
&\Gamma^{(\rm SM)}_{h \to \gamma\gamma} = \frac{\alpha_{\rm EM}^2 g_2^2 m_h^3}{1024\pi^3m_W^2} \left|g^{\rm SM}_{h\gamma\gamma}\right|^2,
\\
&\Gamma^{(\rm SM)}_{h \to Z\gamma} = \frac{\alpha_{\rm EM} g_2^4 m_h^3}{2048\pi^4m_W^2} \left(1-\frac{m_Z^2}{m_h^2} \right)^3 \left|g^{\rm SM}_{hZ\gamma}\right|^2,
\end{align}
where the fine-structure constant $\alpha_{\rm EM}$ should be taken at the scale $q^2 = 0$ (since the photons are on-shell), and
\begin{align*}
g_{h\gamma\gamma}^{\rm SM} =& \sum_f N^c_f Q_f^2 \beta^{1/2}_{\gamma\gamma}\left(\frac{4m_f^2}{m_h^2}\right) + \beta^1_{\gamma\gamma}\left(\frac{4m_W^2}{m_h^2}\right),
\\
g_{h\gamma\gamma} =& \sum_f N^c_f Q_f^2 g_{f\bar{f}}^h \beta^{1/2}_{\gamma\gamma}\left(\frac{4m_f^2}{m_h^2}\right) + g_{WW}^h \beta^1_{\gamma\gamma}\left(\frac{4m_W^2}{m_h^2}\right) + \frac{\lambda_{h\Delta^+\Delta^-} v}{2m_{\Delta^\pm}^2} \beta^0_{\gamma\gamma}\left(\frac{4m_{\Delta^\pm}^2}{m_h^2}\right),
\\
g_{hZ\gamma}^{\rm SM} =& \sum_f \frac{2N^c_f}{\cos\theta_w} Q_f \left(I_{3f} - 2Q_f \sin^2\theta_w \right) \beta^{1/2}_{Z\gamma}\left(\frac{4m_f^2}{m_h^2}, \frac{4m_f^2}{m_Z^2}\right) + \beta^1_{Z\gamma}\left(\frac{4m_W^2}{m_h^2}, \frac{4m_W^2}{m_Z^2}\right),
\\
g_{hZ\gamma} =& \sum_f \frac{2N^c_f}{\cos\theta_w} Q_f \left(I_{3f} - 2Q_f \sin^2\theta_w \right) g_{f\bar{f}}^h \beta^{1/2}_{Z\gamma}\left(\frac{4m_f^2}{m_h^2}, \frac{4m_f^2}{m_Z^2}\right) + g_{WW}^h \beta^1_{Z\gamma} \left(\frac{4m_W^2}{m_h^2}, \frac{4m_W^2}{m_Z^2}\right)
\\
& - \left(\sin^2\beta + \frac{2\cos^2\theta_w}{\cos 2\theta_w} \cos^2\beta\right) \frac{\lambda_{h\Delta^+\Delta^-} v}{2m_{\Delta^\pm}^2} \beta^0_{Z\gamma} \left(\frac{4m_{\Delta^\pm}^2}{m_h^2}, \frac{4m_{\Delta^\pm}^2}{m_Z^2}\right)
\end{align*}
Here $Q_f$ and $I_{3f}$ are the electric charge and the third component of the electroweak isospin of the SM fermion $f$\footnote{In our notation, $I_{3f}=\pm \frac{1}{2} (0)$ for the left(right)-handed fermions in weak isodoublets (isosinglets).} running in the loop, respectively, $N^c_f$ is the number of colours (3 for quarks and 1 for leptons). We also defined
\begin{align}
\lambda_{h\Delta^+\Delta^-} =& \sin\alpha \left\{ \frac{1}{2}\lambda_\Delta v_\Delta \cos^2\beta + \left(\lambda_{\Phi\Delta} v_\Delta + \frac{1}{2}A\right) \sin^2\beta \right\} \nonumber
\\
& + \cos\alpha \left\{ \frac{1}{2}\lambda_\Phi v_\Phi \sin^2\beta + \lambda_{\Phi\Delta} v_\Phi \cos^2\beta -\frac{1}{2}A \sin 2\beta \right\};
\label{eq:coupling_hHH}
\end{align}
the functions $\beta^{0,1/2,1}_{\gamma\gamma,Z\gamma}$ are collected in Appendix~\ref{app:LoopFunc}, and the Higgs coupling to fermions normalised to its SM value is
\begin{align}
&g_{f\bar{f}}^h = \cos\alpha,
\end{align}
The other SM Higgs signal strengths ($b\bar b$, $c\bar c$, $\tau \tau$, $gg$, and $ZZ^*$) are given by $\cos^4\alpha$, and the one for $WW^*$ is given by $\cos^2\alpha(g^h_{WW})^2$.

\begin{figure}[htb!]
\centering
\includegraphics[width=0.49\columnwidth]{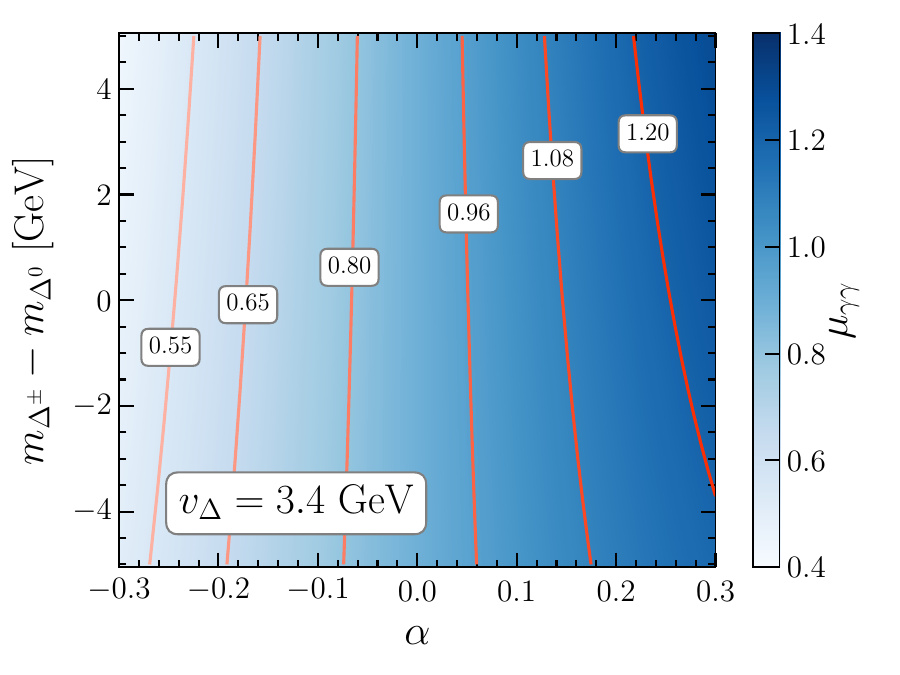} 
\includegraphics[width=0.49\columnwidth]{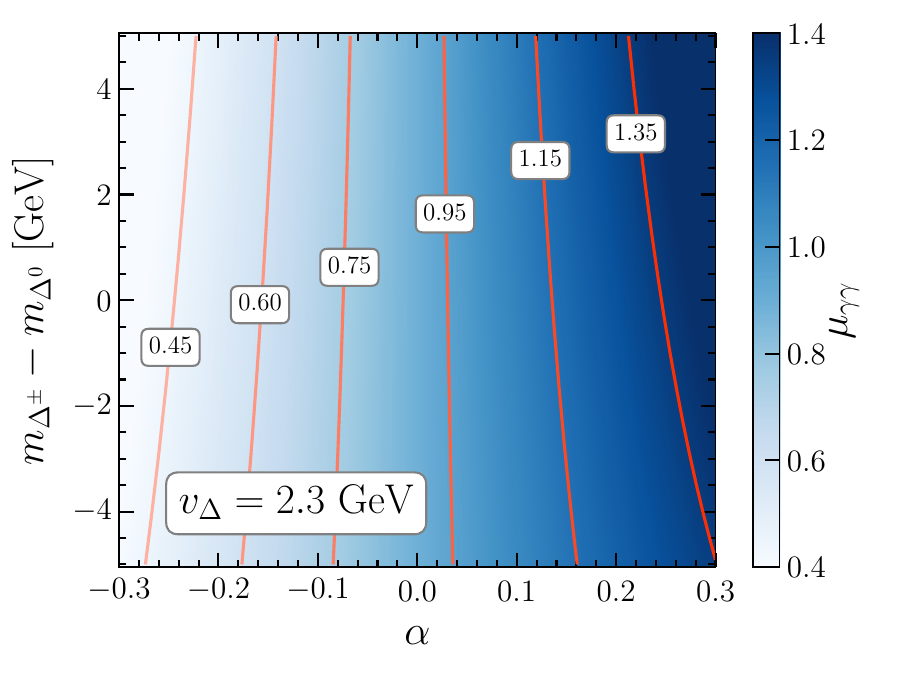} 
\caption{$h \to \gamma\gamma$ signal strength in the $\alpha$ vs $m_{\Delta^\pm} - m_{\Delta^0}$ plane for $m_{\Delta^0} = 152$\,GeV and $v_\Delta = 3.4 $\,GeV (left) and $v_\Delta = 2.3$\,GeV (right).}
\label{fig:haa}
\end{figure}

\begin{figure}[htb!]
\centering
\includegraphics[width=0.49\columnwidth]{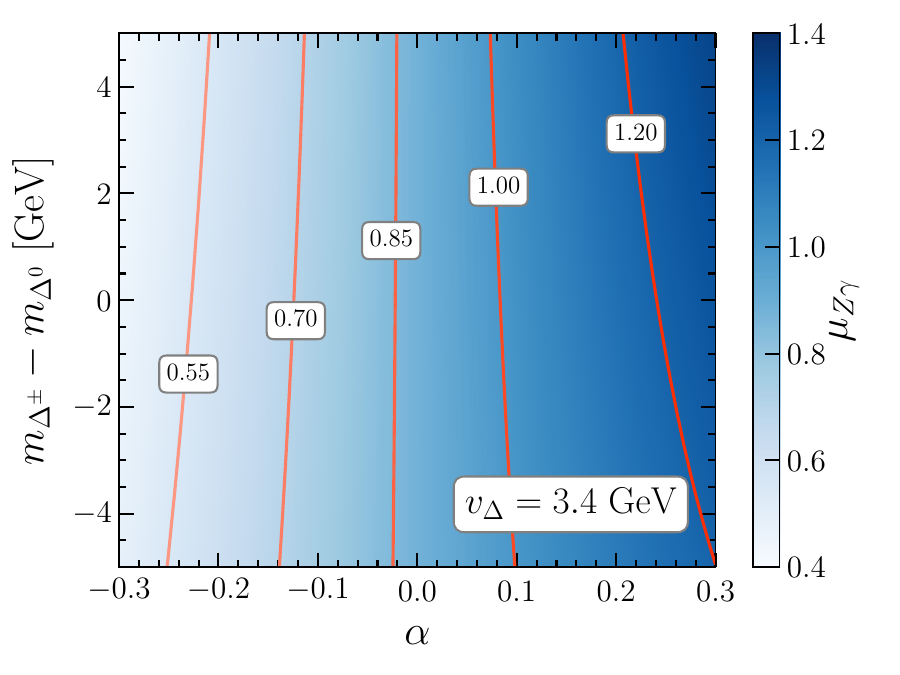} 
\includegraphics[width=0.49\columnwidth]{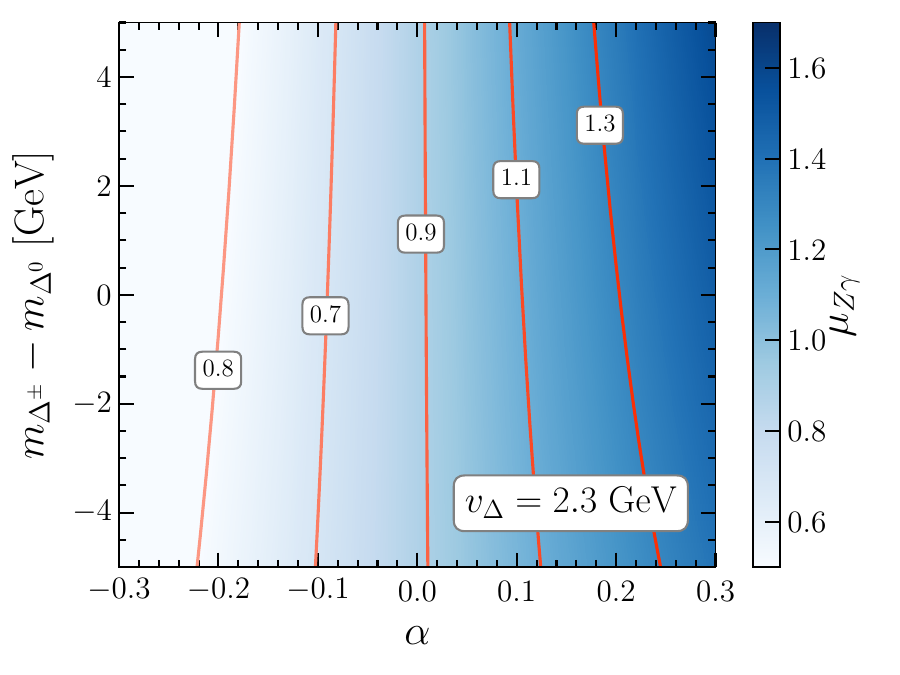} 
\caption{$h \to Z\gamma$ signal strength in the $\alpha$ vs $m_{\Delta^\pm} - m_{\Delta^0}$ plane for $m_{\Delta^0} = 152$\,GeV and $v_\Delta = 3.4 $\,GeV (left) and $v_\Delta = 2.3$\,GeV (right).}
\label{fig:hZa}
\end{figure}

These expressions must be compared to the measurements of CMS and ATLAS for the $h \to \gamma\gamma$, $h \to Z\gamma$ and $h \to ZZ^*$ signal strengths using the full run 2 data at the 13 TeV LHC. Combining their most recent $h \to \gamma\gamma$ measurements: $\mu_{\gamma\gamma}^{\rm CMS} = 1.12^{+0.09}_{-0.09}$~\cite{CMS:2021kom} and  $\mu_{\gamma\gamma}^{\rm ATLAS} = 1.04^{+0.10}_{-0.09}$~\cite{ATLAS:2022tnm}, we get the weighted average
\begin{align}
\mu_{\gamma\gamma}^{\rm LHC} = 1.08^{+0.07}_{-0.06}.
\end{align}
On the other hand, the collaborations themselves performed a combined analysis of their $h \to Z\gamma$ measurements~\cite{ATLAS:2020qcv,CMS:2022ahq} in Ref.~\cite{ATLAS:2023yqk} and reported
\begin{align}
\mu_{Z\gamma}^{\rm LHC} = 2.2\pm 0.7.
\end{align}
Note that this measured value is in mild tension ($\approx 1.9\sigma$) with the SM expectation. The PDG average~\cite{ParticleDataGroup:2022pth} of the $h\to ZZ^*$ signal strength based on the LHC measurements~\cite{ATLAS:2016neq,CMS:2022dwd,ATLAS:2020rej} gives
\begin{align}
\mu_{ZZ^*}^{\rm LHC} = 1.02 \pm 0.08.
\end{align}
While the $h\to ZZ^*$ signal strength puts a constraint on the mixing angle
\begin{align}
|\alpha| \lesssim 0.25(0.38)[0.49] {\rm ~at~} 1\sigma(2\sigma)[3\sigma] {\rm ~level},
\end{align}
the $h \to \gamma\gamma$ and $h \to Z\gamma$ signal strengths impose non-trivial constraints on $m_{\Delta^0}$, $m_{\Delta^\pm} - m_{\Delta^0}$ and $\alpha$. Note that their dependence on $m_{\Delta^\pm}-m_{\Delta^0}$ stems from the $h\Delta^+\Delta^-$ coupling, see Eq.~\eqref{eq:coupling_hHH}. Figs.~\ref{fig:haa} and \ref{fig:hZa} show their dependence on $\alpha$ and $m_{\Delta^\pm} - m_{\Delta^0}$ plane for $m_{\Delta^0} = 152$\,GeV and $v_\Delta = 3.4$\,GeV (left) and $v_\Delta = 2.3$\,GeV (right).\footnote{Understandably, the chosen values for $m_{\Delta^0}$ and $v_\Delta$ might seem arbitrary at this point. However, as we see later in Section~\ref{sec:Wmass}, these values for $v_\Delta$ are preferred by the $W$-mass world average, including/excluding the CDF-II measurement. Further, we find later in Section.~\ref{sec:aaX} that ATLAS di-photon data strongly prefer a Higgs triplet with $m_{\Delta^0} = 152$\,GeV.} As we see, $\alpha > 0$ is favored by the $\mu \to \gamma\gamma$ measurements for $m_{\Delta^0} = 152$\,GeV. The same follows for $m_{\Delta^0} > m_h$. The reverse that $\alpha < 0$ is favoured by the $\mu \to \gamma\gamma$ measurements is largely true for $m_{\Delta^0} < m_h$ (corresponding plot not shown for brevity). As expected, a positive correlation between $\mu_{\gamma\gamma}$ and $\mu_{Z\gamma}$ in the $\alpha$ vs $m_{\Delta^\pm} - m_{\Delta^0}$ plane can be seen. That is the parameter space predicting a larger-than-expected $\mu_{\gamma\gamma}$ predicts a larger-than-expected $\mu_{Z\gamma}$, and vice-versa. However, since the latter is beset with quite a large error, it is not difficult to simultaneously accommodate both measurements.

\subsection{Decays of the triplet-like Higgses $\Delta^0$ and $\Delta^\pm$}
\label{sec:tripletDecays}
The triplet-like Higgs states $\Delta^0$ and $\Delta^\pm$ have two dominant classes of decays:
\begin{enumerate}[label=$(\roman*)$]
\item $\Delta^0$ decays into two SM particles (on-shell or off-shell), either at tree-level to $f\bar f$, $WW^{(*)}$, $ZZ^{(*)}$, $hh^{(*)}$, or at the loop level to $gg$, $\gamma\gamma$, $Z\gamma$. Likewise $\Delta^\pm$ decays to $ff^\prime$, $W^{(*)}Z^{(*)}$, $h^{(*)}W^{(*)}$. The asterisk signals a possible off-shellness.
\item Even though kinematically suppressed due to the small mass splitting, $\Delta^0$ and $\Delta^\pm$ could decay into one another and an off-shell $W$-boson, often referred to as cascade decays.
\end{enumerate}

The tree-level decay rates for the triplet-like Higgs $\Delta^0$ are given by
\begin{align*}
&\Gamma(\Delta^0 \to f\bar{f}) = \frac{N_c m_f^2 m_{\Delta^\pm}}{8\pi v_\Phi^2} \sin^2\alpha \left[\beta\left(\frac{m_f^2}{m_{\Delta^0}^2},\frac{m_f^2}{m_{\Delta^0}^2}\right)\right]^{3/2},
\\
&\Gamma(\Delta^0 \to WW) = \frac{g^4 m_{\Delta^\pm}^3}{256\pi m_W^4} \left(-v_\Phi \sin\alpha + 4v_\Delta\cos\alpha\right)^2 \beta_V\left(\frac{m_W^2}{m_{\Delta^0}^2}\right),
\\
&\Gamma(\Delta^0 \to ZZ) = \frac{g^4 m_{\Delta^\pm}^3}{512\pi m_W^4} \left(-v_\Phi \sin\alpha\right)^2 \beta_V\left(\frac{m_Z^2}{m_{\Delta^0}^2}\right),
\\
&\Gamma(\Delta^0 \to WW^*) = \frac{3 g^6 m_{\Delta^\pm}}{2048\pi^3m_W^2} \left(-v_\Phi \sin\alpha + 4v_\Delta\cos\alpha\right)^2 \beta_V^\prime\left(\frac{m_W^2}{m_{\Delta^0}^2}\right),
\\
&\Gamma(\Delta^0 \to ZZ^*) = \frac{3 g^6 m_{\Delta^\pm}}{2048\pi^3\cos^6\theta_w m_Z^2} \left(-v_\Phi \sin\alpha\right)^2 \left(\frac{7}{12}-\frac{10}{9}\sin^2\theta_W+\frac{40}{27}\sin^4\theta_W\right) \beta_V^\prime\left(\frac{m_Z^2}{m_{\Delta^0}^2}\right),
\\
&\Gamma(\Delta^0 \to hh) = \frac{1}{32\pi m_{\Delta^0}} \lambda_{\Delta^0hh}^2 \left[\beta\left(\frac{m_h^2}{m_{\Delta^0}^2},\frac{m_h^2}{m_{\Delta^0}^2}\right)\right]^{1/2},
\\
&\Gamma(\Delta^0 \to hh^* \to hb\bar{b}) = \frac{3m_b^2 \cos^2\alpha}{32\pi^3 v_\Phi^2 m_H} \lambda_{\Delta^0hh}^2 \beta_S\left(\frac{m_{h}^2}{m_{\Delta^0}^2}\right), 
\\
&\Gamma(\Delta^0 \to hh^* \to h\tau^+\tau^-) = \frac{m_\tau^2 \cos^2\alpha}{32\pi^3 v_\Phi^2 m_H} \lambda_{\Delta^0hh}^2 \beta_S\left(\frac{m_{h}^2}{m_{\Delta^0}^2}\right),
\\
&\Gamma(\Delta^0 \to \Delta^\pm W^{\mp *}) = \frac{9g^2m_{\Delta^0}}{128\pi^3} \lambda_{\Delta^0 \Delta^\pm W^\mp}^2 G\left(\frac{m_{\Delta^\pm}^2}{m_{\Delta^0}^2},\frac{m_W^2}{m_{\Delta^0}^2}\right),
\end{align*}
where
\begin{align*}
&\lambda_{\Delta^0hh} = \left(\lambda_{\Phi\Delta} v_\Delta -\frac{1}{2}A\right) \cos^3\alpha + \left(2\lambda_{\Phi\Delta} -\frac{3}{2}\lambda_\Phi\right) v_\Phi \cos^2\alpha \sin\alpha 
\\
& \hspace{1.5cm} + \left(\frac{3}{2}\lambda_\Delta v_\Delta -2\lambda_{\Phi\Delta} v_\Delta +A\right) \cos\alpha \sin^2\alpha -\lambda_{\Phi\Delta} v_\Phi \sin^3\alpha,
\\
&\lambda_{\Delta^0 \Delta^\pm W^\mp} = -\frac{g}{2}\left(2\cos\alpha \cos\beta - \sin\alpha \sin\beta \right),
\end{align*}
and the loop-induced decays, with at least one massless gauge boson in the final state, are given by
\begin{align*}
&\Gamma({\Delta^0 \to gg}) = \frac{\alpha_s^2 g^2 m_{\Delta^0}^3}{288\pi^3m_W^2} \left|g_{\Delta^0gg}\right|^2,
\\
&\Gamma({\Delta^0 \to \gamma\gamma}) = \frac{\alpha_{\rm EM}^2 g_2^2 m_{\Delta^0}^3}{1024\pi^3m_W^2} \left|g_{\Delta^0 \gamma\gamma}\right|^2,
\\
&\Gamma({\Delta^0 \to Z\gamma}) = \frac{\alpha_{\rm EM} g_2^4 m_{\Delta^0}^3}{2048\pi^4m_W^2} \left(1-\frac{m_Z^2}{m_h^2} \right)^3 \left|g_{\Delta^0Z\gamma}\right|^2,
\end{align*}
where $\alpha_s$ is the strong coupling constant,
\begin{align*}
g_{\Delta^0gg} =& \frac{3}{4} \sum_q g_{f\bar{f}}^{\Delta^0} \beta_{\gamma\gamma}^{1/2}\left(\frac{4m_q^2}{m_{\Delta^0}^2}\right),
\\
g_{\Delta^0\gamma\gamma} =& \sum_f N_c Q_f^2 g_{f\bar{f}}^{\Delta^0} \beta_{\gamma\gamma}^{1/2}\left(\frac{4m_f^2}{m_{\Delta^0}^2}\right) + g_{WW}^{\Delta^0} \beta_{\gamma\gamma}^1\left(\frac{4m_W^2}{m_{\Delta^0}^2}\right) + \frac{\lambda_{\Delta^+\Delta^-\Delta^0} v_{\rm SM}}{2m_{\Delta^\pm}^2}  \beta^0_{\gamma\gamma}\left(\frac{4m_{\Delta^\pm}^2}{m_{\Delta^0}^2}\right),
\\
g_{\Delta^0Z\gamma} =& \sum_f \frac{2N^c_f}{\cos\theta_w} Q_f \left(I_{3f} - 2Q_f \sin^2\theta_w \right) g_{f\bar{f}}^{\Delta^0} \beta^{1/2}_{Z\gamma}\left(\frac{4m_f^2}{m_{\Delta^0}^2}, \frac{4m_f^2}{m_Z^2}\right) + g_{WW}^{\Delta^0} \beta^1_{Z\gamma} \left(\frac{4m_W^2}{m_{\Delta^0}^2}, \frac{4m_W^2}{m_Z^2}\right)
\\
& - \left(\sin^2\beta + \frac{2\cos^2\theta_w}{\cos 2\theta_w} \cos^2\beta\right) \frac{\lambda_{\Delta^0\Delta^+\Delta^-} v}{2m_{\Delta^\pm}^2} \beta^0_{Z\gamma} \left(\frac{4m_{\Delta^\pm}^2}{m_{\Delta^0}^2}, \frac{4m_{\Delta^\pm}^2}{m_Z^2}\right),
\end{align*}
with
\begin{align*}
&\lambda_{\Delta^0\Delta^+\Delta^-} = \cos\alpha \left\{ \frac{1}{2}\lambda_\Delta v_\Delta \cos^2\beta + \left(\lambda_{\Phi\Delta} v_\Delta + \frac{1}{2}A\right) \sin^2\beta \right\}
\\
& \hspace{2cm} - \sin\alpha \left\{ \frac{1}{2}\lambda_\Phi v_\Phi \sin^2\beta + \lambda_{\Phi\Delta} v_\Phi \cos^2\beta -\frac{1}{2}A \sin 2\beta \right\},
\\
&g_{f\bar{f}}^{\Delta^0} = -\sin\alpha;
\end{align*}
and all the loop functions or form factors used in the Higgs decays are collected in Section~\ref{app:LoopFunc}.

Finally, for masses slightly below the $t\bar t$ threshold, $\Delta^0$ can decay into one on–shell and one off-shell top quarks, $\Delta^0 \to tt^* \to tbW$. As in the MSSM or 2HDM, the below-threshold branching ratios can be significant only very close to the $t\bar t$ threshold. Therefore, this decay is negligible in the mass region of our interest and thus not considered further for brevity.

While we provide the leading order decay rates here, QCD corrections can be sizable. To estimate them, we use the higher-order corrections reported in the CERN Yellow Report 3~\cite{LHCHiggsCrossSectionWorkingGroup:2013rie} for $h \to b\bar{b},c\bar{c},\tau\tau,WW,ZZ$ and appropriately apply them to $\Delta^0$ decays for our numerical estimation.

The dominant decay rates for the triplet-like charged Higgs $\Delta^\pm$ are given by~\cite{Rizzo:1980gz,Keung:1984hn,Djouadi:1997rp,Djouadi:2005gi,Djouadi:2005gj}
\begin{align*}
&\Gamma(\Delta^\pm \to ff^\prime) = \frac{N_cm_{\Delta^\pm}^3\sin^2\beta}{8\pi v_\Phi^2} \beta_{ff'}\left(\frac{m_f^2}{m_{\Delta^\pm}^2},\frac{m_{f'}^2}{m_{\Delta^\pm}^2}\right),
\\
&\Gamma(\Delta^\pm \to t^*\bar{b}/\bar{t^*}b \to W^\pm b\bar{b}) = \frac{3m_t^4 m_{\Delta^\pm} \sin^2\beta}{128\pi^3 v_\Phi^4} \beta_t\left(\frac{m_t^2}{m_{\Delta^\pm}^2},\frac{m_W^2}{m_{\Delta^\pm}^2}\right)
\\
&\Gamma(\Delta^\pm \to W^\pm Z) =\frac{\lambda_{\Delta^\pm W^\mp Z}^2}{16\pi m_{\Delta^\pm}}\left[\beta\left(\frac{m_W^2}{m_{\Delta^\pm}^2},\frac{m_Z^2}{m_{\Delta^\pm}^2}\right)\right]^{1/2}\left[2+\frac{m_{\Delta^\pm}^4}{4m_W^2m_Z^2}\left(1-\frac{m_W^2}{m_{\Delta^\pm}^2}-\frac{m_Z^2}{m_{\Delta^\pm}^2}\right)^2\right],
\\
&\Gamma(\Delta^\pm \to W^\pm Z^*) = \frac{9g^2\lambda_{\Delta^\pm W^\mp Z}^2}{128\pi^3\cos^2\theta_w m_{\Delta^\pm}} \left(\frac{7}{12}-\frac{10}{9}\sin^2\theta_W+\frac{40}{27}\sin^4\theta_W\right) H\left(\frac{m_W^2}{m_{\Delta^\pm}^2},\frac{m_Z^2}{m_{\Delta^\pm}^2}\right),
\\
&\Gamma(\Delta^\pm \to W^{\pm*} Z) = \frac{9g^2\lambda_{\Delta^\pm W^\mp Z}^2}{256\pi^3 m_{\Delta^\pm}} H\left(\frac{m_Z^2}{m_{\Delta^\pm}^2},\frac{m_W^2}{m_{\Delta^\pm}^2}\right),
\\
&\Gamma(\Delta^\pm \to h W^\pm) = \frac{m_{\Delta^\pm}^3}{16\pi m_W^2} \lambda_{\Delta^\pm hW^\mp}^2 \left[\beta\left(\frac{m_W^2}{m_{\Delta^\pm}^2},\frac{m_{h}^2}{m_{\Delta^\pm}^2}\right)\right]^{3/2},
\\
&\Gamma(\Delta^\pm \to h W^{\pm *}) = \frac{9g^2m_{\Delta^\pm}}{128\pi^3} \lambda_{\Delta^\pm hW^\mp}^2 G\left(\frac{m_{h}^2}{m_{\Delta^\pm}^2},\frac{m_W^2}{m_{\Delta^\pm}^2}\right),
\\
&\Gamma(\Delta^\pm \to \Delta^0 W^{\pm *}) = \frac{9g^2m_{\Delta^\pm}}{128\pi^3} \lambda_{\Delta^0 \Delta^\pm W^\mp}^2 G\left(\frac{m_{\Delta^0}^2}{m_{\Delta^\pm}^2},\frac{m_W^2}{m_{\Delta^\pm}^2}\right),
\end{align*}
where
\begin{align*}
&\lambda_{\Delta^\pm W^\mp Z} = -\frac{g^2}{2\cos\theta_w}\left(2v_\Delta\cos^2\theta_w\cos\beta-v_\Phi\sin^2\theta_w\sin\beta\right),
\\
&\lambda_{\Delta^\pm hW^\mp} = -\frac{g}{2}\left(2\sin\alpha \cos\beta + \cos\alpha \sin\beta \right).
\end{align*}

Having provided all the decay rate expressions, a brief discussion on their dependence on the free model parameters is in order. Note that, for the mass range of our interest, any of the heavy SM particles can be off-shell. However, for brevity, we omit the asterisk sign signaling this hereinafter. The tree-level decays of $\Delta^0$ depend on three parameters, namely $m_{\Delta^0}$, $v_\Delta$ and $\alpha$, while those of $\Delta^\pm$ depend on $m_{\Delta^\pm}$ and $v_\Delta$ only, except for $\Delta^\pm \to hW^\pm$ which also depends on $\alpha$. The cascade and loop-induced decays depend additionally on the mass-splitting $m_{\Delta^\pm}-m_{\Delta^0}$. As discussed in Sec.~\ref{sec:stability_unitarity}, the requirement of vacuum stability and perturbative unitarity, together with the EW precision data, restricts the mass-splitting to a few GeV. Therefore, on account of this small mass-splitting, the cascade decays are generally suppressed. Further, the dependence of the tree-level decays on $v_\Delta$ drops out as long as $v_\Delta \sim \mathcal{O}(1)$\,GeV. Consequently, for the $v_\Delta$-range favoured by $W$-mass measurements (see Sec.~\ref{sec:Wmass}), the tree-level decays are reasonably independent of $v_\Delta$. However, the loop-induced decays crucially depend on $v_\Delta$.

\begin{figure}[htb!]
\centering
\includegraphics[width=0.49\columnwidth]{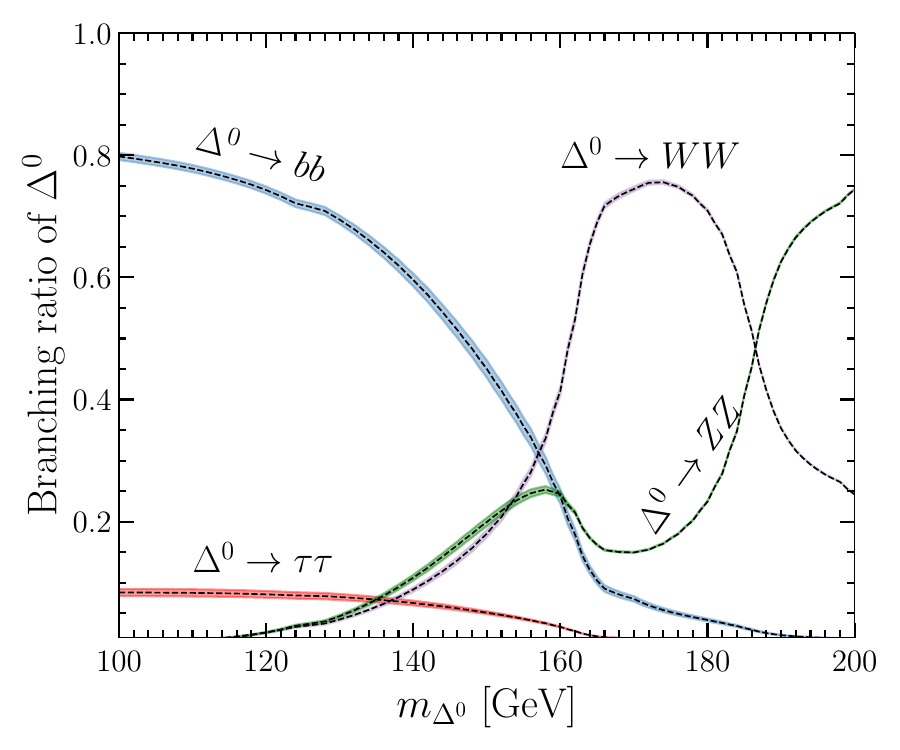} 
\includegraphics[width=0.49\columnwidth]{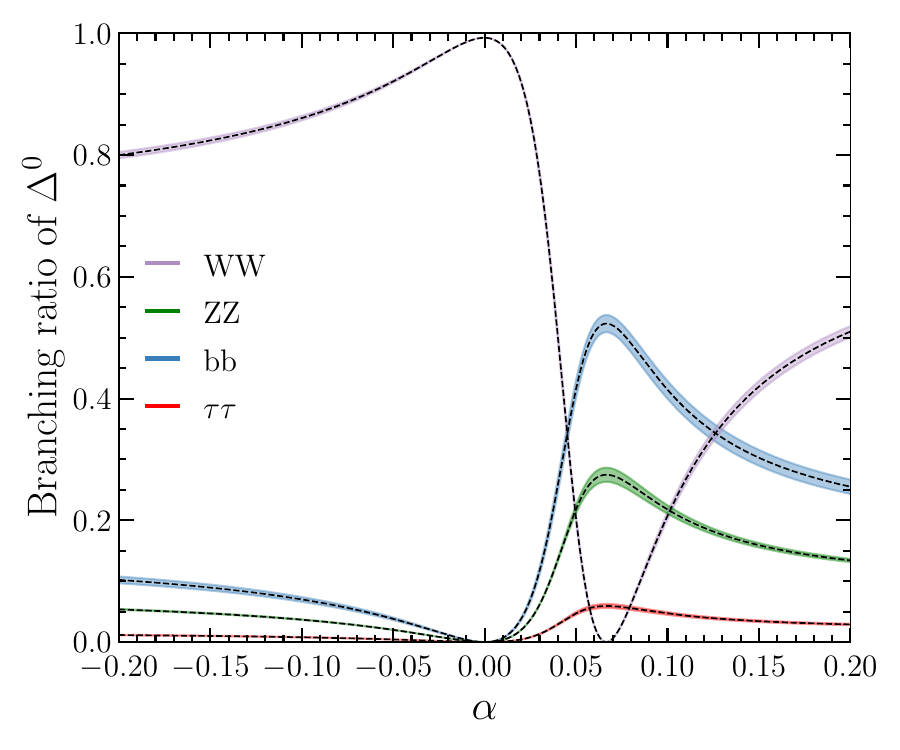} 
\caption{Dominant branching ratios of $\Delta^0$, including the uncertainties estimated from Ref.~\cite{LHCHiggsCrossSectionWorkingGroup:2013rie}, with $\alpha = 0.1$ for the left plot, and $m_{\Delta^0} = 152$\,GeV for the right one.}
\label{fig:BrH}
\end{figure}

In Fig.~\ref{fig:BrH}, we show the variation of the dominant branching ratios of the triplet-like Higgs $\Delta^0$, including the corresponding uncertainties, as a function of its mass $m_{\Delta^0}$ (left) and the mixing angle $\alpha$ (right). The uncertainties are estimated by propagating the errors on the $\tau\tau$, $b\bar b$, $WW^*$ and $ZZ^*$ decays of a hypothetical SM-like Higgs with mass $m_{\Delta^0}$ reported in the CERN Yellow Report~\cite{LHCHiggsCrossSectionWorkingGroup:2013rie}. For definiteness, we take $v_{\Delta} = 3.4$ GeV. Further, we take $\alpha = 0.1$ for the left plot and $m_{\Delta^0} = 152$\,GeV for the right one. While $\Delta^0$ exclusively decays into $WW$ for $\alpha \approx 0$, the other modes become relevant for $\alpha \neq 0$. In particular, for $\alpha > 0$, the $b\bar b$ mode dominates over the rest for $m_{\Delta^0}$ up to close to the $WW$ kinematic threshold, beyond which the gauge-boson modes $WW$ and $ZZ$ become the primary ones. While for $\alpha < 0$ (corresponding plot not shown for brevity), the $b\bar b$ and $WW$ modes are relevant, with the latter dominating over the former much before the $WW$ threshold. However, the $WW$ mode vanishes for $\tan\alpha \approx \frac{4v_\Delta}{v}$; see the right plot. Before moving further, a brief comment on the $\Delta^0 \to hh$ decay is in order. Though this mode is not suppressed by $\alpha$, it is suppressed by the kinematic phase space for $m_{\Delta^0} < 2m_h$, thus not relevant for the mass range of our interest.

\begin{figure}[htb!]
\centering
\includegraphics[width=0.49\columnwidth]{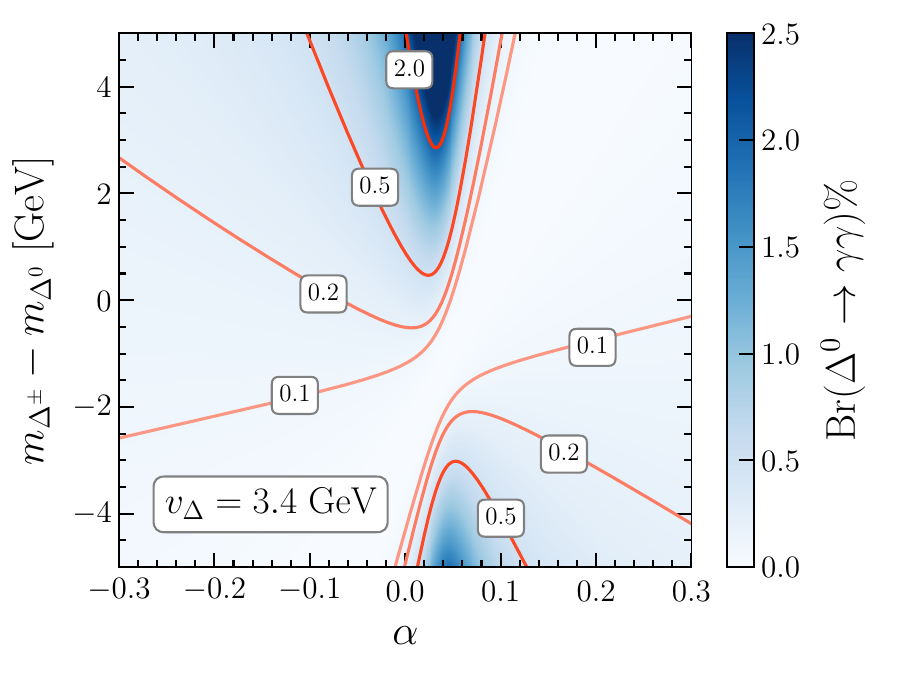} 
\includegraphics[width=0.49\columnwidth]{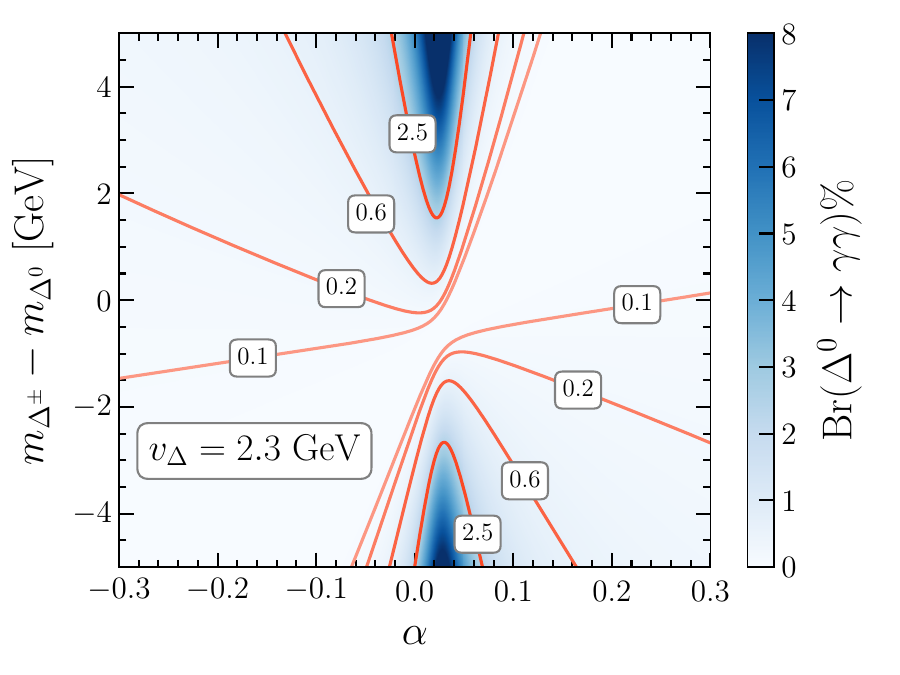} 
\caption{Br$(\Delta^0 \to \gamma\gamma)$ as a function of $\alpha$ vs $m_{\Delta^\pm} - m_{\Delta^0}$ for $m_{\Delta^0} = 152$\,GeV and $v_\Delta = 3.4$\,GeV (left) and $v_\Delta = 2.3$\,GeV (right).}
\label{fig:Haa}
\end{figure}

\begin{figure}[htb!]
\centering
\includegraphics[width=0.49\columnwidth]{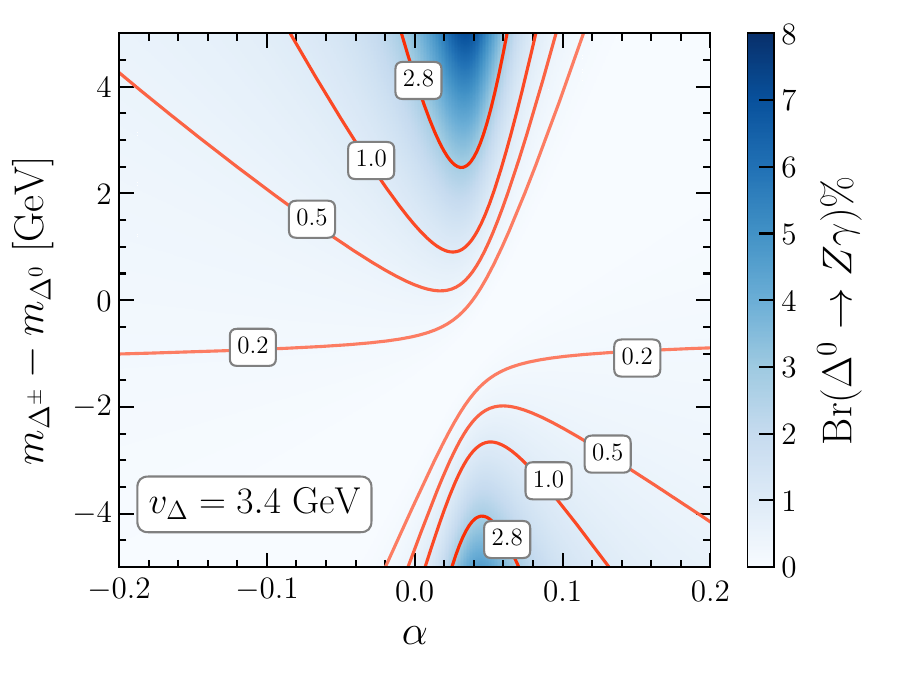} 
\includegraphics[width=0.49\columnwidth]{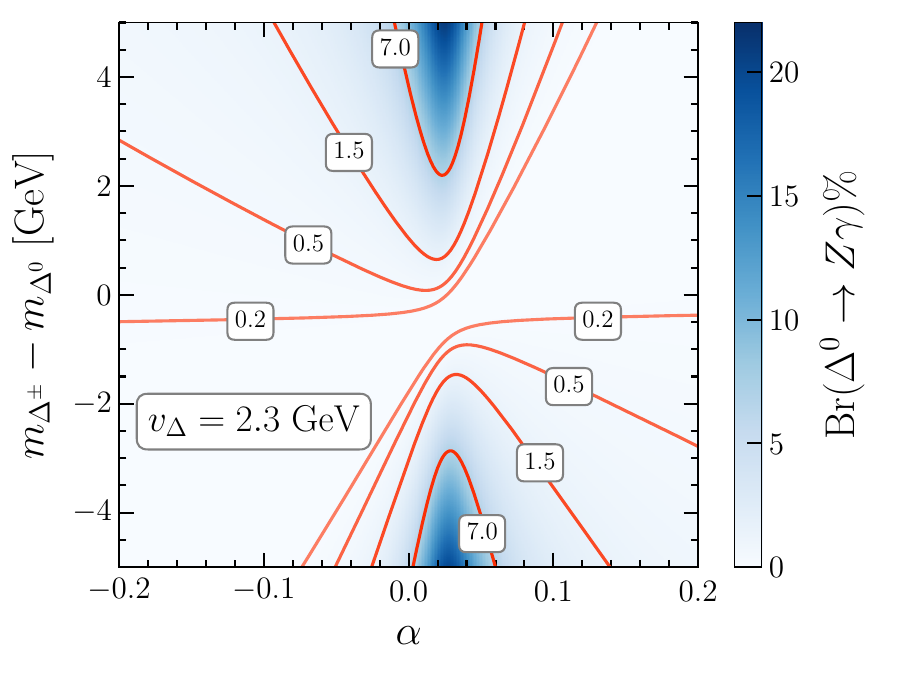} 
\caption{Br$(\Delta^0 \to Z\gamma)$ as a function of $\alpha$ vs $m_{\Delta^\pm} - m_{\Delta^0}$ for $m_{\Delta^0} = 152$\,GeV and $v_\Delta = 3.4$\,GeV (left) and $v_\Delta = 2.3$\,GeV (right).}
\label{fig:HZa}
\end{figure}

Figs.~\ref{fig:Haa} and \ref{fig:HZa} show Br$(\Delta^0 \to \gamma\gamma)$ and Br$(\Delta^0 \to Z\gamma)$ in the $\alpha$ vs $m_{\Delta^\pm} - m_{\Delta^0}$ plane for $m_{\Delta^0} = 152$\,GeV and $v_\Delta = 3.4$\,GeV (left) and $v_\Delta = 2.3$\,GeV (right). As we see, in the vicinity of degenerate mass-spectrum for the triple-like Higgs states, both the $\Delta^0 \to \gamma\gamma$ and $\Delta^0 \to Z\gamma$ branching ratios are $\sim \mathcal{O}(0.1)$\%--$\mathcal{O}(1)$\%.

\begin{figure}[htb!]
\centering
\includegraphics[width=0.65\columnwidth]{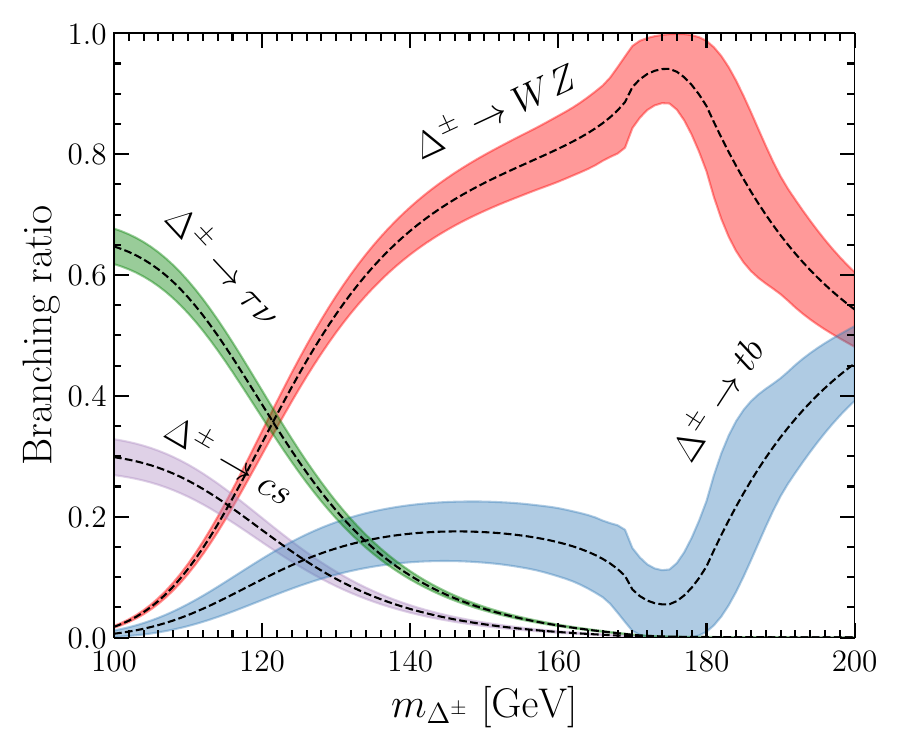} 
\caption{Dominant branching ratios of $\Delta^\pm$, including the uncertainties estimated from Ref.~\cite{LHCHiggsCrossSectionWorkingGroup:2013rie}, as a function of its mass. Note that the dependence on $v_\Delta$ drops out.}
\label{fig:BrHpm}
\end{figure}

In Fig.~\ref{fig:BrHpm}, the dominant branching ratios of the charged Higgs $\Delta^\pm$, including the corresponding uncertainties, are shown as a function of its mass $m_{\Delta^\pm}$. Once again, the uncertainties are estimated by propagating the errors on the $\tau\tau$, $c\bar c$, $t\bar t$ and $ZZ^*$ decays of a hypothetical SM-like Higgs with mass $m_{\Delta^\pm}$ reported in the CERN Yellow Report~\cite{LHCHiggsCrossSectionWorkingGroup:2013rie}. For definiteness, we take $v_{\Delta} = 3.4$ GeV and $\alpha = 0$. For low $m_{\Delta^\pm}$, the most dominant decay mode is $\tau \nu$. While for intermediate $m_{\Delta^\pm}$, which is also the mass range of our interest, the $WZ$ mode dominates over the rest, with the $t\bar b$ mode dominating for high $m_{\Delta^\pm}$. 

Finally, a brief comment on the $\Delta^\pm \to hW^\pm$ decay is in order. As mentioned earlier, among all decays of $\Delta^\pm$, only the $hW^\pm$ mode depends on $\alpha$. This decay is suppressed by $\alpha$ as well as the kinematic phase space through the coupling $\lambda_{\Delta^\pm hW^\mp}$, and thus not shown in the plot. In fact, it vanishes for $\tan\alpha \approx \frac{v_\Delta}{v}$. That said, this can be relevant for sizable $\alpha$ and $m_{\Delta^0} > m_h$.

\section{Phenomenology}
\label{sec:pheno}

\subsection{$W$ mass}
\label{sec:Wmass}
The $W$ mass $m_W = 80.4335(94)\,$GeV as measured by the CDF~II collaboration~\cite{CDF:2022hxs} is significantly larger than the ATLAS measurement $m_W = 80.360(16)$~\cite{ATLAS:2023fsi} as well as the LHCb measurement $m_W = 80.354(32)$~\cite{LHCb:2021bjt}. Combining these measurements with those from the D0 experiment~\cite{Abazov:2009cp,D0:2013jba,LHCb:2021bjt,Aaboud:2017svj} at the Tevatron, and the ALEPH, DELPHI, L3 and OPAL experiments at the LEP~\cite{ALEPH:2013dgf}, the LHC-TeV MW Working Group~\cite{LHC-TeVMWWorkingGroup:2023zkn} has obtained a world average
\begin{align}
m_W^{\rm avg} = 80394.6 \pm 11.5{\rm\,MeV}.
\end{align}
Comparing this with the SM prediction of $m_W^{\rm SM} = 80.3499(56)\,$GeV~\cite{deBlas:2022hdk}, the discrepancy of $44.7\,$MeV amounts to $3.5\sigma$. Since there is a tension between the most precise CDF-II measurement and the rest, removing the former increases the compatibility within the $m_W$ fit and leads to a lower average of
\begin{align}
m_W^{\rm avg~(w/o~CDF~II)} = 80369.2 \pm 13.3{\rm\,MeV},
\end{align}
is obtained. Therefore, the world average, including the CDF-II measurement, requires $v_\Delta = 3.4 \pm 1.0$\,GeV, while excluding CDF-II, one finds $v_\Delta = 2.3 \pm 1.7$\,GeV.\footnote{There is a recent update on the measurement of $m_W$ by ATLAS which finds a slightly heavier $m_W = 80.3665(159)$\,GeV~\cite{ATLAS:2024erm} compared to their previous analysis~\cite{ATLAS:2023fsi} based on the same dataset. However, the said combination did not consider the updated ATLAS measurement. Furthermore, after this combination was performed, CMS published a new measurement $m_W = 80360.2 \pm 9.9$\,GeV~\cite{CMS:2024nau}, which agrees well with the world average excluding the CDF-II measurement. Including these updates might significantly reduce the world average and thus would imply a slightly lower value for $v_\Delta$. However, the specific value for $v_\Delta$ is immaterial for our analysis as long as $v_\Delta \sim \mathcal{O}(1)$\,GeV.} In the rest of this work, we fix $v_\Delta = 3.4\,$GeV or $2.3\,$GeV.
\subsection{Multi-lepton signature from DY production}
\label{sec:multilep}

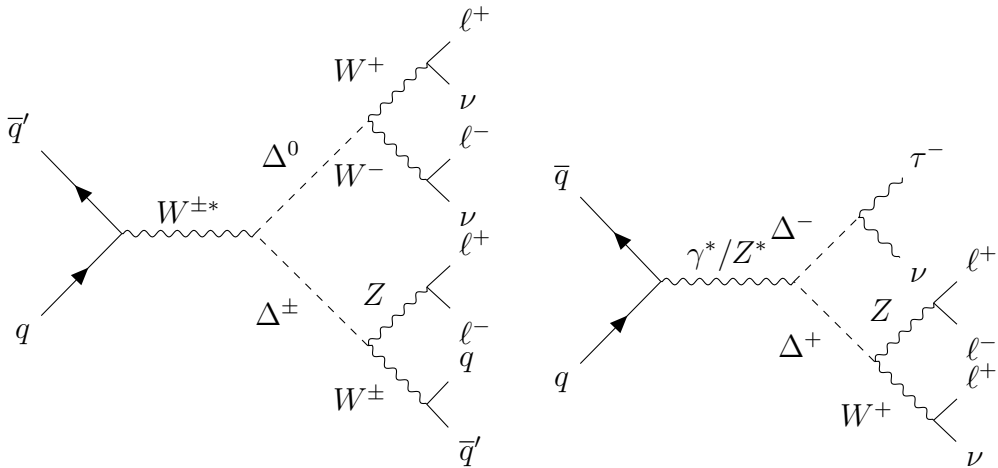
\begin{figure}[htb!]
\centering
\begin{tikzpicture}
\begin{feynman}
\vertex (a);
\vertex [above left=1.5cm of a] (c) {$\overline{q}^\prime$};
\vertex [below left=1.5cm of a] (d) {$q$};
\vertex [right=1.75cm of a] (b);
\vertex [above right=2.1cm of b] (e);
\vertex [below right=2.1cm of b] (f);
\vertex [above right=1.1cm of e] (i);
\vertex [below right=1.1cm of e] (j);
\vertex [above right=1.1cm of f] (k);
\vertex [below right=1.1cm of f] (l);
\vertex [above right=0.4cm of i] (i1) {$\ell^+$};
\vertex [below right=0.4cm of i] (i2) {$\nu$};
\vertex [above right=0.4cm of j] (j1) {$\ell^-$};
\vertex [below right=0.4cm of j] (j2) {$\nu$};
\vertex [above right=0.4cm of k] (k1) {$\ell^+$};
\vertex [below right=0.4cm of k] (k2) {$\ell^-$};
\vertex [above right=0.4cm of l] (l1) {$q$};
\vertex [below right=0.4cm of l] (l2) {$\overline{q}^\prime$};
\diagram{
(d) -- [fermion] (a) -- [fermion] (c);
(a) -- [boson, edge label=$W^{\pm *}$] (b);
(f) -- [scalar, edge label=$\Delta^\pm$] (b) -- [scalar, edge label=$\Delta^0$] (e);
(j) -- [boson, edge label=$W^-$] (e) -- [boson, edge label=$W^+$] (i);
(l) -- [boson, edge label=$W^\pm$] (f) -- [boson, edge label=$Z$] (k);
(i1) -- (i) -- (i2);
(j1) -- (j) -- (j2);
(l1) -- (l) -- (l2);
(k1) -- (k) -- (k2);
};
\end{feynman}
\end{tikzpicture}
\quad
\begin{tikzpicture}
\begin{feynman}
\vertex (a);
\vertex [above left=1.5cm of a] (c) {$\overline{q}$};
\vertex [below left=1.5cm of a] (d) {$q$};
\vertex [right=1.75cm of a] (b);
\vertex [above right=1.2cm of b] (e);
\vertex [below right=1.5cm of b] (f);
\vertex [above right=0.75cm of e] (i) {$\tau^-$};
\vertex [below right=0.75cm of e] (j) {$\nu$};
\vertex [above right=1.1cm of f] (k);
\vertex [below right=1.1cm of f] (l);
\vertex [above right=0.4cm of k] (k1) {$\ell^+$};
\vertex [below right=0.4cm of k] (k2) {$\ell^-$};
\vertex [above right=0.4cm of l] (l1) {$\ell^+$};
\vertex [below right=0.4cm of l] (l2) {$\nu$};

\diagram{
(d) -- [fermion] (a) -- [fermion] (c);
(a) -- [boson, edge label=$\gamma^*/Z^*$] (b);
(f) -- [scalar, edge label=$\Delta^+$] (b) -- [scalar, edge label=$\Delta^-$] (e);
(j) -- [boson] (e) -- [boson] (i);
(l) -- [boson, edge label=$W^+$] (f) -- [boson, edge label=$Z$] (k);
(l1) -- (l) -- (l2);
(k1) -- (k) -- (k2);
};
\end{feynman}
\end{tikzpicture}
\caption{Representative Feynman diagrams for the Drell-Yan production of the triplet-like Higgses and their decays to SM particles leading to multi-lepton final states at the LHC.}
\label{fig:Feyn_multilep}
\end{figure}

Drell-Yan production of the triplet-like Higgs bosons, $pp\to \Delta^0\Delta^\pm$ and $pp\to \Delta^+\Delta^-$, with their decays $\Delta^0 \to W^+W^-, ZZ, b\bar b, \tau^+\tau^-$ and $\Delta^\pm \to W^\pm Z, t\bar b, c\bar s, \tau^\pm\nu$ lead to final states with multiple charged leptons. Two representative Feynman diagrams for these processes are shown in Fig.~\ref{fig:Feyn_multilep}. 

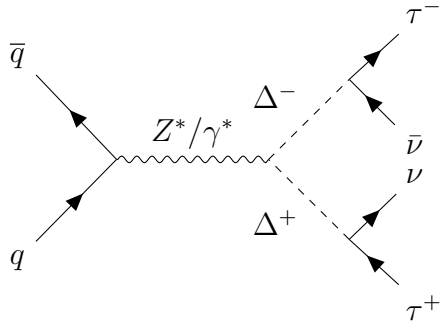
\begin{figure}[htb!]
\centering
\begin{tikzpicture}[baseline=(current bounding box.center)]%[baseline=(b.base)]
\begin{feynman}
\vertex (a);
\vertex [above left=1.5cm of a] (c) {$\overline{q}$};
\vertex [below left=1.5cm of a] (d) {$q$};
\vertex [right=2.0cm of a] (b) ;
\vertex [above right=1.5cm of b] (e);
\vertex [below right=1.5cm of b] (f);
\vertex [above right=0.85cm of e] (i) {$\tau^-$};
\vertex [below right=0.85cm of e] (j) {$\Bar{\nu}$};
\vertex [above right=0.85cm of f] (k) {$\nu$};
\vertex [below right=0.85cm of f] (l) {$\tau^+$};
\diagram{
(d) -- [fermion] (a) -- [fermion] (c);
(a) -- [boson, edge label=$Z^{*}/\gamma^{*}$] (b);
(f) -- [scalar, edge label=$\Delta^+$] (b) -- [scalar, edge label=$\Delta^-$] (e);
(j) -- [fermion] (e) -- [fermion] (i);
(l) -- [fermion] (f) -- [fermion] (k);
};
\end{feynman}
\end{tikzpicture}
\caption{Pair production and subsequent decays of $\Delta^\pm$ leading to the stau-like $\tau\tau\nu\nu$ final state.}
\label{fig:FeynStau}
\end{figure}

As discussed in Sec.~\ref{sec:tripletDecays}, low-mass triplet-like charged Higgs $\Delta^\pm$ dominantly decays to $\tau\nu$; see Fig.~\ref{fig:BrHpm}. Consequently, their pair production and subsequent decays, as shown in Fig.~\ref{fig:FeynStau}, leads to a final state with a pair of $\tau$-leptons and missing transverse momentum, which is the identical collider signature as supersymmetric partners of $\tau$-leptons ($\widetilde \tau$) promptly decaying into a $\tau$ and a (massless) neutralino. This has been searched for by CMS and ATLAS collaborations~\cite{ATLAS:2019gti,CMS:2019eln,CMS:2022rqk,ATLAS:2024fub}.

Using the full run 2 dataset at the 13\,TeV LHC, CMS~\cite{CMS:2022rqk} has obtained 95\% confidence level (CL) upper limits on the cross-section as a function of the $\widetilde \tau$ mass. On the contrary, ATLAS~\cite{ATLAS:2024fub} observed a stronger limit than expected; however, they do not provide upper limits on the cross-section. We thus use the CMS result in Ref.~\cite{CMS:2022rqk} to constrain the triplet-like charged Higgs $\Delta^\pm$. 

\begin{figure}[htb!]
\centering
\includegraphics[width=0.65\columnwidth]{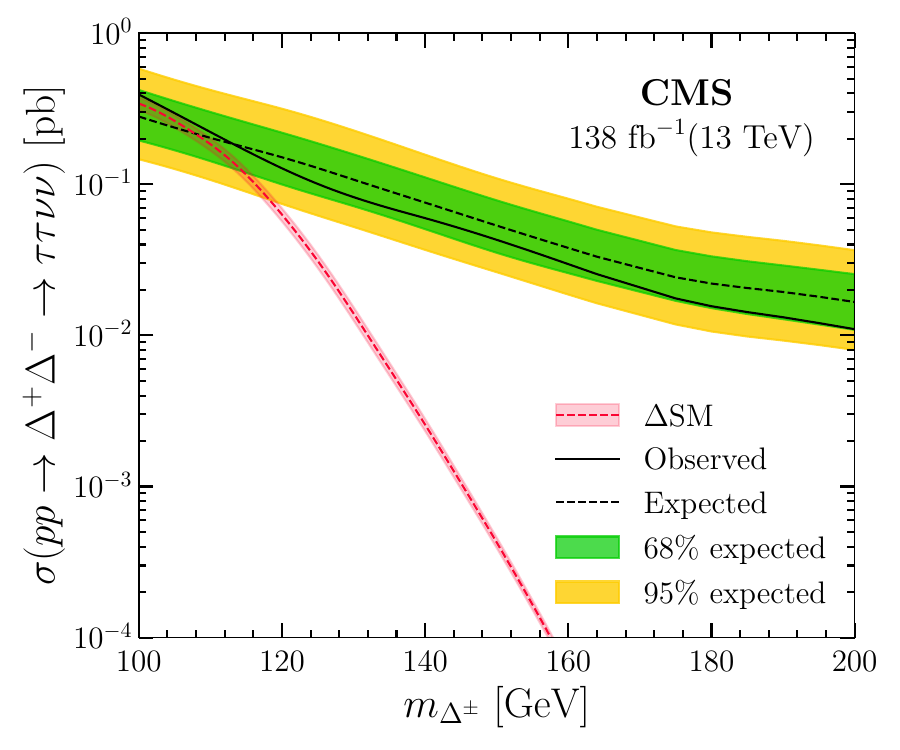} 
\caption{Expected and observed 95\% CL upper limits on the cross-section $\sigma(pp \to \Delta^+\Delta^- \to \tau\tau\nu\nu)$ taken from CMS~\cite{CMS:2022rqk}. The red line and band indicate the model prediction for the signal cross-section, including the NLO+NNLL QCD corrections and uncertainties.}
\label{fig:stau}
\end{figure}

The CMS expected and observed 95\% CL upper limits on the signal cross-section $\sigma(pp \to \Delta^+\Delta^- \to \tau\tau\nu\nu)$, taken from Ref.~\cite{CMS:2022rqk}, are shown in Fig.~\ref{fig:stau}. The inner green and outer yellow bands around the dotted line indicate the 68\% and 95\% CL regions. The red line and the thin-shaded band indicate the model prediction for the signal cross-section, including the NLO+NNLL QCD corrections and uncertainties. The QCD corrections are taken from Refs.~\cite{Ruiz:2015zca,Ajjath:2023ugn}, and the uncertainties are estimated by propagating the ones on the cross-section and branching ratios reported in Refs.~\cite{Ruiz:2015zca,Ajjath:2023ugn} and Ref.~\cite{LHCHiggsCrossSectionWorkingGroup:2013rie}, respectively. This excludes $\Delta^\pm$ with masses below 110\,GeV at 95\% CL.

For higher masses of the triplet-like Higgses, i.e.~$\gtrsim 120$\,GeV, the decays to $WW$ and $ZW$ become relevant. The most relevant signatures stem from the subsequent decays to leptons, which, compared to hadronic final states, have a considerably smaller SM background and, thus, superior constraining power. Since there is no dedicated search for this signature of our model,\footnote{Several multi-lepton searches have been performed using the full run 2 data by CMS and ATLAS, see {\it e.g.} Refs.~\cite{CMS:2019lwf,CMS:2021zkl,ATLAS:2021yyr,ATLAS:2021wob,ATLAS:2022nmi,ATLAS:2022yhd,ATLAS:2022pbd}. However, most of these searches probe specific models and, like for instance Ref.~\cite{ATLAS:2022yhd} and Ref.~\cite{ATLAS:2021yyr} consider events with high lepton invariant masses and/or high missing transverse momenta. Being tuned for specific models, these searches exploit very distinctive event features, such as resonances in invariant-mass distributions and high missing transverse momenta, and are thus not expected to be sensitive in probing the $\Delta$SM.} Ref.~\cite{Butterworth:2023rnw} used several SM measurements implemented in the {\tt Contur} toolkit, and find that the four-lepton search by ATLAS~\cite{ATLAS:2021kog} is the most constraining one. They excluded $\Delta^0$ with mass in the 180--200 GeV range for small $\lambda_{\phi\Delta}$; while for sizeable $\lambda_{\phi\Delta}$, the constraints are easily avoided. Here, we will make use of the model-independent multi-lepton search by ATLAS~\cite{ATLAS:2021wob}, which entails inclusive event selection and thus covers a large phase space of signatures. This search considers 22 signal regions (SRs) and analyses them model-independently by putting limits on the visible cross-sections. 

The three and four charged-lepton events are categorised into several orthogonal SRs based on the number of leptons, the presence/absence of a lepton pair presumably originating from a $Z$-boson decay (on-$Z$/off-$Z$ lepton pair\footnote{A pair of same-flavour and oppositely charged leptons with di-lepton invariant mass within the 10\,GeV window of the $Z$-boson mass is referred to as on-$Z$ lepton pair. A lepton that does not form an on-$Z$ lepton pair with any other lepton in the event is called off-$Z$.}) and the missing transverse momentum: 
\begin{enumerate}[label=$\roman*)$,itemsep=1pt,parsep=1pt,topsep=1pt,partopsep=1pt]
\item $3\ell$ on-$Z$ $E_{T}^{\rm miss} < 50$ GeV, 
\item $3\ell$ on-$Z$ $E_{T}^{\rm miss} > 50$ GeV, 
\item $3\ell$ off-$Z$ $E_{T}^{\rm miss} < 50$ GeV,
\item $3\ell$ off-$Z$ $E_{T}^{\rm miss} > 50$ GeV,
\item $4\ell$ on-$Z$ $E_{T}^{\rm miss} < 50$ GeV,
\item $4\ell$ on-$Z$ $E_{T}^{\rm miss} > 50$ GeV,
\item $4\ell$ off-$Z$,
\end{enumerate}
with each $3\ell$($4\ell$) SRs further divided into four(two) bins of the invariant mass of the leptons: 0\,GeV--200\,GeV, 200\,GeV--400\,GeV, 400\,GeV--600\,GeV and $>600$\,GeV (0\,GeV--400\,GeV and $>400$\,GeV), thereby resulting in 22~SRs in total.

We recast this search to constrain the $\Delta$SM by simulating the processes $pp\to W^{\pm *} \to \Delta^0 \Delta^\pm$ and $pp \to \gamma^*/Z^* \to \Delta^+\Delta^-$ with $\Delta^0 \to W^+W^-, \, ZZ$ and $\Delta^\pm \to W^\pm Z, t\bar b, c\bar s, \tau^\pm\nu$ for $m_\Delta$ between 120\,GeV and 200\,GeV using the UFO modules generated from {\tt SARAH}~\cite{Staub:2013tta,Staub:2015kfa}\footnote{We also build the UFO model files at NLO using {\tt FeynRules}~\cite{Degrande:2011ua,Alloul:2013bka, Degrande:2014vpa}. While it is desirable to perform the simulations at NLO, in order to reduce the computational resources needed for our analysis, we perform them at LO and then naively scale the production cross-section to account for the NLO+NNLL QCD corrections as discussed in Sec.~\ref{sec:prodDecay}.} in {\tt MadGraph5\_aMC\_v3.5.3}~\cite{Alwall:2014hca,Frederix:2018nkq} with the {\tt NNPDF23\_nlo\_as\_0118\_qed} parton distribution function~\cite{Ball:2013hta}. The simulated parton-level events are then passed through the regular chain of tools, namely {\tt Pythia 8.3}~\cite{Sjostrand:2014zea} and {\tt Delphes 3.5.0}~\cite{deFavereau:2013fsa}, to simulate the effect of subsequent decays of the unstable particles, radiations, showering, fragmentation and hadronisation, various detector effects, and particle-level object reconstruction. 

For the selection of various objects, {\it viz.} photons, leptons (electrons and muons) and jets (including $\tau$-tagged and $b$-tagged jets), we meticulously follow the said ATLAS search~\cite{ATLAS:2021wob}. In particular, we appropriately tune the {\tt Delphes} card for the ATLAS detector to take into account various object reconstruction, isolation and selection requirements and jet tagging efficiencies. We use the anti-$k_T$ algorithm~\cite{Cacciari:2008gp} implemented in {\tt FastJet 3.3.4}~\cite{Cacciari:2011ma} to reconstruct jets. The missing transverse momentum (with magnitude $E_{\rm T}^{\rm miss}$) is estimated from the momentum imbalance in the transverse direction associated with all reconstructed objects in an event. Finally, we apply the event selections and categorise the selected events according to the signal regions in Refs.~\cite{ATLAS:2021wob}.

While the branching ratio of the dominant $\Delta^\pm$ modes depends only on its mass, the partial widths of $\Delta^0\to WW,ZZ,bb,\tau\tau$ depend on $\alpha$ in addition (see Sec.~\ref{sec:tripletDecays}). While $\Delta^0$ exclusively decays to $WW$ for $\alpha \sim 0$, the $bb$ and $ZZ$ modes become relevant for non-zero $\alpha$. In fact, for $\alpha = \tan^{-1} (4 v_\Delta / v_\Phi)$, the rate to $WW$ can even be zero, as can be seen in Fig.~\ref{fig:BrH}. However, in this case, the decay to $ZZ$, which also leads to a sizable number of leptons, can be sizable. We will therefore consider the two limiting cases of Br$(\Delta^0\to WW) =1$ and Br$(\Delta^0\to WW)=0$, where in the latter case non-zero Br$(\Delta^0\to ZZ)$ is considered. We then interpolate the two limiting cases as a function of the triplet mass.

\subsubsection*{Results}
\begin{figure}[htb!]
\centering
\includegraphics[width=0.49\columnwidth]{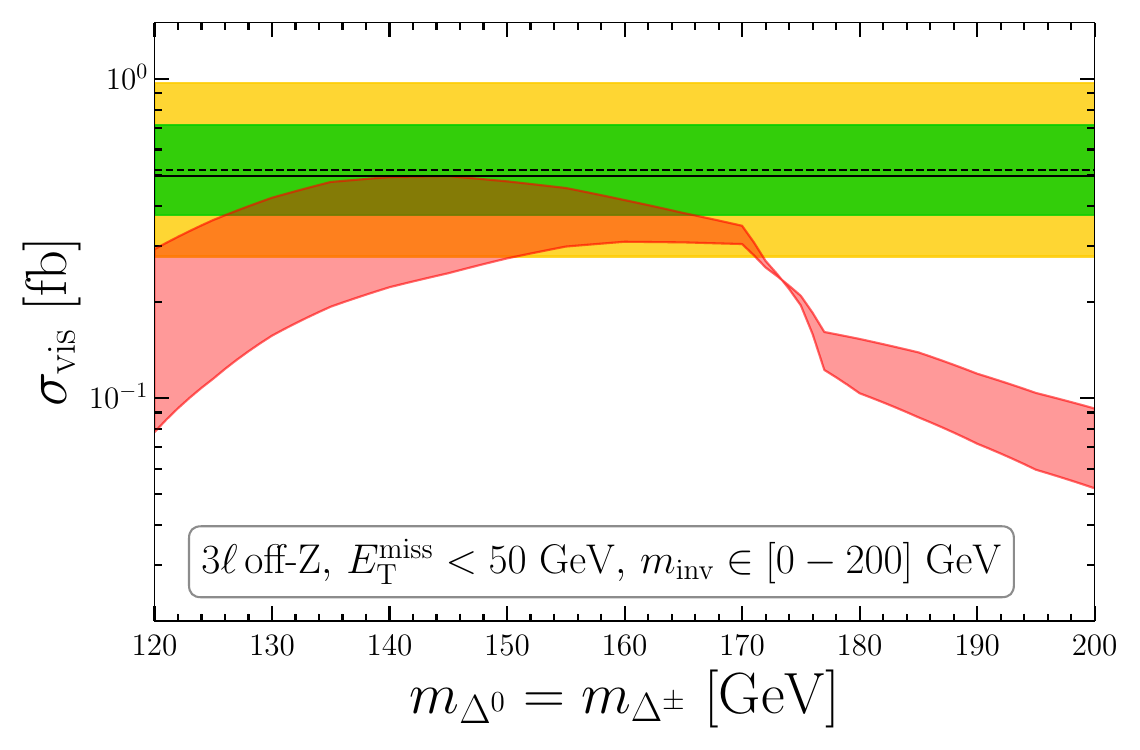}
\includegraphics[width=0.49\columnwidth]{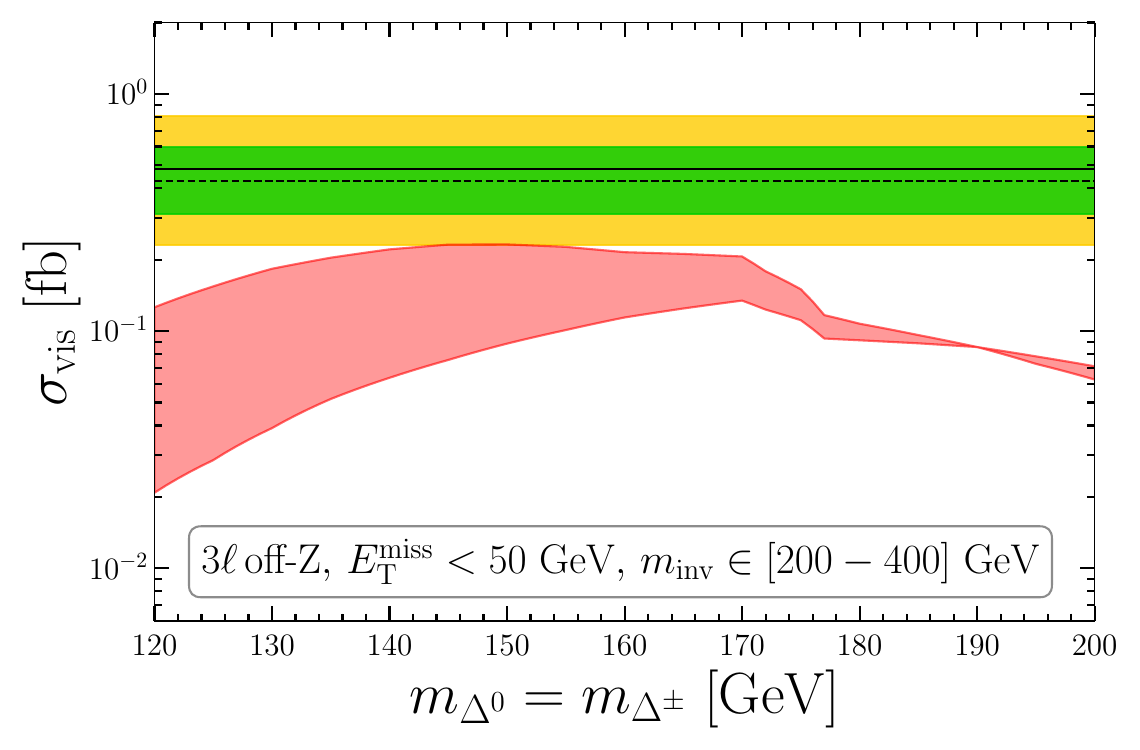}
\includegraphics[width=0.49\columnwidth]{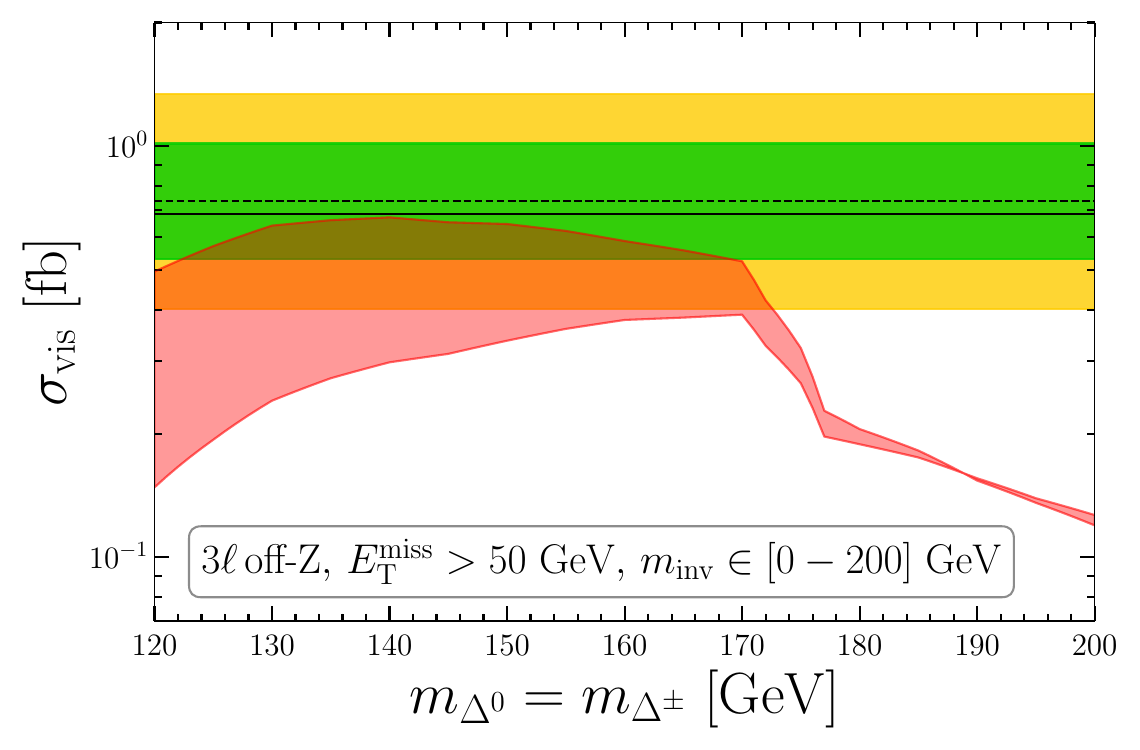}
\includegraphics[width=0.49\columnwidth]{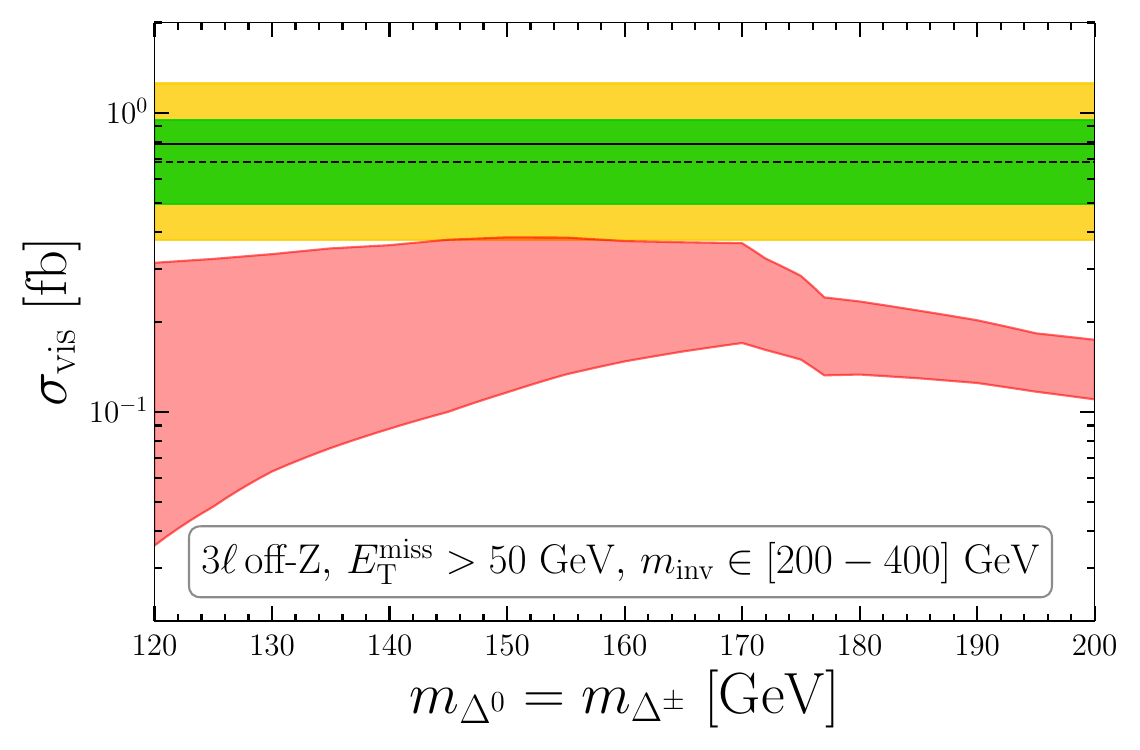}
\includegraphics[width=0.49\columnwidth]{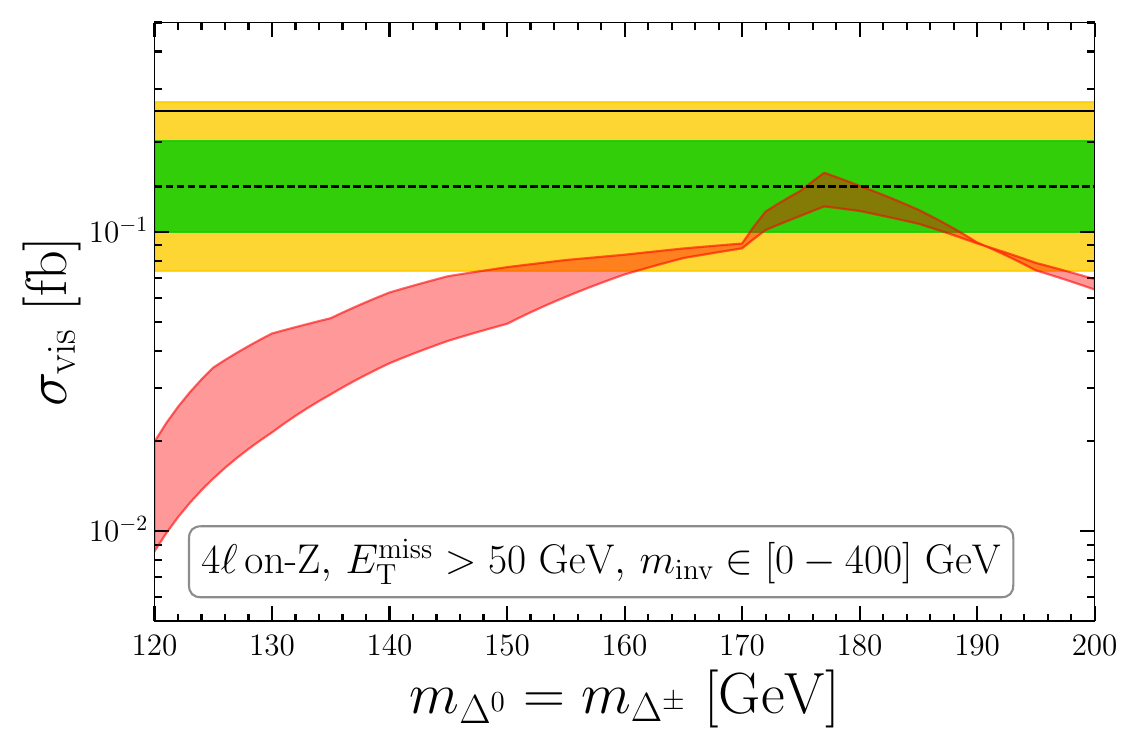}
\includegraphics[width=0.49\columnwidth]{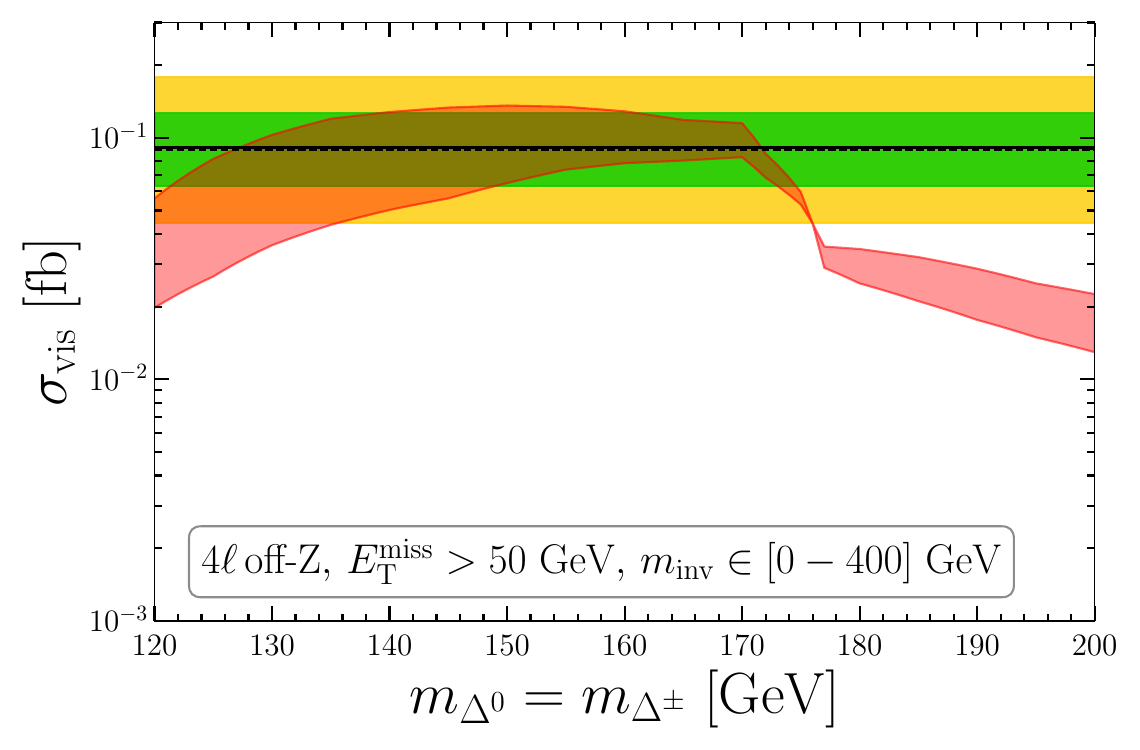}
\caption{The black solid (dashed) lines show the observed (expected) 95\% CL upper limits from ATLAS on the visible cross-section for different signal regions considered in Ref.~\cite{ATLAS:2021wob} and the green (yellow) band the expected $1\sigma$ ($2\sigma$) range. Our model prediction is shown as a red band, where the upper (lower) line of the band is obtained by simulating the processes $p p \to \Delta^\pm \Delta^\mp$ and $p p \to \Delta^\pm \Delta^0$ with Br$(\Delta^0 \to W W) = 100\% \, (0\%)$ and $\Delta^\pm$ decays according to the branching ratios shown in Fig.~\ref{fig:BrHpm}.}
\label{fig:multilepSR}
\end{figure}

In Fig.~\ref{fig:multilepSR}, we show the upper limits on the visible signal cross-section observed by ATLAS~\cite{ATLAS:2021wob} and the $\Delta$SM model predictions for the SRs which give the most relevant constraints. The dashed and solid lines, respectively, indicate the expected and observed limits. The inner green and outer yellow bands around the dashed lines indicate the regions containing 68\% and 95\% of the distribution of limits expected under the background-only hypothesis. The red-shaded bands indicate the $\Delta$SM model predictions for the corresponding signal cross-section, with the upper (lower) line of each band corresponding to a 100\%(0\%) branching ratio for $\Delta^0 \to WW$. One can see that while the predicted effect is, in many cases, close to the observed limits, the $\Delta$SM cannot be excluded with this model-independent multi-lepton search by ATLAS. However, run 3 and high-luminosity data~\cite{Cepeda:2019klc} will be able to test the $\Delta$SM in the mass region below 200\,GeV.

\subsection{Associated production of the triplet-like Higgs and resonant di-photon signatures: $\gamma\gamma + X$}
\label{sec:aaX}

\begin{figure}[htb!]
\centering
\begin{tikzpicture}[baseline=(current bounding box.center)]
\begin{feynman}
\vertex (a);
\vertex [above left=1.5cm of a] (c) {$\overline{q}^\prime$};
\vertex [below left=1.5cm of a] (d) {$q$};
\vertex [right=1.75cm of a] (b) ;
\vertex [above right=1.5cm of b] (e);
\vertex [below right=1.5cm of b] (f);
\vertex [above right=0.75cm of e] (i) {$\gamma$};
\vertex [below right=0.75cm of e] (j) {$\gamma$};
\vertex [above right=0.75cm of f] (k) {$x$};
\vertex [below right=0.75cm of f] (l) {$y$};
\diagram{
(d) -- [fermion] (a) -- [fermion] (c);
(a) -- [boson, edge label=$W^{\pm *}$] (b);
(f) -- [scalar, edge label=$\Delta^\pm$] (b) -- [scalar, edge label=$\Delta^0$] (e);
(j) -- [boson] (e) -- [boson] (i);
(l) -- (f) --  (k);
};
\end{feynman}
\end{tikzpicture}%\quad
\caption{The Drell-Yan production of $\Delta^0$ and $\Delta^\pm$ followed by their prompt decays $\Delta^0 \to \gamma\gamma$ and $\Delta^\pm \to xy$ ($xy \in WZ,tb,cs,\tau\nu$) leading to $\gamma\gamma+X$ final states.}
\label{fig:FeynaaX}
\end{figure}
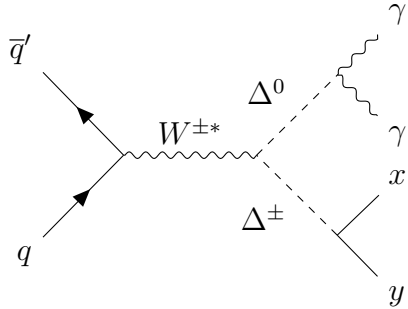

Drell-Yan production of the triplet-like Higgs states via an off-shell $W$-boson, i.e.~$pp\to W^*\to \Delta^0\Delta^\pm$ with $\Delta^0 \to \gamma\gamma$ leads to final states with a photon pair (di-photon) and additional particles and/or missing transverse momentum from the decay of  $\Delta^\pm$. This $\gamma\gamma+X$ signature is shown in Fig.~\ref{fig:FeynaaX}. The mass of $\Delta^0$ can be reconstructed from the di-photon invariant mass ($m_{\gamma\gamma}$) distribution.

In the $\Delta$SM model, $\Delta^0 \to \gamma\gamma$ has a branching ratio of $\sim \mathcal{O}(0.1)$\%--$\mathcal{O}(1)$\% (see Fig.~\ref{fig:Haa} and the relevant discussion in Sec.~\ref{sec:tripletDecays}). Though this decay rate is much lower than the ones for several other channels such as $WW,ZZ,bb,\tau\tau$ (see Fig.~\ref{fig:BrH}), the sensitivity of the di-photon final state is enhanced by the clearness of the corresponding detector signal as well as the smaller SM background.

An analysis of $\gamma\gamma+X$ has been performed by ATLAS, targeting the SM Higgs~\cite{ATLAS:2023omk}. This search considers 22 SRs categorized by $X$, the objects produced in association with the di-photon; see Table~3 of Ref.~\cite{ATLAS:2023omk} for the definitions of the SR. In addition, ATLAS has released a search for non-resonant di-Higgs production, which includes $\gamma\gamma+\tau$~\cite{ATLAS:2024lhu} not covered in the previous analysis. Both these searches provide $m_{\gamma\gamma}$ distributions in the 105\,GeV--160\,GeV range, thereby covering part of the mass range of our interest for the $\Delta$SM.

Because the ATLAS analyses performed a model-independent analysis of the associated production of the SM Higgs, no channels were combined, and the hypothesis of a BSM Higgs was not considered. However, the sidebands can be used to perform a search for BSM Higgses; within the $\Delta$SM, all signal regions are correlated so that they can be combined. 

For this, we simulate the process $pp\to W^{\pm *} \to (\Delta^0 \to \gamma\gamma)(\Delta^\pm \to WZ,tb,cs,\tau\nu)$ for 50 values of $m_\Delta$ in the 105\,GeV--157\,GeV range. As we did for the multi-lepton search in Sec.~\ref{sec:multilep}, we use the same set of tools to recast this search. Once again, we appropriately tune the delphes card for the ATLAS detector to take into account various object reconstruction, isolation and selection requirements. Following that, we apply the event selections and categorise the selected $\gamma\gamma+X$ events into several SRs as in Ref.~\cite{ATLAS:2023omk,ATLAS:2024lhu}.

\subsubsection*{Effects in the relevant SRs}
Given the dominant decay modes of $\Delta^\pm$ ($WZ,tb$ and $\tau\nu$) and their subsequent decays to leptons, quarks and neutrinos, we expect that the SRs targeting leptons ($\ell$), taus ($\tau$), multiple jets ($j$) and $E_T^{\rm miss}$ are the most sensitives ones. Those targeting top quarks are expected to become relevant for larger $m_{\Delta}$. Among the three SRs related to top quarks, the one targeting hadronically decaying top quarks is not sensitive to the signal of our model for the mass range of our interest because it is targeting hadronically decaying top quarks from $t\bar t$ production.\footnote{ATLAS uses a boosted decision tree (BDT) for top-quark reconstruction and requires a tight cut of 0.9 on the BDT score. The corresponding signal in the $\Delta$SM model consisting of a bottom quark and an (off-shell) top quark being quite different, the resulting efficiency is expected to be very small, thus rendering this SR irrelevant for the mass range of our interest.} 

Before proceeding, a few comments on SRs are in order. The $1\ell$ SR has been considered by both the Ref.~\cite{ATLAS:2023omk} and Ref.~\cite{ATLAS:2024lhu}. Since the latter applies a $b$-jet veto, this SR is nearly uncorrelated with the $\ell b$ SR considered in Ref.~\cite{ATLAS:2023omk}. This leads us to use the $1\ell$ SR of Ref.~\cite{ATLAS:2024lhu}. Furthermore, for both the $1\ell$ and $1\tau_{\rm had}$ SRs of Ref.~\cite{ATLAS:2024lhu}, they use a BDT to further categorise the events into three categories based on the BDT score. Since the corresponding weight files are not available yet, it would be a highly non-trivial task to appropriately implement the same BDTs for a little too gain in sensitivity. Therefore, instead of implementing the BDTs, we take a conservative approach to adding the event yields of all three BDT categories to recover the yields that would have resulted after the event selections (and before the BDTs). Concerning the $3\ell$ SR, since ATLAS has observed no event, we treat the entire $m_{\gamma\gamma}$ range as a single bin. Similarly, we do not show the $E_T^{\rm miss}>300$\,GeV SR since very few events have been observed and it is correlated to the other MET categories.\footnote{We checked that for our best-fit point for 152\,GeV, $\approx2$ events are predicted while $\approx$1 NP event is observed, showing the consistency.} Finally, while scaling by an overall $k$-factor of 1.15 at the cross-section level accounts for the NLO+NNLL QCD correction in the production of the triplet Higgs states, additional correction accrues from enhanced selection efficiencies. This is particularly important for the SRs targeting high jet activity because of gluon radiation. To estimate this, we simulate the corresponding signal at NLO and find an additional correction factor 1.2 for the $4j$ SR.

After simulating the effect in all signal regions given by ATLAS, taking into account the consideration above, it turns out that the 10 SRs listed in Table~\ref{tab:SRs} are relevant in the $\Delta$SM model for the mass range of our interest. This means that the other SRs lead to such weak constraints on Br$(\Delta^0\to \gamma\gamma)$ that disregarding them in a combined analysis has virtually no impact on the result.

\begin{table}[htb!]
\centering
\scalebox{0.8}{
\begin{tabular}{m{3.7cm}m{3cm}m{6.4cm}m{4.2cm}}
\toprule
\\[-.2cm]
Target & Signal region & Detector level selections & Correlation
\\[.2cm]
\midrule
\\[-.2cm]
High jet activity~\cite{ATLAS:2023omk} & $4j$ & $n_{\rm jet} \ge 4$, $|\eta_{\rm jet}| < 2.5$ & --
\\[.2cm]
\midrule
\\[-.2cm]
\multirow{2}{3.7cm}{Top~\cite{ATLAS:2023omk}} & $\ell b$ & $n_{\ell=e,\mu} \ge 1$, $n_{b\text{-jet}} \ge 1$ & --
\\[.2cm]
& $t_{\rm lep}$ & $n_{\ell=e,\mu} = 1$, $n_{\rm jet} = n_{b\text{-jet}} = 1$ & --
\\[.2cm]
\midrule
\\[-.2cm]
\multirow{3}{3.7cm}{Lepton~\cite{ATLAS:2023omk,ATLAS:2024lhu}} & $2\ell$ & $ee,\mu\mu$ or $e\mu$ & $ < 26\%$ ($1\ell$)
\\[.2cm]
& $3\ell$ & $n_{\ell=e,\mu} \geq 3$ &
\\[.2cm]
& $1\ell$ & $n_{\ell=e,\mu} = 1$, $n_{\tau_{\rm had}} = 0$, $n_{b\text{-jet}} = 0$, $E_{\rm T}^\text{\rm miss} > 35$~GeV (only for $e$-channel) & $ < 26\%$ ($2\ell$)
\\[.4cm]
\midrule
\\[-.4cm]
Tau \cite{ATLAS:2024lhu} & 1$\tau_{\rm had}$  & $n_{\ell=e,\mu} = 0$, $n_{\tau_{\rm had}} =1 $, $n_{b\text{-jet}}=0$, $E_{\rm T}^{\rm miss} > 35$~GeV & $-$
\\[.4cm]
\midrule
\\[-.4cm]
\multirow{3}{3.7cm}{$E_{\rm T}^{\rm miss}$~\cite{ATLAS:2023omk}} & $E_{\rm T}^{\rm miss} > 100$~GeV & $E_{\rm T}^{\rm miss} > 100$~GeV & $29\%$ ($E_{\rm T}^{\rm miss} > 200$ GeV)
\\[.2cm]
& $E_{\rm T}^{\rm miss} > 200$~GeV  & $E_{\rm T}^{\rm miss} > 200$~GeV & $29\%$ ($E_{\rm T}^{\rm miss} > 100$ GeV)
\\[.2cm]
& $E_{\rm T}^{\rm miss} > 300$~GeV  & $E_{\rm T}^{\rm miss} > 300$~GeV & --
\\[.2cm]
\bottomrule
\end{tabular}
}
\caption{The signal regions of the ATLAS analyses~\cite{ATLAS:2023omk,ATLAS:2024lhu} which are sensitive to the Drell-Yan production of the scalar triplet within our mass range of interest. $n_\ell$, $n_j$, $n_{b\text{-jet}}$, $n_{\tau_{\text{had}}}$ denotes the number of leptons, jets, $b$-tagged jets and $\tau$-tagged jets, respectively; and $\eta_{\rm jet}$ stands for the jet rapidity. Also, correlations ($>5$\%) among overlapping SRs are quoted.}
\label{tab:SRs}
\end{table}

\subsubsection*{Background (re)fitting}
To signals arising from BSM production, as in our case, the background consists of two components: resonant and continuum. The resonant component arises from the SM Higgs boson production and thus manifests itself as a narrow peak in the $m_{\gamma\gamma}$ spectrum. The continuum component, depending on the SRs, arises from the production of two initial/final-state photons, the
misidentification of jets and the production of EW bosons and top quarks. The continuum backgrounds are estimated from data by fitting the $m_{\gamma\gamma}$ spectrum in the mass range 105\,GeV--160\,GeV with an analytic function with free parameters. Note that those estimated by ATLAS in Refs.~\cite{ATLAS:2023omk,ATLAS:2024lhu} are obtained assuming there is only a single resonance at 125\,GeV, i.e. the SM Higgs. On the contrary, in the $\Delta$SM model, there is another resonance at $m_{\Delta^0}$, i.e. the triplet-like Higgs, albeit with different signal strength. This resonance manifests itself as a narrow peak in the $m_{\gamma\gamma}$ spectrum, much like the one at 125\,GeV. Consequently, we need to (re)estimate the continuum background from a (re)fit to the data. ATLAS follows the procedure {\it spurious-signal test}~\cite{ATLAS:2018hxb} to select a background function for a given SR from a number of candidate functions such as power-law functions, Bernstein polynomials, and exponential functions (of a polynomial), to ensure small potential bias in the extracted signal yield compared to the experimental precision. However, for the sake of simplicity, we use the following common function for all SRs (except the $3\ell$ SR, for which the yields are too small to fit a distribution):
\begin{equation}
\left(1 - \frac{m_{\gamma\gamma}}{\sqrt{s}}\right)^{b} \left(\frac{m_{\gamma\gamma}}{\sqrt{s}}\right)^{a_0 +  a_1\log (m_{\gamma\gamma}/\sqrt{s})},
\label{eq:refit}
\end{equation}
where $a_0, a_1$ and $b$ are the free parameters which are independent across SRs, $\sqrt{s} = 13$\,TeV is the LHC run 2 centre-of-mass energy. This function is fitted with the data subtracted by the resonances---SM Higgs at 125\,GeV and triplet-like Higgs at $m_{\Delta^0}$---to estimate the (re)fitted continuum background.

\subsubsection*{Statistical model}
The data are interpreted as briefly described below. A likelihood function is built from the number of observed and (re)fitted $\Delta$SM signal-plus-background events. Assuming the bins in the $m_{\gamma\gamma}$ spectrum for a given SR as independent number-counting experiments, the likelihood is modelled as
\begin{eqnarray}
\mathcal{L}_{\rm SR}(\mu) = \prod_{i} {\rm Poiss}\left(n^i|\mu s^i + b^i\right)
\end{eqnarray}
where $i$ runs over the 22 bins in the $m_{\gamma\gamma}$ spectrum in the 105\,GeV--160\,GeV range and $\mu$ represents the parameter of interest or signal strength: $\Delta^0 \to \gamma\gamma$ branching ratio, Br$(\Delta^0\to\gamma\gamma)$ in our case (this choice to be justified shortly). Here, as is typically done for number-counting experiments, the Poisson distribution is used to compare the measured data $n$ with the modelled expectation comprising of a signal yield $s$ and a background yield $b$. Then, the global likelihood $\mathcal{L}(\mu)$ for the $\gamma\gamma+X$ measurements is obtained as the product of the likelihood functions for the relevant SRs. Finally, the profile likelihood ratio test statistics~\cite{Cowan:2010js} is given by
\begin{align*}
\Lambda(\mu) = -2\log\frac{\mathcal{L}(\mu)}{\mathcal{L}\left(\hat\mu\right)}
\end{align*}
where $\hat\mu$ refers to the value of $\mu$ that maximises the likelihood. This test statistics asymptotically follows a $\chi^2$ distribution such that approximate 68\% and 95\% CL ($1\sigma$ and $2\sigma$) intervals can be easily constructed by requiring $\Lambda(\mu) = 1$ and $\Lambda(\mu) = 4$, respectively. Note that we allow for an unphysical negative $\mu$ to take into account the effect of downward fluctuations of the background.

\subsubsection*{Results}

\begin{figure}[htb!]
\centering
\includegraphics[width=0.49\columnwidth]{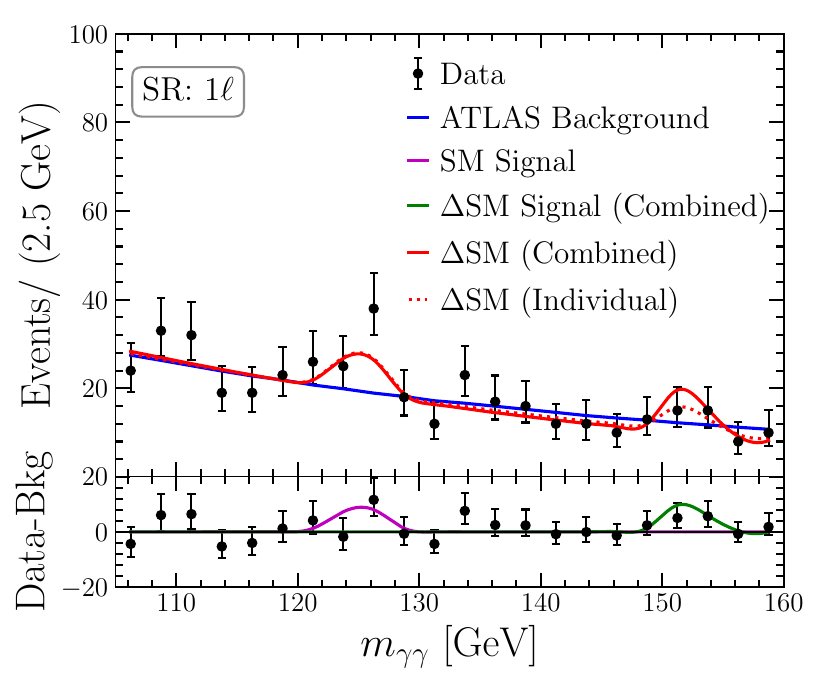}
\includegraphics[width=0.49\columnwidth]{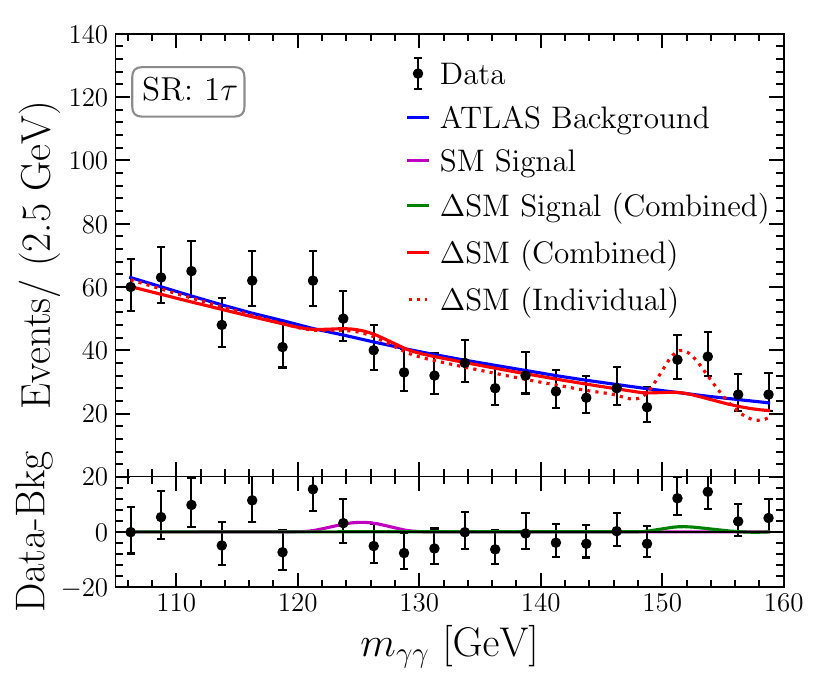}
\includegraphics[width=0.49\columnwidth]{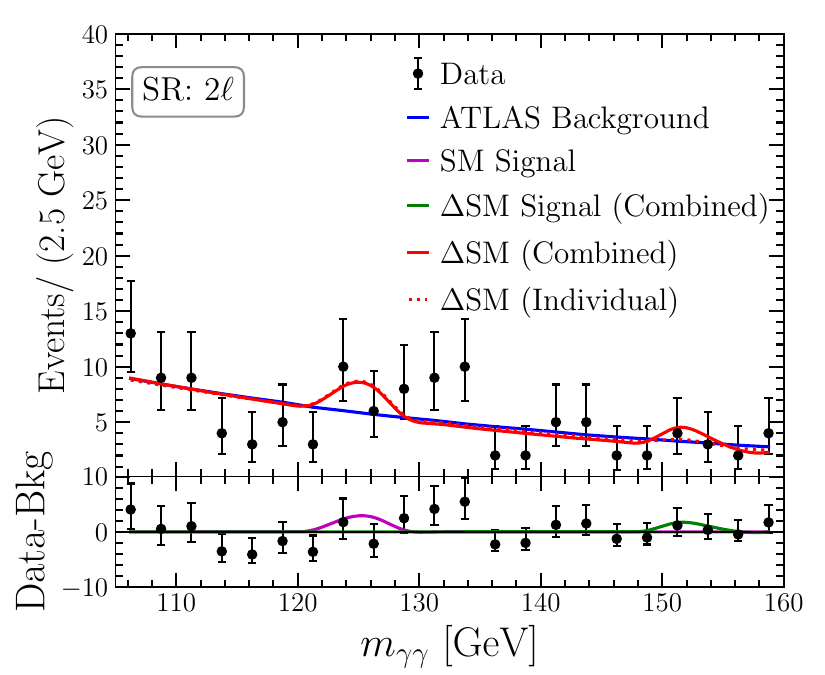}
\includegraphics[width=0.49\columnwidth]{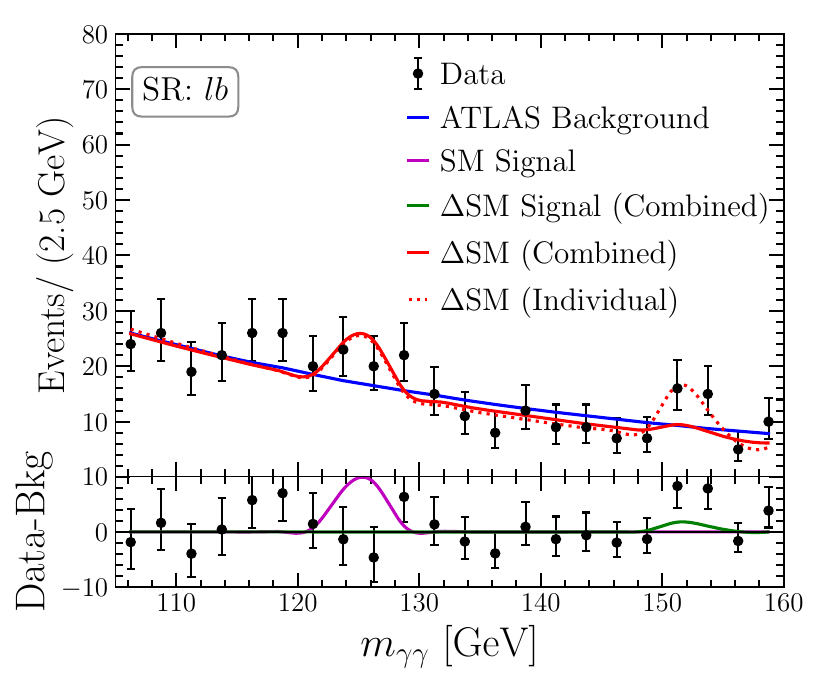}
\includegraphics[width=0.49\columnwidth]{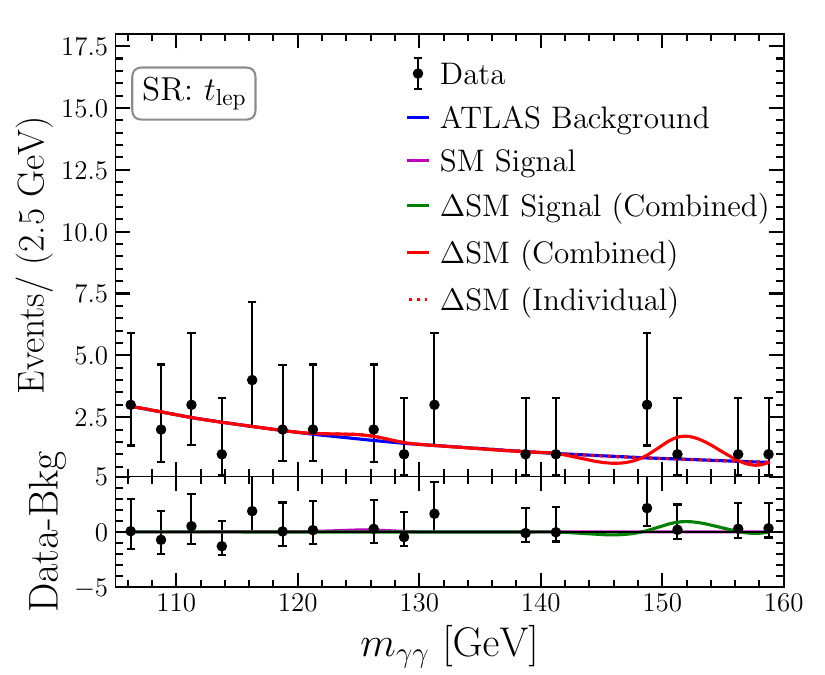}
\includegraphics[width=0.49\columnwidth]{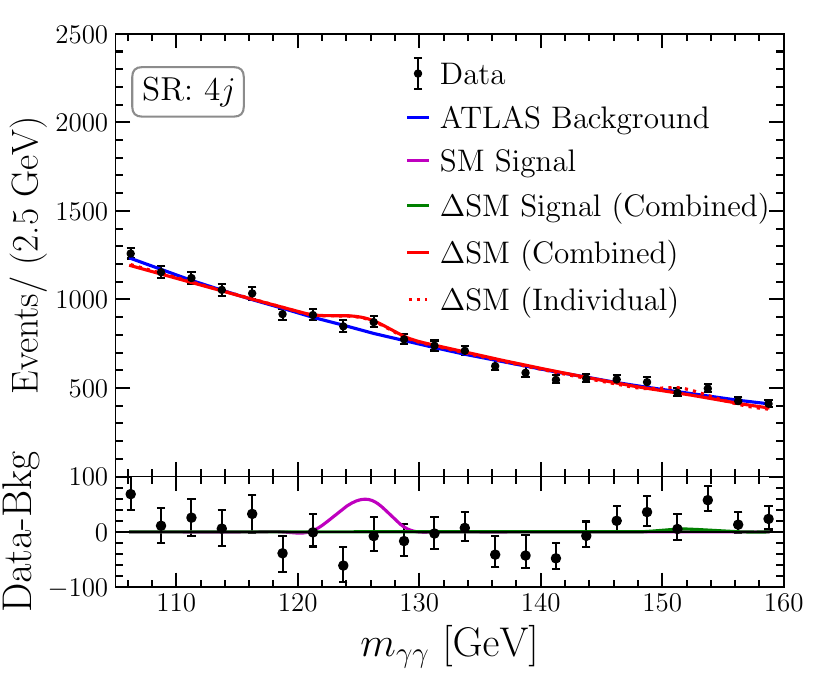}
\caption{Di-photon invariant mass distributions for the relevant signal regions. The data and corresponding uncertainties (indicated by black points and vertical bars) are shown together with the continuum background (blue line) from the ATLAS analyses and the total $\Delta$SM events (red line). The latter is comprised of the (re)fitted background (not shown for brevity), the SM Higgs signal at 125\,GeV (magenta line) and the $\Delta$SM triplet Higgs signal at 152\,GeV (green line).}
\label{fig:aaX_maa1}
\end{figure}

\begin{figure}[htb!]
\centering
\includegraphics[width=0.49\columnwidth]{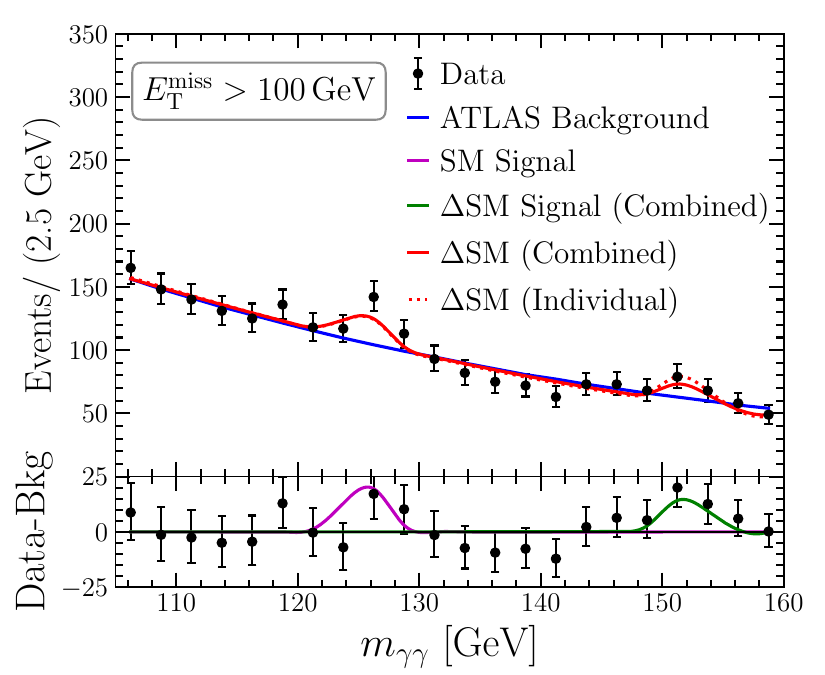}
\includegraphics[width=0.49\columnwidth]{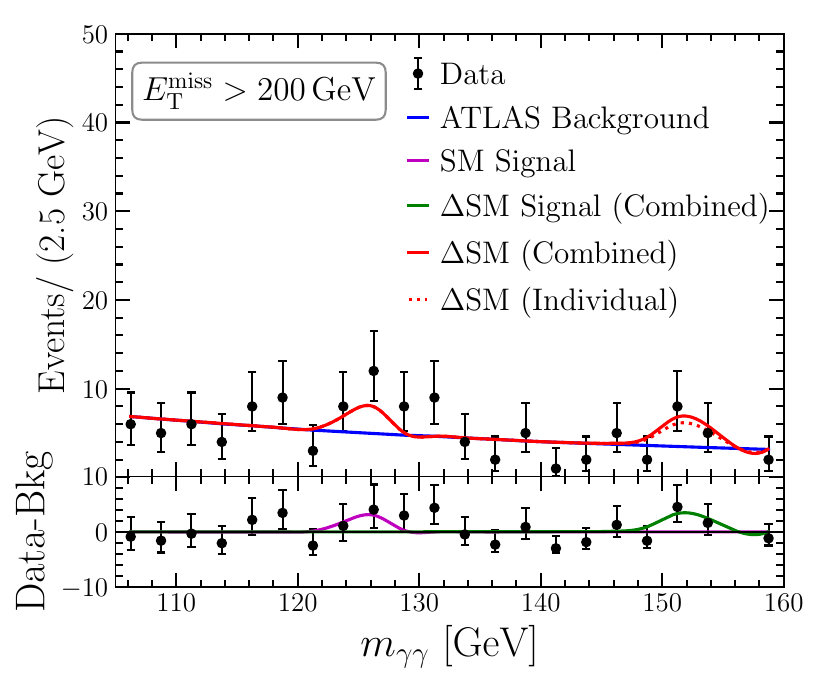}
\caption{Di-photon invariant mass distributions for the relevant signal regions (continued). The description is the same as of Fig.~\ref{fig:aaX_maa1}.}
\label{fig:aaX_maa2}
\end{figure}

In Figs.~\ref{fig:aaX_maa1} and \ref{fig:aaX_maa2}, we show the fit to the di-photon invariant mass distributions for the relevant signal regions listed in Table~\ref{tab:SRs}. The data and continuum backgrounds taken from the ATLAS analyses are shown in black (points with error bars) and blue, respectively. Also shown are the 125 GeV SM Higgs signals (magenta) and 152 GeV $\Delta$SM triplet Higgs signals with Br$(\Delta^0\to\gamma\gamma) = 0.7\%$ (green). For brevity, we do not show the (re)fitted continuum backgrounds; the (re)fitted $\Delta$SM signal-plus-backgrounds are shown in red. Note that the $\Delta$SM benchmark chosen here is occasioned by our findings later in this section that there is a strong preference for Br$(\Delta^0\to\gamma\gamma) \approx 0.7\%$ at 152\,GeV when all relevant SRs are statistically combined.

\begin{figure}[htb!]
\centering
\includegraphics[width=0.9\columnwidth]{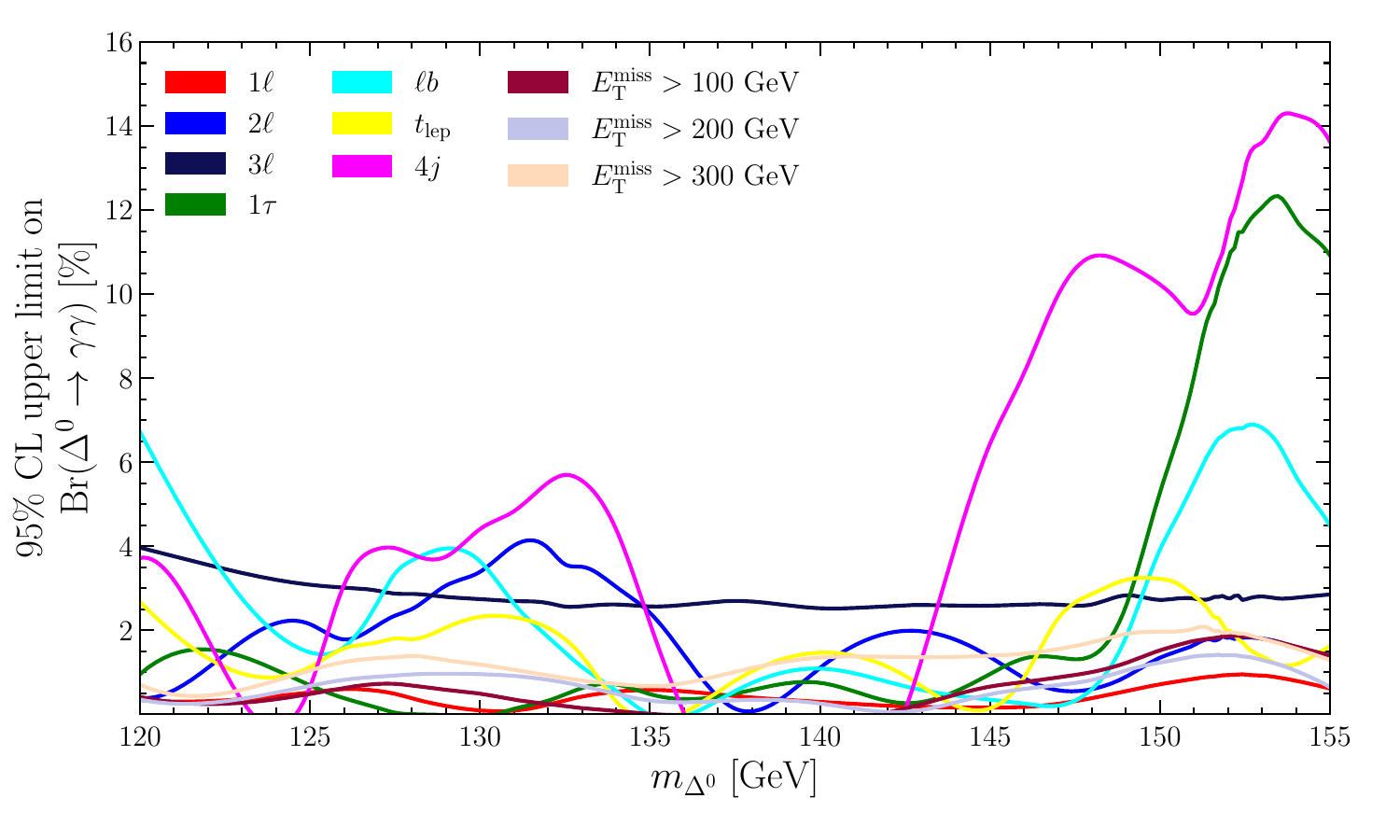}
\caption{95\% CL upper limits on Br$(\Delta^0\to\gamma\gamma)$ as a function of $m_{\Delta^0}$ for various SRs.}
\label{fig:aaX_95CL}
\end{figure}

Now, we put upper limits on the parameter of interest, Br$(\Delta^0\to\gamma\gamma)$. While this is trivially calculable, as discussed in Sec.~\ref{sec:tripletDecays}, it crucially depends on the triplet mass $m_{\Delta^0}$, the mixing angle $\alpha$, the mass-splitting $m_{\Delta^\pm}-m_{\Delta^0}$ (through the $\Delta^0\Delta^\pm\Delta^\mp$ coupling) and the triplet VEV $v_\Delta$. Hence, rather than varying these parameters, we subsume them within a single observable Br$(\Delta^0\to\gamma\gamma)$, which is of great interest as far as the LHC search program for additional Higgs is concerned. 95\% CL upper limits on Br$(\Delta^0\to\gamma\gamma)$ as a function of the triplet Higgs mass $m_{\Delta^0}$ for various SRs are shown in Fig.~\ref{fig:aaX_95CL}. For brevity, we only show the SRs that are most constraining, i.e.~provide relevant bounds on Br$\Delta^0\to\gamma\gamma$.

\begin{figure}[htb!]
\centering
\includegraphics[width=0.49\columnwidth]{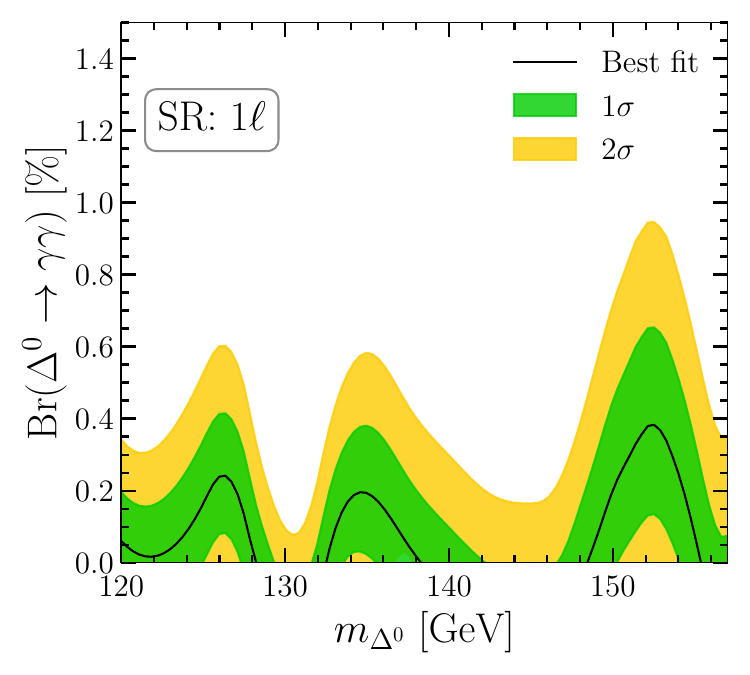}
\includegraphics[width=0.49\columnwidth]{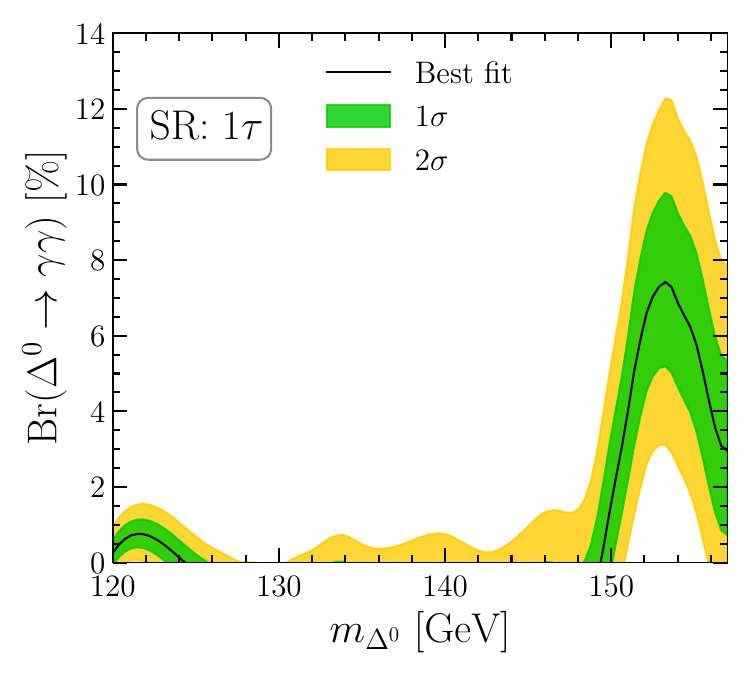}
\includegraphics[width=0.49\columnwidth]{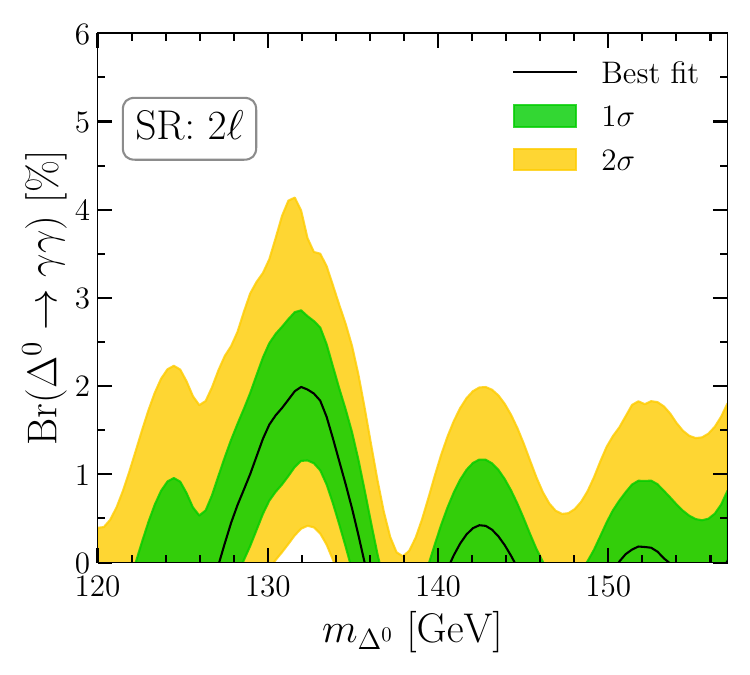}
\includegraphics[width=0.49\columnwidth]{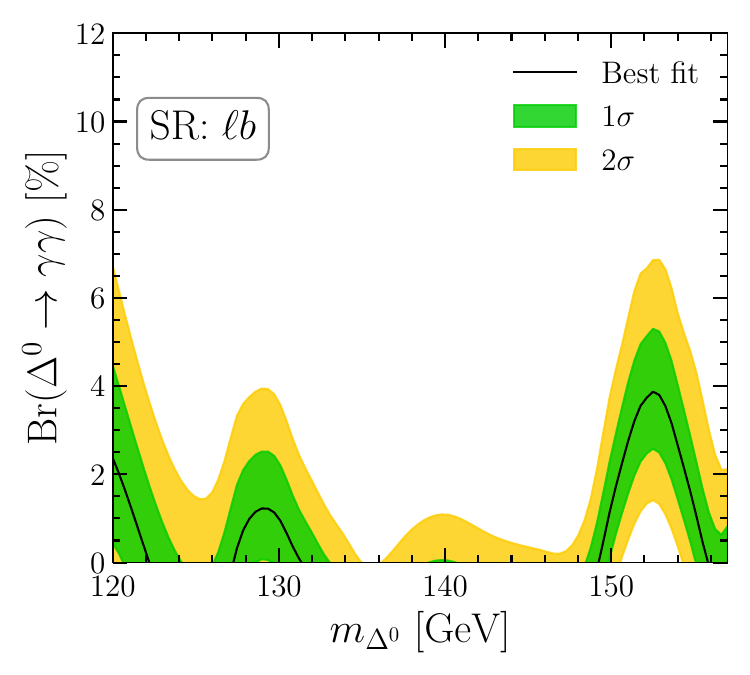}
\includegraphics[width=0.49\columnwidth]{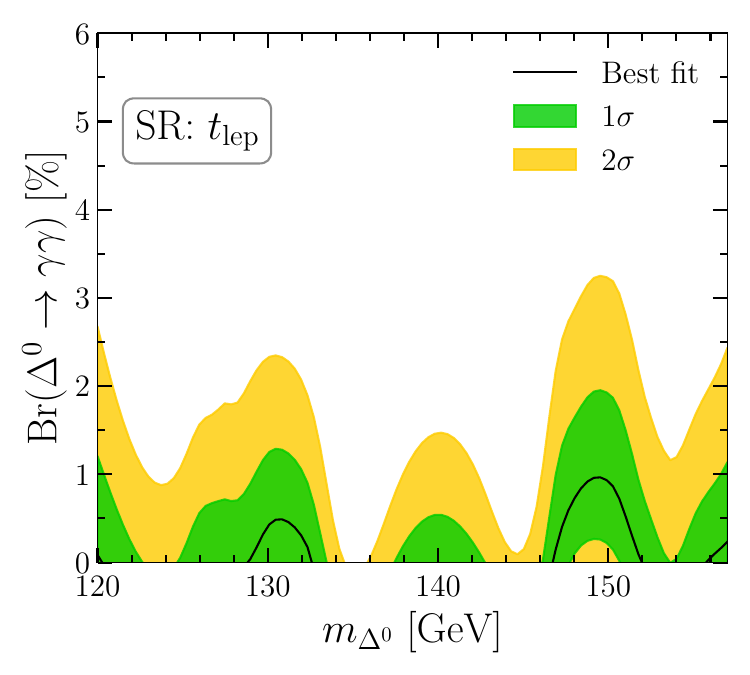}
\includegraphics[width=0.49\columnwidth]{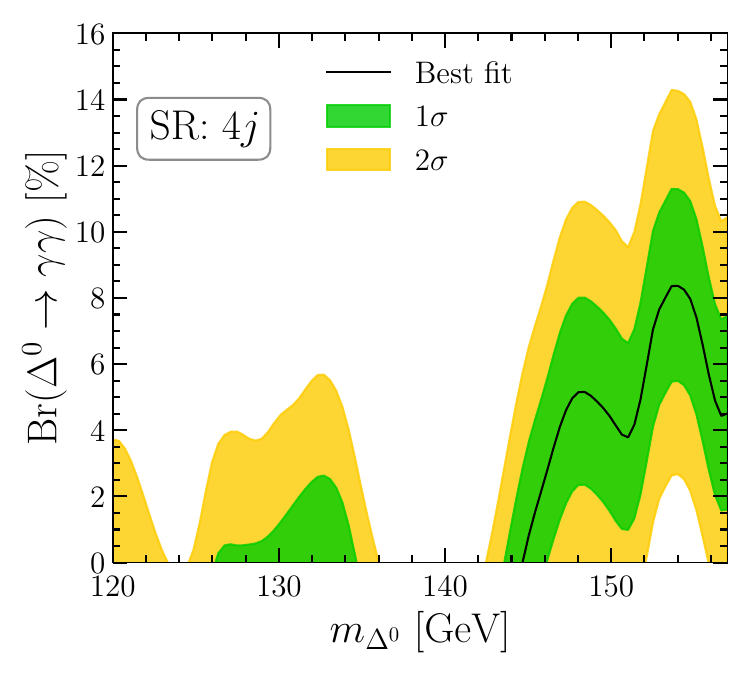}
\caption{Preferred ranges for Br$(\Delta^0\to \gamma\gamma)$ as a function of $m_{\Delta^0}$ for the relevant SRs. Green corresponds to $1\sigma$ and yellow to $2\sigma$, and the best fit is shown as a solid black line.}
\label{fig:aaX_best1}
\end{figure}

\begin{figure}[htb!]
\centering
\includegraphics[width=0.49\columnwidth]{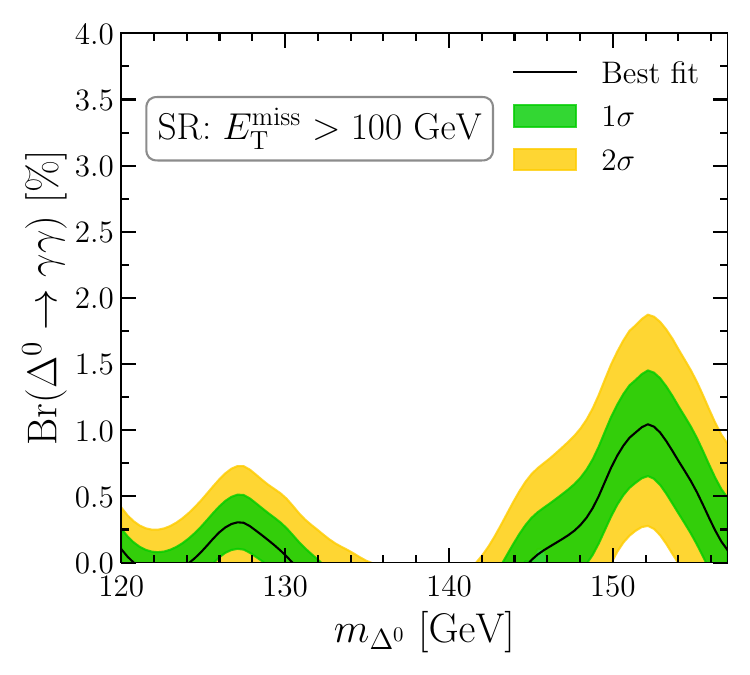}
\includegraphics[width=0.49\columnwidth]{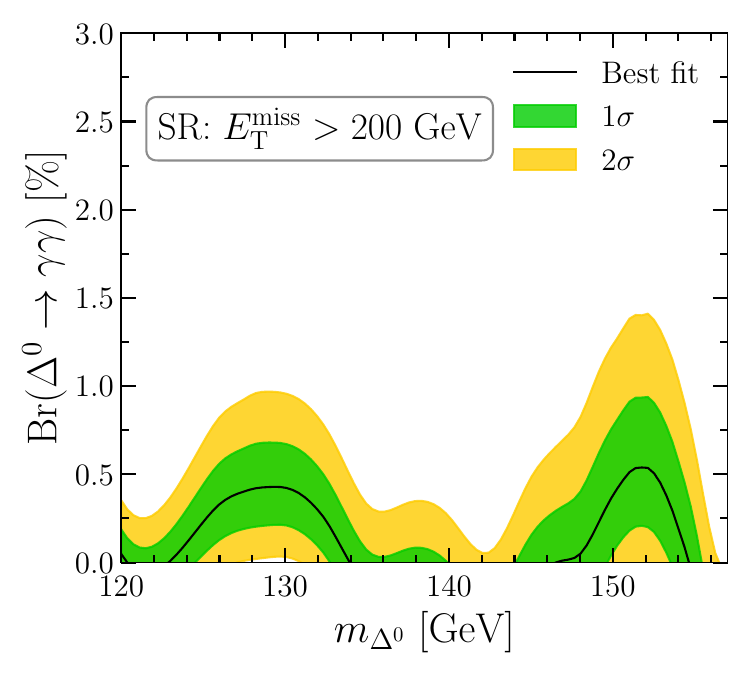}
\caption{Preferred ranges for Br$(\Delta^0\to \gamma\gamma)$ as a function of $m_{\Delta^0}$ for the relevant SRs  (continued). Green corresponds to $1\sigma$ and yellow to $2\sigma$, and the best fit is shown as a solid black line.}
\label{fig:aaX_best2}
\end{figure}

We then proceed to find the preferred ranges for Br$(\Delta^0\to \gamma\gamma)$ as a function of $m_{\Delta^0}$ for the relevant SRs. The obtained best-fit values, along with the $1\sigma$ and $2\sigma$ ranges, are shown in Figs.~\ref{fig:aaX_best1} and \ref{fig:aaX_best2}. For reasons mentioned earlier, here, we do not show the corresponding plots for $3\ell$ and $E_T^{\rm miss}>300$\,GeV SRs. Interestingly, all relevant SRs display a preference for a non-zero $\Delta^0 \to \gamma\gamma$ decay rate around 152\,GeV. 

\begin{figure}[htb!]
\centering
\includegraphics[width=0.7\columnwidth]{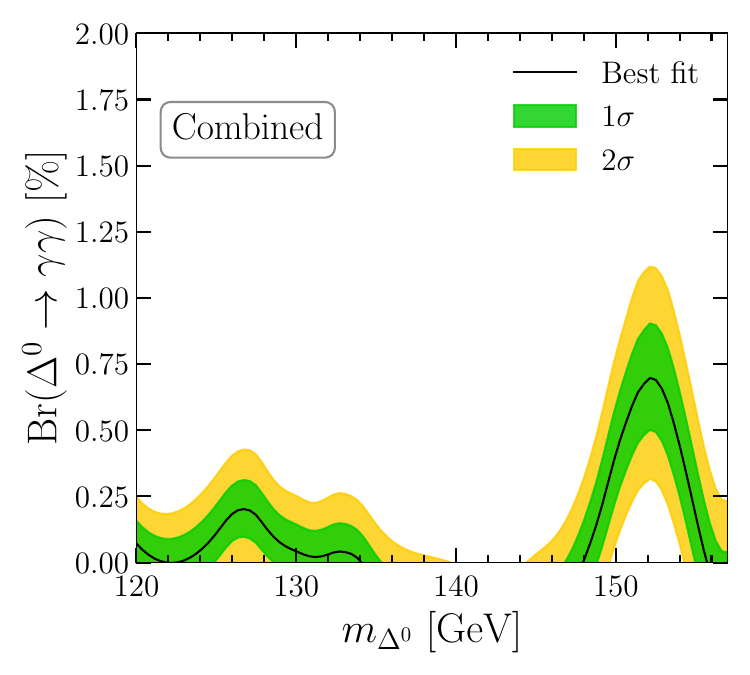}
\caption{Preferred ranges for Br$(\Delta^0\to \gamma\gamma)$ as a function of $m_{\Delta^0}$ for the statistical combination of the relevant SRs.}
\label{fig:aaX_best_comb}
\end{figure}

Finally, we combine the 10 relevant SRs, including their correlations, to obtain the combined preferred ranges for Br$(\Delta^0\to \gamma\gamma)$ as a function of $m_\Delta$, as shown in Fig.~\ref{fig:aaX_best_comb}.\footnote{Note that there are moderate correlations among some SRs, in particular, among the $E_T^{\rm miss}$ ones. However, we checked that removing the $E_T^{\rm miss} > 300$\,GeV SR from the fit has a very small impact on the final result.} We see a strong preference for a non-zero Br$(\Delta^0\to \gamma\gamma)$ in the 150\,GeV--155\,GeV range. This is most pronounced at 152\,GeV with Br$(\Delta^0\to \gamma\gamma) \approx 0.7\%$, and the corresponding significance amounts to $3.9\sigma$.

\begin{figure}[htb!]
\centering
\includegraphics[width=1\columnwidth]{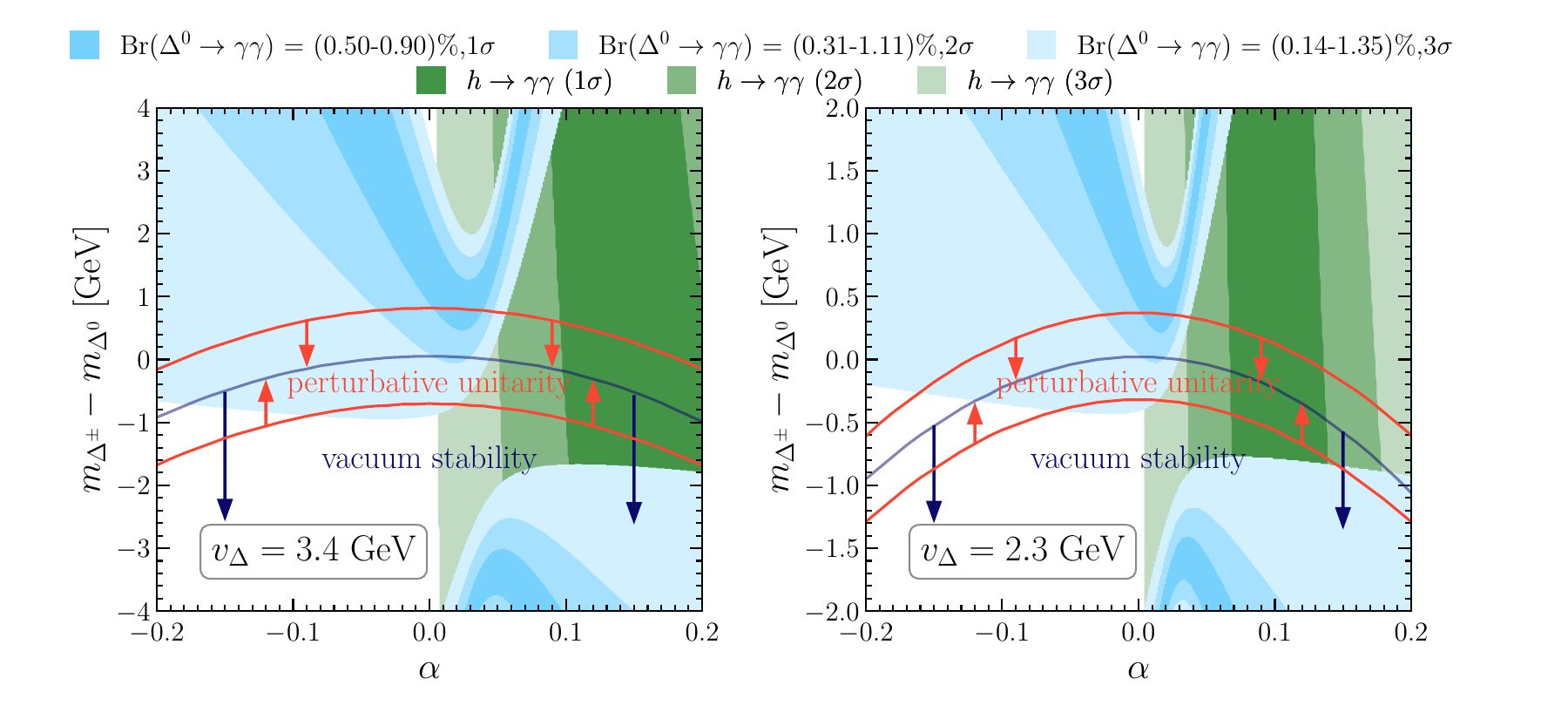} 
\caption{Preferred regions in the $\alpha$ vs $m_{\Delta^\pm}-m_{\Delta^0}$ plane for $m_{\Delta^0} = 152$\,GeV and two values of $v_\Delta$: 4.1\,GeV (left) and 2.7\,GeV (right). The band between the two orange lines satisfies perturbative unitarity, and the region below the blue line leads to a stable vacuum at the EW scale. The green regions are allowed by the SM Higgs to di-photon signal strength at $1\sigma$, $2\sigma$ and $3\sigma$ levels. The $1\sigma$, $2\sigma$ and $3\sigma$ regions for Br$(\Delta^0\to \gamma\gamma)$ preferred by the ATLAS $\gamma\gamma+X$ data are shown in violet.}
\label{fig:BrH152aa}
\end{figure}

For the sake of phenomenological interest, we have subsumed the free parameters of the $\Delta$SM model---the triplet mass $m_{\Delta^0}$, the mixing angle $\alpha$, the mass-splitting $m_{\Delta^\pm}-m_{\Delta^0}$ and the triplet VEV $v_\Delta$---into a single observable Br$[\Delta^0\to\gamma\gamma]$. Before closing this section, we closely look at how the preferred di-photon decay rate can be obtained. For a given $m_{\Delta^0}$ and $v_\Delta$, Br$(\Delta^0\to\gamma\gamma)$ depends on $\alpha$ and $m_{\Delta^\pm}-m_{\Delta^0}$. In Fig.~\ref{fig:BrH152aa}, we show the preferred regions in the $\alpha$ vs $m_{\Delta^\pm}-m_{\Delta^0}$ plane for $m_{\Delta^0} = 152$ GeV and two values of $v_\Delta$: 4.1\,GeV (left) and 2.7\,GeV (right), corresponding to the best-fit values obtained from the world $W$-mass fit with and without including the CDF-II measurement. The band between the two orange lines satisfies perturbative unitarity, and the region below the blue line leads to a stable vacuum at the EW scale. The green regions are allowed by the SM Higgs to di-photon signal strength~\cite{CMS:2021kom,ATLAS:2022tnm} at $1\sigma$ (1.02-1.15), $2\sigma$ (0.96-1.22) and $3\sigma$ (0.90-1.29) levels. The $1\sigma$ (0.50--0.90\%), $2\sigma$ (0.31--1.11\%) and $3\sigma$ (0.14--1.35\%) regions for Br$(\Delta^0\to \gamma\gamma)$ are shown in violet. We see that the $1\sigma$ region (and also part of the $2\sigma$ region) preferred by the ATLAS $\gamma\gamma+X$ searches are not in accordance with the vacuum stability and perturbative unitarity constraints. This demonstrates that while we have growing evidence for a 152\,GeV triplet-like Higgs produced in association with leptons (including $\tau$-leptons), quarks (including $b$-quarks) and neutrinos, the $\Delta$SM should be superseded by BSM models with additional fields at or above the EW scale so as to restore the stability of the vacuum and perturbativity of the theory.

\subsection{Higgs to $Z\gamma$ search in the $\ell^+\ell^-\gamma$ final state}
\label{sec:Za}

\begin{figure}[htb!]
\centering
\includegraphics[width=0.6\columnwidth]{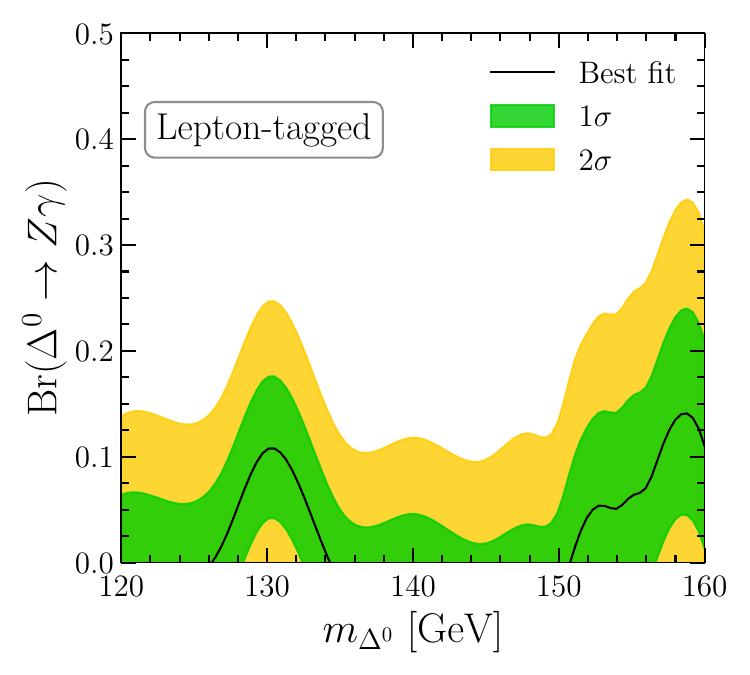}
\caption{Preferred ranges for Br$(\Delta^0\to Z\gamma)$ as a function of $m_{\Delta^0}$. Green corresponds to $1\sigma$ and yellow to $2\sigma$, and the best fit is shown as a solid black line.}
\label{fig:za_lepton}
\end{figure}

Similar to the di-photon mode, $\Delta^0$ can have a branching ratio of about $\mathcal{O}(0.1)$\%--$\mathcal{O}(1)$\% (see Fig.~\ref{fig:HZa} and the relevant discussion in Sec.~\ref{sec:tripletDecays}). Experimentally, the final state resulting from the leptonic decay of $Z$ ($Z\to \ell^+\ell^-$ with $\ell=e,\mu$) is the most accessible since the leptons are highly distinctive. CMS has recently performed such a pertinent search with the full run 2 data in Ref.~\cite{CMS:2022ahq}. This search categorises the events into three broad mutually exclusive SRs: Lepton-tagged, Dijet, and Untagged, further categorising the latter two into 3 and 4 SRs. We meticulously recast this search in the context of the $\Delta$SM and find that only the Lepton-tagged SR is relevant. In Fig.~\ref{fig:za_lepton}, we show the preferred ranges for Br$(\Delta^0\to Z\gamma)$ as a function of $m_{\Delta^0}$. Though the current bounds are not competitive with di-photon searches, they are expected to become relevant with high-luminosity LHC data~\cite{Cepeda:2019klc}.

\section{Summary and Outlook}
\label{sec:summary}
Though the ongoing LHC search program for new particles has not led to any discovery yet, it has, nonetheless, unveiled several hints for new particles in the form of ``anomalies''~\cite{Crivellin:2023zui}. This includes the multi-lepton anomalies~\cite{vonBuddenbrock:2019ajh,Fischer:2021sqw} which points towards the existence of a new Higgs with a mass of $150 \pm 5$\, GeV which dominantly decays to $WW$~\cite{vonBuddenbrock:2017gvy}. Furthermore, the re-analyses of the sidebands of the ATLAS and CMS Higgs boson searches indicate the presence of a new narrow resonance with a mass around 152$\,$GeV produced in association with leptons, quarks, and neutrinos~\cite{Crivellin:2021ubm,Bhattacharya:2023lmu}. 

In this context, it has been shown that the $\gamma\gamma+X$ excesses are consistent with the Drell-Yan production of a $152\,$GeV Higgs from a scalar triplet with $Y=0$, with a significance of $\approx4\sigma$~\cite{Ashanujjaman:2023etj,Crivellin:2024uhc}, which predicts a positive-definite shift in the $W$ mass, in agreement with the global fit. Furthermore, its neutral component decays dominantly to $WW$, as suggested by the multi-lepton anomalies.

Motivated by this, we have detailed the $Y=0$ Higgs triplet model and its phenomenology in this work. In particular, we have studied the tree-level perturbative unitarity and bounded-from-below constraints on the scalar potential. Further, we have discussed various possible vacua configurations and the stability of the neutral ones against the charge-breaking ones, which is discussed in Appendix~\ref{app:globality}. We showed that the absence of tachyonic modes in the Higgs sector suffices to ensure that the desired EW vacuum corresponds to the global minimum of the potential. Combined, these constraints delineate the theoretically allowed parameter space within the perturbative regime and, together with the constraints from the $\rho$ parameter, drive the triplet-like neutral and charged Higgs states to be quasi-mass-degenerate.

We have discussed the production and decays of the Higgs states, including the $h\to \gamma\gamma$, $h\to Z\gamma$ and $h \to ZZ^*$ decays, which put non-trivial constraints on the model parameter space. Furthermore, we have detailed their phenomenology at the LHC based on three types of final states: stau-like signature, multi-lepton and $\gamma\gamma+X$ final states. The findings are summarised below.
\begin{enumerate}[label=$\roman*)$,itemsep=1pt,parsep=1pt,topsep=1pt,partopsep=1pt]
\item  Low-mass ($\lesssim 120$ GeV) triplet-like charged Higgs dominantly decays to $\tau\nu$. Therefore, the process $p p \to \Delta^+\Delta^-$ leads to final states involving a pair of $\tau$-leptons and missing transverse momentum, which is akin to the collider signature of stau. Our reinterpretation of the corresponding CMS search~\cite{CMS:2022rqk} excludes triplet-like charged Higgs with a mass below 110\,GeV at 95\% CL.
\item For triplet-like Higgs bosons with higher masses ($\gtrsim 120$ GeV) the dominant decays modes are $\Delta^\pm \to W^\pm Z, t \bar{b}$ and $\Delta^0 \to W^+ W^-, ZZ$. Therefore, their Drell-Yan production can lead to final states with multiple charged leptons from the leptonic decays of $W$ and $Z$. To probe the $\Delta$SM, we recast a recent model-independent analysis of ATLAS~\cite{ATLAS:2021wob} involving three and four leptons in the final state. Our analysis shows that for some SRs, the predicted effect is close to the observed limit of ATLAS, but $\Delta$ SM cannot be excluded from this search. However, run 3 and high-luminosity LHC data will be able to test the signatures of the $\Delta$SM in multi-lepton final states.

\item The associated production of the triplet-like Higgses and their prompt decays to di-photon lead to $\gamma\gamma+X$ final states. We have performed two such pertinent searches by ATLAS. We have performed a (re)fit to the data to determine the continuum background and a likelihood-based statistical interpretation to find that, out of the 25 signal regions considered, 10 of them put stringent limits on the triplet-like Higgs, particularly on its di-photon branching ratio Br$(\Delta^0\to \gamma\gamma)$. Interestingly, all relevant signal regions show a weakening of the limits around 125\,GeV--130\,GeV and 152\,GeV. More importantly, there is a strong preference for a non-zero di-photon decay rate in the 150\,GeV--155\,GeV range. When all relevant SRs are statistically combined, this is most pronounced at 152\,GeV with the corresponding branching ratio of about 0.7\% and the corresponding significance of about $3.9\sigma$. While such a sizable branching ratio to di-photons is easily achievable in the $\Delta$SM, part of this model parameter space consistent with the di-photon signal strength of the SM Higgs is in tension with the vacuum stability and perturbative unitarity.
\end{enumerate}

The last point, i.e. the preferred range for the di-photon decay rate is in tension with the vacuum stability and perturbative unitarity, demonstrates that while we have growing evidence for a 152\,GeV Higgs produced in association with leptons, quarks and neutrinos, the $\Delta$SM is not expected to be the final theory superseding the SM. In fact, with additional fields at or above the EW scale, the problem of the large di-photon rate in unstable field configurations can be solved in (at least) three non-exclusive ways. Additional field(s) can
\begin{enumerate}[label=$\roman*)$,itemsep=1pt,parsep=1pt,topsep=1pt,partopsep=1pt]
\item modify the (effective) scalar potential such that the vacuum stability is restored at larger field values,
\item lead to loop contributions to $\Delta^0\to \gamma\gamma$,
\item allow for a larger VEV of the triplet by cancelling its effect in the $\rho$ parameter as in the Georgi-Machacek model~\cite{Georgi:1985nv}, thereby enlarging the region with the stable EW vacuum.
\end{enumerate}
$(i)$ and $(ii)$ are possible, e.g.~by introducing additional charged Higgs bosons; an interesting UV-realisation of this is the $\Delta$2HDMS: the $\Delta$SM extended with a second Higgs doublet and a singlet~\cite{Coloretti:2023yyq}. In addition to explaining the di-photon excess in associated production~\cite{Banik:2024ftv} this model can address the significant tensions in differential top quark distributions~\cite{ATLAS:2023gsl} as shown in Ref.~\cite{Banik:2023vxa} and lead to baryogenesis via a two-step phase transition~\cite{Inoue:2015pza}. 

In summary, the $\Delta$SM, analysed in detail in this article, represents a step towards the establishment of a model that can simultaneously explain the multi-lepton anomalies and the narrow resonance in associated production with a mass of 152\,GeV. However, more data and improved SM predictions are necessary to arrive at the conclusive model superseding the SM.

\appendix
\section{Feynman rules for the $\Delta$SM} 
\label{app:Feynman}

Here, we provide the Feynman rules for vertices involving the triplet fields. All momenta are considered to be incoming. We take the positively charged components of the triplets as particles and the negatively charged ones as antiparticles. Consequently, particle flow arrows point in the direction of particle momentum, while they point opposite to the momentum direction for antiparticles. We use the shorthand notations $s_\beta = \sin \beta$ and $c_\beta = \cos \beta$.
\vspace{0.5cm}
\newline
    \begin{tikzpicture}[baseline=(current bounding box.center)]
        \begin{feynman}
            \vertex (a);
            \vertex [below right=1cm of a] (b) {$\Bar{f}$};
            \vertex [above right=1cm of a] (c) {$f$};
            \vertex [left=1cm of a] (d) {$\Delta^0$};
            \diagram{
                (b) -- [fermion] (a) -- [fermion] (c);
                (a) -- [scalar] (d);
            };
            \vertex [right=1.5cm of a] (e){$i\dfrac{m_f}{v_\Phi} s_\alpha$}; 
        \end{feynman}
    \end{tikzpicture}
    \\[.5cm]
    \begin{tikzpicture}[baseline=(current bounding box.center)]
        \begin{feynman}
            \vertex (a);
            \vertex [below right=1cm of a] (b) {$\ell, \nu$};
            \vertex [above right=1cm of a] (c) {$\Bar{\nu}, \Bar{\ell}$};
            \vertex [left=1cm of a] (d) {$\Delta^\pm$};
            \diagram{
                (b) -- [fermion] (a) -- [fermion] (c);
                (a) -- [scalar] (d);
            };
            \vertex [right=1.5cm of a] (e){$i\dfrac{\sqrt{2} m_{\ell}}{v_\Phi}s_\beta P_{R,L}$}; 
        \end{feynman}
    \end{tikzpicture}
    \\[.5cm]
    \begin{tikzpicture}[baseline=(current bounding box.center)]
        \begin{feynman}
            \vertex (a);
            \vertex [below right=1cm of a] (b) {$d_j,u_j$};
            \vertex [above right=1cm of a] (c) {$\Bar{u}_i,\Bar{d}_i$};
            \vertex [left=1cm of a] (d) {$\Delta^\pm$};
            \diagram{
                (b) -- [fermion] (a) -- [fermion] (c);
                (a) -- [scalar] (d);
            };
            \vertex [right=1.5cm of a] (e){$-i \dfrac{\sqrt{2}}{v_\Phi}s_\beta~V_{u_i,d_j}(m_{u_i} P_{L} - m_{d_j} P_{R}), \quad i \dfrac{\sqrt{2}}{v_\Phi}s_\beta~V^{*}_{u_j,d_i}(m_{d_i} P_{L} - m_{u_j} P_{R}) $}; 
        \end{feynman}
    \end{tikzpicture}
    \\[.5cm]
    \begin{tikzpicture}[baseline=(current bounding box.center)]
        \begin{feynman}
            \vertex (a);
            \vertex [below right=1cm of a] (b) {$W^{+}_{\nu}$};
            \vertex [above right=1cm of a] (c) {$W^{-}_{\mu}$};
            \vertex [left=1cm of a] (d) {$\Delta^0$};
            \diagram{
                (b) -- [boson] (a) -- [boson] (c);
                (a) -- [scalar] (d);
            };
            \vertex [right=1.5cm of a] (e){$-\dfrac{i}{2} g^2 \left(v_\Phi s_\alpha - 4v_\Delta c_\alpha\right)g_{\mu\nu}$};
        \end{feynman}
    \end{tikzpicture}
\\[.5cm]
    \begin{tikzpicture}[baseline=(current bounding box.center)]
        \begin{feynman}
            \vertex (a);
            \vertex [below right=1cm of a] (b) {$Z_{\nu}$};
            \vertex [above right=1cm of a] (c) {$Z_{\mu}$};
            \vertex [left=1cm of a] (d) {$\Delta^0$};
            \diagram{
                (b) -- [boson] (a) -- [boson] (c);
                (a) -- [scalar] (d);
            };
            \vertex [right=1.5cm of a] (e){$-\dfrac{i}{2} \dfrac{g^2}{c_{w}^2} v_\Phi s_\alpha~g_{\mu\nu}$};
        \end{feynman}
    \end{tikzpicture}
\\[.5cm]
    \begin{tikzpicture}[baseline=(current bounding box.center)]
        \begin{feynman}
            \vertex (a);
            \vertex [below right=1cm of a] (b) {$Z_{\nu}$};
            \vertex [above right=1cm of a] (c) {$W_{\mu}^{\mp}$};
            \vertex [left=1cm of a] (d) {$\Delta^\pm$};
            \diagram{
                (b) -- [boson] (a) -- [boson] (c);
                (a) -- [scalar] (d);
            };
            \vertex [right=1.5cm of a] (e){$-\dfrac{i}{2} \dfrac{g^2}{c_{w}} \left( 2v_\Delta c_{w}^2 c_{\beta} - v_\Phi s_{w}^2 s_\beta\right) g_{\mu\nu}$};
        \end{feynman}
    \end{tikzpicture}
\\[.5cm]
    \begin{tikzpicture}[baseline=(current bounding box.center)]
        \begin{feynman}
            \vertex (a);
            \vertex [below right=1cm of a] (b) {$\gamma_{\mu}$};
            \vertex [above right=1cm of a] (c) {$W_{\nu}^\mp$};
            \vertex [left=1cm of a] (d) {$\Delta^\pm$};
            \diagram{
                (b) -- [boson] (a) -- [boson] (c);
                (a) -- [scalar] (d);
            };
            \vertex [right=1.5cm of a] (e){$-\dfrac{i}{2} g^2 s_w \left( 2v_\Delta c_{\beta} + v_\Phi s_\beta\right) g_{\mu\nu}$};
        \end{feynman}
    \end{tikzpicture}
\\[.5cm]
    \begin{tikzpicture}[baseline=(current bounding box.center)]
        \begin{feynman}
            \vertex (a);
            \vertex [below right=1cm of a] (b) {$\Delta^0$};
            \vertex [above right=1cm of a] (c) {$W^{\mp}_{\mu}$};
            \vertex [left=1cm of a] (d) {$\Delta^\pm$};
            \diagram{
                (b) -- [scalar] (a) -- [boson] (c);
                (a) -- [scalar] (d);
            };
            \vertex [right=1.5cm of a] (e){$\mp \dfrac{i}{2} g \left(s_\alpha s_\beta - 2c_\alpha c_{\beta} \right) \left(p^{\Delta^0}_\mu - p^{\Delta^\pm}_\mu\right)$};
        \end{feynman}
    \end{tikzpicture}
\\[.5cm]    
    \begin{tikzpicture}[baseline=(current bounding box.center)]
        \begin{feynman}
            \vertex (a);
            \vertex [below right=1cm of a] (b) {$h$};
            \vertex [above right=1cm of a] (c) {$W_{\mu}^{\mp}$};
            \vertex [left=1cm of a] (d) {$\Delta^\pm$};
            \diagram{
                (b) -- [scalar] (a) -- [boson] (c);
                (a) -- [scalar] (d);
            };
            \vertex [right=1.5cm of a] (e){$\mp\dfrac{i}{2} g \left(2s_\alpha c_{\beta} +c_\alpha s_\beta \right) \left(- p^{h}_\mu + p^{\Delta^\pm}_\mu\right)$};
        \end{feynman}
    \end{tikzpicture}
\\[.5cm]
    \begin{tikzpicture}[baseline=(current bounding box.center)]
        \begin{feynman}
            \vertex (a);
            \vertex [below right=1cm of a] (b) {$\gamma_{\mu}$};
            \vertex [above right=1cm of a] (c) {$\Delta^-$};
            \vertex [left=1cm of a] (d) {$\Delta^+$};
            \diagram{
                (b) -- [scalar] (a) -- [boson] (c);
                (a) -- [scalar] (d);
            };
            \vertex [right=1.5cm of a] (e){$-i e \left(p^{\Delta^-}_\mu - p^{\Delta^+}_\mu \right)$};
        \end{feynman}
    \end{tikzpicture}
    \\[.5cm]
    \begin{tikzpicture}[baseline=(current bounding box.center)]
        \begin{feynman}
            \vertex (a);
            \vertex [below right=1cm of a] (b) {$Z_{\mu}$};
            \vertex [above right=1cm of a] (c) {$\Delta^-$};
            \vertex [left=1cm of a] (d) {$\Delta^+$};
            \diagram{
                (b) -- [scalar] (a) -- [boson] (c);
                (a) -- [scalar] (d);
            };
            \vertex [right=1.5cm of a] (e){$-\dfrac{i}{2}\dfrac{g}{c_{w}}(c_{\beta}^2 + c_{2w}) \left(p^{\Delta^-}_\mu - p^{\Delta^+}_\mu \right)$};
        \end{feynman}
    \end{tikzpicture}
    \\[.5cm]
        \begin{tikzpicture}[baseline=(current bounding box.center)]
        \begin{feynman}
            \vertex (a);
            \vertex [above left=1cm of a] (b) {$\Delta^0$};
            \vertex [below left=1cm of a] (c) {$\Delta^0$};
            \vertex [above right=1cm of a] (e) {$W_{\mu}^-$};
            \vertex [below right=1cm of a] (f) {$W_{\nu}^+$};
            \diagram{
                (c) -- [scalar] (a) -- [scalar] (b);
                (f) -- [boson] (a) -- [boson] (e);
            };
            \vertex [right=1.5cm of a] (e){$\dfrac{i}{2} g^2 \left(1 + 3c_{\alpha}^2\right)g_{\mu\nu}$};
        \end{feynman}
    \end{tikzpicture}
    \\[.5cm]
    \begin{tikzpicture}[baseline=(current bounding box.center)]
        \begin{feynman}
            \vertex (a);
            \vertex [above left=1cm of a] (b) {$\Delta^0$};
            \vertex [below left=1cm of a] (c) {$\Delta^0$};
            \vertex [above right=1cm of a] (e) {$Z_{\mu}$};
            \vertex [below right=1cm of a] (f) {$Z_{\nu}$};
            \diagram{
                (c) -- [scalar] (a) -- [scalar] (b);
                (f) -- [boson] (a) -- [boson] (e);
            };                  
            \vertex [right=1.5cm of a] (e){$\dfrac{i}{2} \dfrac{g^2}{c_{w}^2} s_{\alpha}^2 g_{\mu\nu}$};

        \end{feynman}
    \end{tikzpicture}
\\[.5cm]
    \begin{tikzpicture}[baseline=(current bounding box.center)]
        \begin{feynman}
            \vertex (a);
            \vertex [above left=1cm of a] (b) {$\Delta^\pm$};
            \vertex [below left=1cm of a] (c) {$\Delta^0$};
            \vertex [above right=1cm of a] (e) {$W_{\mu}^{\mp}$};
            \vertex [below right=1cm of a] (f) {$Z_{\nu}$};
            \diagram{
                (c) -- [scalar] (a) -- [scalar] (b);
                (f) -- [boson] (a) -- [boson] (e);
            };
            \vertex [right=1.5cm of a] (e){$-\dfrac{i}{2} \dfrac{g^2}{c_{w}} \left(2c_{w}^2 c_\alpha c_{\beta} + s_{w}^2 s_\alpha s_\beta \right) g_{\mu\nu}$};
        \end{feynman}
    \end{tikzpicture}
\\[.5cm]
    \begin{tikzpicture}[baseline=(current bounding box.center)]
        \begin{feynman}
            \vertex (a);
            \vertex [above left=1cm of a] (b) {$\Delta^\pm$};
            \vertex [below left=1cm of a] (c) {$\Delta^0$};
            \vertex [above right=1cm of a] (e) {$W_{\nu}^{\mp}$};
            \vertex [below right=1cm of a] (f) {$\gamma_{\mu}$};
            \diagram{
                (c) -- [scalar] (a) -- [scalar] (b);
                (f) -- [boson] (a) -- [boson] (e);
            };
            \vertex [right=1.5cm of a] (e){$- \dfrac{i}{2} g^2 s_w \left(2c_\alpha c_{\beta} - s_\alpha s_\beta  \right) g_{\mu\nu}$};
        \end{feynman}
    \end{tikzpicture}
\\[.5cm]    
    \begin{tikzpicture}[baseline=(current bounding box.center)]
        \begin{feynman}
            \vertex (a);
            \vertex [above left=1cm of a] (b) {$\Delta^\pm$};
            \vertex [below left=1cm of a] (c) {$\Delta^\pm$};
            \vertex [above right=1cm of a] (e) {$W_{\mu}^{\mp}$};
            \vertex [below right=1cm of a] (f) {$W_{\nu}^{\mp}$};
            \diagram{
                (c) -- [scalar] (a) -- [scalar] (b);
                (f) -- [boson] (a) -- [boson] (e);
            };
            \vertex [right=1.5cm of a] (e){$-2ig^2 c_{\beta}^2 g_{\mu\nu}$};
        \end{feynman}
    \end{tikzpicture}
    \\[.5cm]
    \begin{tikzpicture}[baseline=(current bounding box.center)]
        \begin{feynman}
            \vertex (a);
            \vertex [above left=1cm of a] (b) {$\Delta^+$};
            \vertex [below left=1cm of a] (c) {$\Delta^-$};
            \vertex [above right=1cm of a] (e) {$W_{\mu}^-$};
            \vertex [below right=1cm of a] (f) {$W_{\nu}^+$};
            \diagram{
                (c) -- [scalar] (a) -- [scalar] (b);
                (f) -- [boson] (a) -- [boson] (e);
            };
            \vertex [right=1.5cm of a] (e){$\dfrac{i}{2} g^2 \left(1+c_{\beta}^2\right) g_{\mu\nu}$};
        \end{feynman}
    \end{tikzpicture}
    \\[.5cm]
    \begin{tikzpicture}[baseline=(current bounding box.center)]
        \begin{feynman}
            \vertex (a);
            \vertex [above left=1cm of a] (b) {$\Delta^+$};
            \vertex [below left=1cm of a] (c) {$\Delta^-$};
            \vertex [above right=1cm of a] (e) {$Z_{\mu}$};
            \vertex [below right=1cm of a] (f) {$Z_{\nu}$};
            \diagram{
                (c) -- [scalar] (a) -- [scalar] (b);
                (f) -- [boson] (a) -- [boson] (e);
            };
            \vertex [right=1.5cm of a] (e){$\dfrac{i}{2} \dfrac{g^2}{c_w^2 } (4 c_{\beta}^2 c_{w}^4 + s_{\beta}^2 c_{2w}^2) g_{\mu\nu}$};
        \end{feynman}
    \end{tikzpicture}
    \\[.5cm]
    \begin{tikzpicture}[baseline=(current bounding box.center)]
        \begin{feynman}
            \vertex (a);
            \vertex [above left=1cm of a] (b) {$\Delta^+$};
            \vertex [below left=1cm of a] (c) {$\Delta^-$};
            \vertex [above right=1cm of a] (e) {$\gamma_{\mu}$};
            \vertex [below right=1cm of a] (f) {$\gamma_{\nu}$};
            \diagram{
                (c) -- [scalar] (a) -- [scalar] (b);
                (f) -- [boson] (a) -- [boson] (e);
            };
            \vertex [right=1.5cm of a] (e){$2ie^2 g_{\mu\nu}$};
        \end{feynman}
    \end{tikzpicture}
    \\[.5cm]
    \begin{tikzpicture}[baseline=(current bounding box.center)]
        \begin{feynman}
            \vertex (a);
            \vertex [above left=1cm of a] (b) {$\Delta^+$};
            \vertex [below left=1cm of a] (c) {$\Delta^-$};
            \vertex [above right=1cm of a] (e) {$\gamma_{\mu}$};
            \vertex [below right=1cm of a] (f) {$Z_{\nu}$};
            \diagram{
                (c) -- [scalar] (a) -- [scalar] (b);
                (f) -- [boson] (a) -- [boson] (e);
            };
            \vertex [right=1.5cm of a] (e){$i\dfrac{e g}{c_{w}} \left(c_{\beta}^2 + c_{2w} \right) g_{\mu\nu}$};

        \end{feynman}
    \end{tikzpicture}
    \\[.5cm]
    \begin{tikzpicture}[baseline=(current bounding box.center)]
        \begin{feynman}
            \vertex (a);
            \vertex [below right=1cm of a] (b) {$\Delta^0$};
            \vertex [above right=1cm of a] (c) {$\Delta^0$};
            \vertex [left=1cm of a] (d) {$\Delta^0$};
            \diagram{
                (b) -- [scalar] (a) -- [scalar] (c);
                (a) -- [scalar] (d);
            };
            \vertex [right=1.5cm of a] (e){$\begin{array}{ll}
                -\dfrac{3}{2} i [-A c_{\alpha } s_{\alpha }^2-2 c_{\alpha }^2 s_{\alpha } v_{\Phi } \lambda _{\Phi \Delta}+c_{\alpha }^3 \lambda _{\Delta }
   v_{\Delta }+s_{\alpha } s_{2 \alpha } v_{\Delta } \lambda _{\Phi \Delta}-\lambda _{\Phi } s_{\alpha }^3 v_{\Phi }]
            \end{array}$}
            ;
        \end{feynman}
    \end{tikzpicture}
    % \red{Charges don`t match}
    \\[.5cm]
    \begin{tikzpicture}[baseline=(current bounding box.center)]
        \begin{feynman}
            \vertex (a);
            \vertex [below right=1cm of a] (b) {$\Delta^0$};
            \vertex [above right=1cm of a] (c) {$h$};
            \vertex [left=1cm of a] (d) {$\Delta^0$};
            \diagram{
                (b) -- [scalar] (a) -- [scalar] (c);
                (a) -- [scalar] (d);
            };
            \vertex [right =1.5cm of a] (e){  $
            \begin{array}{ll}
                  -\dfrac{1}{8} i [&2 s_{\alpha } \left(3 c_{2 \alpha } \left(A+v_{\Delta } \left(\lambda _{\Delta }-2 \lambda _{\Phi \Delta}\right)\right)+ A-2
   v_{\Delta } \lambda _{\Phi \Delta}+3 \lambda _{\Delta } v_{\Delta }\right) \\[.2cm]
   &-3 c_{3 \alpha } v_{\Phi } \left(\lambda _{\Phi }-2 \lambda _{\Delta
    \Phi }\right)+c_{\alpha } v_{\Phi } \left(2 \lambda _{\Phi \Delta}+3 \lambda _{\Phi }\right)]
            \end{array}
            $};            
        \end{feynman}
    \end{tikzpicture}
    \\[.5cm]
    \begin{tikzpicture}[baseline=(current bounding box.center)]
        \begin{feynman}
            \vertex (a);
            \vertex [below right=1cm of a] (b) {$h$};
            \vertex [above right=1cm of a] (c) {$h$};
            \vertex [left=1cm of a] (d) {$\Delta^0$};
            \diagram{
                (b) -- [scalar] (a) -- [scalar] (c);
                (a) -- [scalar] (d);
            };
            \vertex [right=1.5cm of a] (e){$ \begin{array}{ll}
               \dfrac{1}{8} i [&c_{\alpha } \left(A-2 v_{\Delta } \lambda _{\Phi \Delta}-3 \lambda _{\Delta } v_{\Delta }\right)+3 c_{3 \alpha } \left(A-2
   v_{\Delta } \lambda _{\Phi \Delta}+\lambda _{\Delta } v_{\Delta }\right) \\[.2cm]
   &+3 s_{3 \alpha } v_{\Phi } \left(\lambda _{\Phi }-2 \lambda _{\Delta 
   \Phi }\right)+s_{\alpha } v_{\Phi } \left(2 \lambda _{\Phi \Delta}+3 \lambda _{\Phi }\right)]
            \end{array}$};
        \end{feynman}
    \end{tikzpicture}
        % \red{Charges don`t match}
    \\[.5cm]
    \begin{tikzpicture}[baseline=(current bounding box.center)]
        \begin{feynman}
            \vertex (a);
            \vertex [below right=1cm of a] (b) {$\Delta^+$};
            \vertex [above right=1cm of a] (c) {$\Delta^-$};
            \vertex [left=1cm of a] (d) {$h$};
            \diagram{
                (b) -- [scalar] (a) -- [scalar] (c);
                (a) -- [scalar] (d);
            };
            \vertex [right=1.5cm of a] (e){$ \begin{array}{ll}
            -\dfrac{1}{2} i [&-2 A c_{\alpha } c_{\beta } s_{\beta }+s_{\beta }^2 \left(s_{\alpha } \left(A+2 v_{\Delta } \lambda _{\Delta  \Phi
   }\right)+c_{\alpha } \lambda _{\Phi } v_{\Phi }\right) \\[.2cm]
   &+c_{\beta }^2 \left(2 c_{\alpha } v_{\Phi } \lambda _{\Phi \Delta}+\lambda _{\Delta }
   s_{\alpha } v_{\Delta }\right)]
            \end{array}$};
        \end{feynman}
    \end{tikzpicture}
    \\[.5cm]    
    \begin{tikzpicture}[baseline=(current bounding box.center)]
        \begin{feynman}
            \vertex (a);
            \vertex [below right=1cm of a] (b) {$\Delta^+$};
            \vertex [above right=1cm of a] (c) {$\Delta^-$};
            \vertex [left=1cm of a] (d) {$\Delta^0$};
            \diagram{
                (b) -- [scalar] (a) -- [scalar] (c);
                (a) -- [scalar] (d);
            };
            \vertex [right=1.5cm of a] (e){$\begin{array}{ll}
            -\dfrac{1}{2} i [&s_{\alpha } \left(A s_{2 \beta }-2 c_{\beta }^2 v_{\Phi } \lambda _{\Phi \Delta}-\lambda _{\Phi } s_{\beta }^2 v_{\Phi
   }\right) \\[.2cm]
   &+c_{\alpha } \left(s_{\beta }^2 \left(A+2 v_{\Delta } \lambda _{\Phi \Delta}\right)+c_{\beta }^2 \lambda _{\Delta } v_{\Delta
   }\right)]
            \end{array}$};
        \end{feynman}
    \end{tikzpicture}
        % \red{Charges don`t match}
    \\[.5cm]
    \begin{tikzpicture}[baseline=(current bounding box.center)]
        \begin{feynman}
            \vertex (a);
            \vertex [above left=1cm of a] (b) {$\Delta^0$};
            \vertex [below left=1cm of a] (c) {$\Delta^0$};
            \vertex [above right=1cm of a] (e) {$\Delta^0$};
            \vertex [below right=1cm of a] (f) {$\Delta^0$};
            \diagram{
                (c) -- [scalar] (a) -- [scalar] (b);
                (f) -- [scalar] (a) -- [scalar] (e);
            };
            \vertex [right=1.5cm of a] (e){
            $-\dfrac{3}{2} i \left(c_{\alpha }^4 \lambda _{\Delta }+s_{2 \alpha }^2 \lambda _{\Phi \Delta}+\lambda _{\Phi } s_{\alpha }^4\right)$
            };
        \end{feynman}
    \end{tikzpicture}
    \\[.5cm]
    \begin{tikzpicture}[baseline=(current bounding box.center)]
        \begin{feynman}
            \vertex (a);
            \vertex [above left=1cm of a] (b) {$\Delta^0$};
            \vertex [below left=1cm of a] (c) {$\Delta^0$};
            \vertex [above right=1cm of a] (e) {$\Delta^0$};
            \vertex [below right=1cm of a] (f) {$h$};
            \diagram{
                (c) -- [scalar] (a) -- [scalar] (b);
                (f) -- [scalar] (a) -- [scalar] (e);
            };
            \vertex [right=1.5cm of a] (e){
            $-\dfrac{3}{8} i s_{2 \alpha } [c_{2 \alpha } \left(-4 \lambda _{\Phi \Delta}+\lambda _{\Delta }+\lambda _{\Phi }\right)+\lambda _{\Delta
   }-\lambda _{\Phi }]$
            };
        \end{feynman}
    \end{tikzpicture}
    \\[.5cm]
    \begin{tikzpicture}[baseline=(current bounding box.center)]
        \begin{feynman}
            \vertex (a);
            \vertex [above left=1cm of a] (b) {$\Delta^0$};
            \vertex [below left=1cm of a] (c) {$\Delta^0$};
            \vertex [above right=1cm of a] (e) {$h$};
            \vertex [below right=1cm of a] (f) {$h$};
            \diagram{
                (c) -- [scalar] (a) -- [scalar] (b);
                (f) -- [scalar] (a) -- [scalar] (e);
            };
            \vertex [right=1.5cm of a] (e){
            $-\dfrac{1}{16} i [-3 c_{4\alpha} \left(-4 \lambda _{\Phi \Delta}+\lambda _{\Delta }+\lambda _{\Phi }\right)+3 \left(\lambda _{\Delta }+\lambda
   _{\Phi }\right)+4 \lambda _{\Phi \Delta}]$
            };
        \end{feynman}
    \end{tikzpicture}
    \\[.5cm]
    \begin{tikzpicture}[baseline=(current bounding box.center)]
        \begin{feynman}
            \vertex (a);
            \vertex [above left=1cm of a] (b) {$\Delta^0$};
            \vertex [below left=1cm of a] (c) {$h$};
            \vertex [above right=1cm of a] (e) {$h$};
            \vertex [below right=1cm of a] (f) {$h$};
            \diagram{
                (c) -- [scalar] (a) -- [scalar] (b);
                (f) -- [scalar] (a) -- [scalar] (e);
            };
            \vertex [right=1.5cm of a] (e){$ \dfrac{3}{8} i s_{2 \alpha } [c_{2 \alpha } \left(-4 \lambda _{\Phi \Delta}+\lambda _{\Delta }+\lambda _{\Phi }\right)-\lambda _{\Delta
   }+\lambda _{\Phi }]$};
        \end{feynman}
    \end{tikzpicture}
    \\[.5cm]
    \begin{tikzpicture}[baseline=(current bounding box.center)]
        \begin{feynman}
            \vertex (a);
            \vertex [above left=1cm of a] (b) {$\Delta^0$};
            \vertex [below left=1cm of a] (c) {$\Delta^0$};
            \vertex [above right=1cm of a] (e) {$\Delta^-$};
            \vertex [below right=1cm of a] (f) {$\Delta^+$};
            \diagram{
                (c) -- [scalar] (a) -- [scalar] (b);
                (f) -- [scalar] (a) -- [scalar] (e);
            };
            \vertex [right=1.5cm of a] (e){$ -\dfrac{1}{2} i [c_{\alpha }^2 \left(c_{\beta }^2 \lambda _{\Delta }+2 s_{\beta }^2 \lambda _{\Phi \Delta}\right)+s_{\alpha }^2 \left(2
   c_{\beta }^2 \lambda _{\Phi \Delta}+\lambda _{\Phi } s_{\beta }^2\right)]$};
        \end{feynman}
    \end{tikzpicture}
    \\[.5cm]
    \begin{tikzpicture}[baseline=(current bounding box.center)]
        \begin{feynman}
            \vertex (a);
            \vertex [above left=1cm of a] (b) {$h$};
            \vertex [below left=1cm of a] (c) {$h$};
            \vertex [above right=1cm of a] (e) {$\Delta^-$};
            \vertex [below right=1cm of a] (f) {$\Delta^+$};
            \diagram{
                (c) -- [scalar] (a) -- [scalar] (b);
                (f) -- [scalar] (a) -- [scalar] (e);
            };
            \vertex [right=1.5cm of a] (e){$ -\dfrac{1}{2} i [c_{\alpha }^2 \left(2 c_{\beta }^2 \lambda _{\Phi \Delta}+\lambda _{\Phi } s_{\beta }^2\right)+s_{\alpha }^2 \left(c_{\beta
   }^2 \lambda _{\Delta }+2 s_{\beta }^2 \lambda _{\Phi \Delta}\right)]$};
        \end{feynman}
    \end{tikzpicture}
    \\[.5cm]
    \begin{tikzpicture}[baseline=(current bounding box.center)]
        \begin{feynman}
            \vertex (a);
            \vertex [above left=1cm of a] (b) {$\Delta^0$};
            \vertex [below left=1cm of a] (c) {$h$};
            \vertex [above right=1cm of a] (e) {$\Delta^-$};
            \vertex [below right=1cm of a] (f) {$\Delta^+$};
            \diagram{
                (c) -- [scalar] (a) -- [scalar] (b);
                (f) -- [scalar] (a) -- [scalar] (e);
            };
            \vertex [right=1.5cm of a] (e){$ -\dfrac{1}{8} i s_{2 \alpha } [c_{2 \beta } \left(-4 \lambda _{\Phi \Delta}+\lambda _{\Delta }+\lambda _{\Phi }\right)+\lambda _{\Delta
   }-\lambda _{\Phi }]$};
        \end{feynman}
    \end{tikzpicture}
    \\[.5cm]
    \begin{tikzpicture}[baseline=(current bounding box.center)]
        \begin{feynman}
            \vertex (a);
            \vertex [above left=1cm of a] (b) {$\Delta^+$};
            \vertex [below left=1cm of a] (c) {$\Delta^-$};
            \vertex [above right=1cm of a] (e) {$\Delta^-$};
            \vertex [below right=1cm of a] (f) {$\Delta^+$};
            \diagram{
                (c) -- [scalar] (a) -- [scalar] (b);
                (f) -- [scalar] (a) -- [scalar] (e);
            };
            \vertex [right=1.5cm of a] (e){$ -i [c_{\beta }^4 \lambda _{\Delta }+s_{2 \beta }^2 \lambda _{\Phi \Delta}+\lambda _{\Phi } s_{\beta }^4]$};
        \end{feynman}
    \end{tikzpicture}
\section{Loop functions}
\label{app:LoopFunc}
In this appendix, we collect all the loop functions or form factors used in the Higgs decays in Sec.~\ref{sec:hDecays} and Sec.~\ref{sec:tripletDecays}. The ones relevant for the loop-induced Higgs decays are
\begin{align*}
&\beta_{\gamma\gamma}^0(x) = -x\left[1-xf(x)\right],
\\
&\beta_{\gamma\gamma}^{1/2}(x) = 2x\left[1+(1-x)f(x)\right],
\\   
&\beta_{\gamma\gamma}^1(x) = -\left[2+3x+3x(2-x)f(x)\right],
\\
&\beta_{Z\gamma}^0(x,y) = \frac{\cos 2\theta_w}{\cos\theta_w} I_1(x,y),
\\
&\beta_{Z\gamma}^{1/2}(x,y) = I_1(x,y)-I_2(x,y),
\\
&\beta_{Z\gamma}^1(x,y) = \cos\theta_w \left[ \left\{ \left(1+\frac{2}{x}\right) \tan^2\theta_w - \left(5+\frac{2}{x} \right) \right\} I_1(x,y) + 4(3-\tan^2\theta_w) I_2(x,y) \right],
\end{align*}
with the functions $f(x), j(x), I_1(x,y)$ and $I_2(x,y)$ defined as
\begin{align*}
&f(x) = \begin{cases} 
\left[\sin^{-1}(\sqrt{1/x})\right]^2 & x \ge 1 
\\   
-\frac{1}{4} \left[\log{\frac{1+\sqrt{1-x}}{1-\sqrt{1-x}}} - i\pi\right]^2 & x < 1 
\end{cases} 
\\
&j(x) = \begin{cases} 
\sqrt{x-1} \sin^{-1}(\sqrt{1/x}) & x \ge 1 
\\   
\frac{1}{2} \sqrt{1-x} \left[\log{\frac{1+\sqrt{1-x}}{1-\sqrt{1-x}}} - i\pi\right] & x < 1
\end{cases}
\\
&I_1(x,y) = \frac{xy}{2(x-y)} + \frac{x^2y^2}{2(x-y)^2} \left[ f(x)-f(y) \right] + \frac{x^2y}{(x-y)^2} \left[ j(x)-j(y) \right],
\\
&I_2(x,y) = -\frac{xy}{2(x-y)}\left[ f(x)-f(y) \right],
\end{align*}
and the ones relevant for the tree-level Higgs decays are
\begin{align*}
&\beta(x,y) = (1-x-y)^2-4xy,
\\
&\beta_{ff^\prime}(x,y) = \left[ (x+y)(1-x-y)-4xy \right] \times \sqrt{\beta(x,y)},
\\
&\beta_S(x) = (x-1)\left(2-\frac{1}{2} \log x\right) + \frac{1-5x}{\sqrt{4x-1}}\left[\tan^{-1}\frac{2x-1}{\sqrt{4x-1}}-\tan^{-1}\frac{1}{\sqrt{4x-1}}\right],
\\
&\beta_V(x) = (1-4x+12x^2) \times \sqrt{\beta(x,x)},
\\
&\beta_V^\prime(x) = \frac{3(1-8x+20x^2)}{\sqrt{4x-1}} \cos^{-1}\frac{3x-1}{2x\sqrt{x}} - \frac{(1-x)}{2x}\left(2-13x+47x^2\right) -\frac{3}{2}(1-6x+4x^2)\log{x},
\\
&\beta_t(x,y) = \frac{y^2}{x^3}(4xy+3x-4y) \log \frac{y(x-1)}{x-y} +(3x^2-4x-3y^2+1) \log\frac{x-1}{x-y} -\frac{5}{2}
\\
&\hspace{1.5cm} +\frac{1-y}{x^2} (3x^3-xy-2xy^2+4y^2) + y\left(4-\frac{3}{2}y\right),
\\
&G(x,y) = \frac{1}{12y}\Bigg[2\left(-1+x\right)^3-9\left(-1+x^2\right)y+6\left(-1+x\right)y^2 -6\left(1+x-y\right)y\sqrt{-\beta(x,y)} \times
\\
&\hspace{1.5cm} \Bigg\{\tan^{-1}\left(\frac{1-x+y}{\sqrt{-\beta(x,y)}}\right) + \tan^{-1}\left(\frac{1-x-y}{\sqrt{-\beta(x,y)}}\right)\Bigg\}-3\left(1+\left(x-y\right)^2-2y\right)y\log x\Bigg],
\\
&H(x,y) = \frac{1}{4x \sqrt{-\beta(x,y)}} \left\{ \tan^{-1}\left(\frac{1-x+y}{\sqrt{-\beta(x,y)}}\right) +\tan^{-1}\left(\frac{1-x-y}{\sqrt{-\beta(x,y)}}\right) \right\} \Big\{-3x^3+(9y+7)x^2
\\
&\hspace{1.5cm} -5(1-y)^2x+(1-y)^3\Big\} +\frac{1}{24xy}\Big\{(-1+x)(2+2x^2+6y^2-4x-9y+39xy)
\\
&\hspace{1.5cm} -3y(1-3x^2+y^2-4x-2y+6xy)\log x\Big\},
\end{align*}
\section{Stability and charge-breaking minima}
\label{app:globality}
The scalar potential in Eq.~\eqref{eq:pot}, in principle, can have several stationary points---both electric charge-conserving and charge-breaking. In this section, we discuss all possible stationary points and obtain the necessary conditions for the desired EW vacuum to be the global minimum of the model by calculating the differences in the potential depths between the desired EW vacuum and others, some of which might not be phenomenologically viable. It might be argued that absolute stability of the vacuum is not a necessary requirement since the vacuum could be meta-stable, i.e.~have, a lifetime longer than the age of the Universe. While the latter is certainly an interesting possibility, it would require a detailed calculation of the tunnelling time of the vacuum into the global minimum, which is beyond the scope of this work. Therefore, we restrict to the stronger requirement that the vacuum is at the global minimum.

Concerning the gauge choice, note that we can always absorb three real scalar component fields via a suitable gauge choice (such as a rotation in field space). For definiteness, we choose the SM unitary gauge, wherein the doublet is reduced to a neutral, real component. As a consequence, the doublet VEV is, without loss of generality, always real and neutral. Further, the triplet being real (see Eq.~\eqref{eq:triplet} and the text thereafter), the corresponding VEVs are also real.

The $\Delta$SM can have several stationary points as listed in Table~\ref{Tab:vacua}. Of these, three are charge-conserving minima: $N_1$, $N_2$ and $N_3$, referred to as {\it normal minima}, wherein the neutral components of the scalar fields get VEVs. On the contrary, there are two possible charge-breaking (CB) minima. Such a scenario exists when electric charge-carrying VEVs appear after spontaneous symmetry breaking. The occurrence of CB vacuum results in a non-zero photon mass, which is incompatible with electromagnetic observations.
 
\begin{table}[b!]
\begin{center}
\begin{tabular}{|c|ccc|}
\hline
N1 & $\langle \Phi^0\rangle = \frac{1}{\sqrt{2}}v_\Phi$ & $\langle \Delta^0\rangle = v_\Delta$ & $\langle \Delta^\pm\rangle = 0$  \\
\hline
N2 & $\langle \Phi^0\rangle = \frac{1}{\sqrt{2}}v_\Phi$ & $\langle \Delta^0\rangle = 0$ & $\langle \Delta^\pm\rangle = 0$  \\
\hline
N3 & $\langle \Phi^0\rangle = 0$ & $\langle \Delta^0\rangle = v_\Delta$ & $\langle \Delta^\pm\rangle = 0$  \\
\hline
CB1 & $\langle \Phi^0\rangle = \frac{1}{\sqrt{2}}v$ & $\langle \Delta^0\rangle = v_0$ & $\langle \Delta^\pm\rangle = v_1$  \\
\hline
CB2 & $\langle \Phi^0\rangle = \frac{1}{\sqrt{2}}v$ & $\langle \Delta^0\rangle =0$ & $\langle \Delta^\pm\rangle = v_1$ \\ 
\hline
\end{tabular}
\end{center}
\caption{Possible stationary points in the $\Delta$SM model.}
\label{Tab:vacua}
\end{table}

\subsection*{{\it N1} stationary point}
The {\it N1} stationary point corresponds to the vacuum structure where both $\Phi^0$ and $\Delta^0$ get VEVs:
\begin{equation}
\langle \Phi\rangle_{N1}  = \frac{1}{\sqrt{2}}\begin{pmatrix} 0 \\ v_\Phi \end{pmatrix}, \quad \langle \Delta\rangle_{N1} = \frac{1}{2} \begin{pmatrix} v_\Delta & 0 \\ 0 & -v_\Delta \end{pmatrix},
\end{equation}
This vacuum structure has been detailed in Section~\ref{sec:model}.

\subsection*{{\it N2} stationary point}
The {\it N2} stationary point corresponds to the vacuum structure where only $\Phi^0$ gets VEV:
\begin{equation}
\langle \Phi\rangle_{N2}  =  \frac{1}{\sqrt{2}} \begin{pmatrix} 0 \\ v_\Phi \end{pmatrix}, \quad \langle \Delta\rangle_{N2} = \frac{1}{2} \begin{pmatrix} 0 & 0 \\ 0 & 0 \end{pmatrix},
\end{equation}
The minimization condition for this vacuum configuration is
\begin{align}
\mu_\Phi^2 &= \frac{1}{4} v_\Phi^2 \lambda_{\Phi}.
\end{align}
To get the correct electroweak symmetry breaking for this vacuum configuration, we need $v_\Phi = 246$\,GeV. Also, to get {\it N2} extremum, we require $A = 0$, thereby restoring the global $O(4)_H \times O(3)_\Delta$ symmetry of the potential in Eq.~\eqref{eq:pot}. The consequence is the degenerate mass spectrum for the triplet scalars: $m_{\Delta^0} = m_{\Delta^\pm}$. The scalar masses are given by, 
\begin{align}
& m_h^2 = \frac{\lambda_{\Phi }v_\Phi^2}{2},
\\
& m_{\Delta^0}^2 = m_{\Delta^\pm}^2 = \frac{\lambda_{\Phi \Delta} v^2_\Phi}{2} - \mu_\Delta^2.
\end{align}

\subsection*{{\it N3} stationary point}
The {\it N3} stationary point corresponds to the vacuum structure where only $\Delta^0$ gets VEV:
\begin{equation}
\langle \Phi\rangle_{N1}  = \frac{1}{\sqrt{2}}\begin{pmatrix} 0 \\ 0 \end{pmatrix}, \quad \langle \Delta\rangle_{N1} = \frac{1}{2} \begin{pmatrix} v_\Delta & 0 \\ 0 & -v_\Delta \end{pmatrix},
\end{equation}
As the absence of a doublet VEV would imply massless fermions, this extremum is in disagreement with observations. Therefore, we do not discuss this further.

\subsection*{{\it CB1} stationary point}
In the CB1 case, both the neutral and charged components of the triplet scalar, $\Delta^0$ and $\Delta^\pm$ get VEVs:
\begin{equation}
\langle \Phi\rangle = \frac{1}{\sqrt{2}}\begin{pmatrix} 0 \\ v \end{pmatrix}, \quad \langle \Delta\rangle = \frac{1}{2} \begin{pmatrix} v_0 &  v_1 \\ v_1 & -v_0 \end{pmatrix}.
\end{equation}
This configuration is the most generic CB vacuum. The minimization condition for this vacuum configuration is
\begin{align}
&\mu_\Phi^2 = \frac{v^2}{4}\lambda_{\Phi } + \left( \frac{v_0^2}{2} + v_1^2 \right) \lambda_{\Phi\Delta } 
\\
&\mu_\Delta^2 = \left( \frac{v_0^2}{4} + \frac{v_1^2}{2} \right) \lambda_{\Delta } + \frac{v^2}{2}\lambda_{\Phi\Delta} 
\\
&A = 0
\end{align}

\subsection*{{\it CB2} stationary point}
In the CB2 case, only the charged component of the triplet scalar, $\Delta^\pm$ gets VEV:
\begin{equation}
\langle \Phi\rangle = \frac{1}{\sqrt{2}}\begin{pmatrix} 0 \\ v \end{pmatrix}, \quad \langle \Delta\rangle = \frac{1}{2} \begin{pmatrix} 0 & v_1 \\ v_1 & 0 \end{pmatrix}
\end{equation}
The minimization condition for this vacuum configuration is
\begin{align}
&\mu_\Phi^2 = \frac{v^2}{4}\lambda_\Phi + v_1^2\lambda_{\Phi\Delta} 
\\
&\mu_\Delta^2 = \frac{v_1^2}{2}\lambda_\Delta + \frac{v^2}{2}\lambda_{\Phi\Delta}
\end{align}

\subsection*{Condition for global minima}
In this section, we will examine the stability of the {\it normal minima}, in particular, the {\it N1} stationary point, which we assume to be the desired global minimum of the model. To this end, assuming that {\it N1} or {\it N2} coexists with the {\it CB} minima in some region of the model parameter space, we calculate the differences in potential depth between the co-existing minima. For {\it N1} to be the global minimum, it is crucial that the potential depth at ${\it N1}$ is lower than those of the co-existing minima.

To calculate the difference in potential depth between two co-existing minima, we follow the bilinear formalism detailed in Refs.~\cite{Ferreira:2004yd,Barroso:2005sm}, see Refs.~\cite{Ferreira:2019hfk,Azevedo:2020mjg,Hundi:2023tdq} for recent works. Note that the potential $V$ in Eq.~\eqref{eq:pot} is the sum of three homogeneous functions of order 2, 3 and 4 in the fields $\Phi$ and $\Delta$: 
\begin{align}
V &= V_2 + V_3 + V_4.
\end{align}
Consequently, up to $V_3$, $V$ can be expressed as a quadratic polynomial on the vector $X$ constructed from the bilinears $x_1 = |\Phi|^2$, $x_2 = |\Delta^0|^2/2$ and $x_3 = |\Delta^\pm|^2/2$. Therefore, we have
\begin{align}
V &= \underbrace{M^T X}_{V_2} + \underbrace{A \Phi^\dag \Delta \Phi}_{V_3} + \underbrace{\frac{1}{2} X^T \Lambda X}_{V_4},
\end{align}
where
\begin{align}
& X = \begin{pmatrix} x_1 \\ x_2 \\ x_3 \end{pmatrix},
\quad
M = \begin{pmatrix} -\mu_\Phi^2 \\ -\mu_\Delta^2  \\ -2 \mu_\Delta^2 \end{pmatrix}, 
\quad
\Lambda  =  \begin{pmatrix} \frac{\lambda_{\Phi}}{2} & \lambda_{\Phi\Delta} & 2\lambda_{\Phi\Delta}  
\\ 
\lambda_{\Phi\Delta} & \frac{\lambda_{\Delta}}{2} & \lambda_\Delta
\\ 
2 \lambda_{\Phi\Delta} & \lambda_\Delta &2 \lambda_\Delta \end{pmatrix}
\end{align}

At any stationary point {\it SP}, the potential $V$ must follow the minimisation conditions, {\it i.e.} $\frac{\partial V}{\partial \phi_i} = 0$ with $\phi_i$ denoting real scalar degree of freedoms of the model. Thus, we have
\begin{align} 
& \sum_i \phi_i\frac{\partial V}{\partial \phi_i}\bigg|_{SP} = 0 
\quad 
\Rightarrow 2(V_2)_{SP} + 3(V_3)_{SP} + 4(V_4)_{SP} = 0,
\end{align}
where the latter follows from Euler's homogeneous function theorem. Therefore, the value of $V$ evaluated at {\it SP} is
\begin{align}
V_{SP} &= \frac{1}{2} (V_2)_{SP} + \frac{1}{4} (V_4)_{SP}.
\end{align}
Now, we define the gradient of $V_2+V_4$ along $X$: $V^\prime = M + \Lambda X$ and compute the product of $V^\prime$ evaluated at a stationary point with $X$ evaluated at a different stationary point:
\begin{align}
&X_{SP1}^T V^\prime_{SP2} = 2V_{SP1} - \frac{1}{2}(V_3)_{SP1} + X_{SP1}^T \Lambda X_{SP2},
\\
&X_{SP2}^T V^\prime_{SP1} = 2V_{SP2} - \frac{1}{2}(V_3)_{SP2} + X_{SP2}^T \Lambda X_{SP1}
\end{align}
Finally, subtracting these two equations, we get
\begin{equation}
V_{SP2} - V_{SP1} = \frac{1}{2} \left(X_{SP2}^T V^\prime_{SP1} - X_{SP1}^T V^\prime_{SP2}\right) + \frac{1}{4}\left((V_3)_{SP2} - (V_3)_{SP1}\right).
\label{eq:diff-pot}
\end{equation}
In what follows, we present the bilinear vector $X$ for all possible stationary points and the differences in potential depth between the co-existing minima using Eq.~\eqref{eq:diff-pot}.
\begin{align}
X_{N1}  = \frac{1}{2}\begin{pmatrix} v_\Phi^2 \\ v_\Delta^2 \\ 0 \end{pmatrix},
\quad
X_{N2}  = \frac{1}{2}\begin{pmatrix} v_\Phi^2 \\ 0 \\ 0 \end{pmatrix},
\quad
X_{CB1}  = \frac{1}{2}\begin{pmatrix} v^2 \\ v_1^2 \\ v_2^2 \end{pmatrix},
\quad
X_{CB2}  = \frac{1}{2}\begin{pmatrix} v^2 \\ 0 \\ v_2^2 \end{pmatrix},
\end{align}

\begin{align}
& V_{CB1} - V_{N1} = \frac{(v_\Delta^2+2 v_1^2+4 v_2^2)v_\Phi^2+4 v_\Delta^2 v^2}{8(4v_\Delta^2 + v_\Phi^2)} m_{\Delta^\pm}^2, \label{eq:CB1_N1}
\\
& V_{CB2} - V_{N1} = \frac{4v_2^2 v_\Phi^2 + v_\Delta^2(v_\Phi^2 + 4v^2)}{8(4v_\Delta^2 + v_\Phi^2)} m_{\Delta^\pm}^2, \label{eq:CB2_N1}
\\
& V_{N2}-V_{N1} = \frac{v_\Delta^2}{2} m_{h}^2, \label{eq:N2_N1}
\\
& V_{CB1} - V_{N2} = \frac{v_1^2 + v_2^2}{4} m_{\Delta^0}^2, \label{eq:CB1_N2}
\\
& V_{CB2}-V_{N2} = \frac{v_2^2}{2} m_{\Delta^0}^2,  \label{eq:CB2_N2}
\end{align}
where $m_h$, $m_{\Delta^0}$ and $m_{\Delta^\pm}$ in Eq.~\eqref{eq:CB1_N1}, Eq.~\eqref{eq:CB2_N1} and Eq.~\eqref{eq:N2_N1} are the physical masses evaluated at the $N1$ stationary point, whereas those in Eq.~\eqref{eq:CB1_N2} and \eqref{eq:CB2_N2} are evaluated at the $N_2$ stationary point. Therefore, as it turns out, the desired EW vacuum $N_1$ is indeed the global minimum of the model for $v_\Delta > 0$ as long as the scalars are not tachyonic, {\it i.e.}~$m_h^2 > 0$, $m_{\Delta^0}^2 > 0$ and $m_{\Delta^\pm}^2 > 0$. While for $v_\Delta = 0$, the normal minimum $N_2$ is the global minimum.

\acknowledgments
S.A.~acknowledges partial support from the National Natural Science Foundation of China under grant No.~11835013. The work of A.C.~is supported by a professorship grant from the Swiss National Science Foundation (No.~PP00P21\_76884). S.P.M.~acknowledges using the SAMKHYA: High-Performance Computing Facility provided by the Institute of Physics, Bhubaneswar.

\bibliographystyle{utphys}
\bibliography{Y0triplet}

\providecommand{\href}[2]{#2}\begingroup\raggedright\begin{thebibliography}{100}

\bibitem{ParticleDataGroup:2022pth}
{\bfseries Particle Data Group} Collaboration, R.~L. Workman {\em et~al.},
  ``{Review of Particle Physics},''
  \href{http://dx.doi.org/10.1093/ptep/ptac097}{{\em PTEP} {\bfseries 2022}
  (2022) 083C01}.

\bibitem{HFLAV:2019otj}
{\bfseries HFLAV} Collaboration, Y.~S. Amhis {\em et~al.}, ``{Averages of
  $b$-hadron, $c$-hadron, and $\tau $-lepton properties as of 2018},''
  \href{http://dx.doi.org/10.1140/epjc/s10052-020-8156-7}{{\em Eur. Phys. J. C}
  {\bfseries 81} no.~3, (2021) 226},
  \href{http://arxiv.org/abs/1909.12524}{{\ttfamily arXiv:1909.12524
  [hep-ex]}}.

\bibitem{ALEPH:2005ab}
{\bfseries ALEPH, DELPHI, L3, OPAL, SLD, LEP Electroweak Working Group, SLD
  Electroweak Group, SLD Heavy Flavour Group} Collaboration, S.~Schael {\em
  et~al.}, ``{Precision electroweak measurements on the $Z$ resonance},''
  \href{http://dx.doi.org/10.1016/j.physrep.2005.12.006}{{\em Phys. Rept.}
  {\bfseries 427} (2006) 257--454},
  \href{http://arxiv.org/abs/hep-ex/0509008}{{\ttfamily arXiv:hep-ex/0509008}}.

\bibitem{Higgs:1964ia}
P.~W. Higgs, ``{Broken symmetries, massless particles and gauge fields},''
  \href{http://dx.doi.org/10.1016/0031-9163(64)91136-9}{{\em Phys. Lett.}
  {\bfseries 12} (1964) 132--133}.

\bibitem{Englert:1964et}
F.~Englert and R.~Brout, ``{Broken Symmetry and the Mass of Gauge Vector
  Mesons},'' \href{http://dx.doi.org/10.1103/PhysRevLett.13.321}{{\em Phys.
  Rev. Lett.} {\bfseries 13} (1964) 321--323}.

\bibitem{Higgs:1964pj}
P.~W. Higgs, ``{Broken Symmetries and the Masses of Gauge Bosons},''
  \href{http://dx.doi.org/10.1103/PhysRevLett.13.508}{{\em Phys. Rev. Lett.}
  {\bfseries 13} (1964) 508--509}.

\bibitem{Guralnik:1964eu}
G.~S. Guralnik, C.~R. Hagen, and T.~W.~B. Kibble, ``{Global Conservation Laws
  and Massless Particles},''
  \href{http://dx.doi.org/10.1103/PhysRevLett.13.585}{{\em Phys. Rev. Lett.}
  {\bfseries 13} (1964) 585--587}.

\bibitem{Aad:2012tfa}
{\bfseries ATLAS} Collaboration, G.~Aad {\em et~al.}, ``{Observation of a new
  particle in the search for the Standard Model Higgs boson with the ATLAS
  detector at the LHC},''
  \href{http://dx.doi.org/10.1016/j.physletb.2012.08.020}{{\em Phys. Lett. B}
  {\bfseries 716} (2012) 1--29},
  \href{http://arxiv.org/abs/1207.7214}{{\ttfamily arXiv:1207.7214 [hep-ex]}}.

\bibitem{Chatrchyan:2012ufa}
{\bfseries CMS} Collaboration, S.~Chatrchyan {\em et~al.}, ``{Observation of a
  New Boson at a Mass of 125 GeV with the CMS Experiment at the LHC},''
  \href{http://dx.doi.org/10.1016/j.physletb.2012.08.021}{{\em Phys. Lett. B}
  {\bfseries 716} (2012) 30--61},
  \href{http://arxiv.org/abs/1207.7235}{{\ttfamily arXiv:1207.7235 [hep-ex]}}.

\bibitem{Crivellin:2023zui}
A.~Crivellin and B.~Mellado, ``{Anomalies in particle physics and their
  implications for physics beyond the standard model},''
  \href{http://dx.doi.org/10.1038/s42254-024-00703-6}{{\em Nature Rev. Phys.}
  {\bfseries 6} no.~5, (2024) 294--309},
  \href{http://arxiv.org/abs/2309.03870}{{\ttfamily arXiv:2309.03870
  [hep-ph]}}.

\bibitem{Chatrchyan:2012jja}
{\bfseries CMS} Collaboration, S.~Chatrchyan {\em et~al.}, ``{Study of the Mass
  and Spin-Parity of the Higgs Boson Candidate Via Its Decays to $Z$ Boson
  Pairs},'' \href{http://dx.doi.org/10.1103/PhysRevLett.110.081803}{{\em Phys.
  Rev. Lett.} {\bfseries 110} no.~8, (2013) 081803},
  \href{http://arxiv.org/abs/1212.6639}{{\ttfamily arXiv:1212.6639 [hep-ex]}}.

\bibitem{Aad:2013xqa}
{\bfseries ATLAS} Collaboration, G.~Aad {\em et~al.}, ``{Evidence for the
  spin-0 nature of the Higgs boson using ATLAS data},''
  \href{http://dx.doi.org/10.1016/j.physletb.2013.08.026}{{\em Phys. Lett. B}
  {\bfseries 726} (2013) 120--144},
  \href{http://arxiv.org/abs/1307.1432}{{\ttfamily arXiv:1307.1432 [hep-ex]}}.

\bibitem{ATLAS:2016neq}
{\bfseries ATLAS, CMS} Collaboration, G.~Aad {\em et~al.}, ``{Measurements of
  the Higgs boson production and decay rates and constraints on its couplings
  from a combined ATLAS and CMS analysis of the LHC $pp$ collision data at
  $\sqrt{s}=7$ and 8 TeV},''
  \href{http://dx.doi.org/10.1007/JHEP08(2016)045}{{\em JHEP} {\bfseries 08}
  (2016) 045}, \href{http://arxiv.org/abs/1606.02266}{{\ttfamily
  arXiv:1606.02266 [hep-ex]}}.

\bibitem{Langford:2021osp}
{\bfseries ATLAS, CMS} Collaboration, J.~M. Langford, ``{Combination of Higgs
  measurements from ATLAS and CMS : couplings and $\kappa$-framework},''
  \href{http://dx.doi.org/10.22323/1.382.0136}{{\em PoS} {\bfseries LHCP2020}
  (2021) 136}.

\bibitem{ATLAS:2021vrm}
{\bfseries ATLAS} Collaboration, ``{Combined measurements of Higgs boson
  production and decay using up to $139$ fb$^{-1}$ of proton-proton collision
  data at $\sqrt{s}= 13$ TeV collected with the ATLAS experiment},''.

\bibitem{Silveira:1985rk}
V.~Silveira and A.~Zee, ``{SCALAR PHANTOMS},''
  \href{http://dx.doi.org/10.1016/0370-2693(85)90624-0}{{\em Phys. Lett. B}
  {\bfseries 161} (1985) 136--140}.

\bibitem{Pietroni:1992in}
M.~Pietroni, ``{The Electroweak phase transition in a nonminimal supersymmetric
  model},'' \href{http://dx.doi.org/10.1016/0550-3213(93)90635-3}{{\em Nucl.
  Phys. B} {\bfseries 402} (1993) 27--45},
  \href{http://arxiv.org/abs/hep-ph/9207227}{{\ttfamily arXiv:hep-ph/9207227}}.

\bibitem{McDonald:1993ex}
J.~McDonald, ``{Gauge singlet scalars as cold dark matter},''
  \href{http://dx.doi.org/10.1103/PhysRevD.50.3637}{{\em Phys. Rev. D}
  {\bfseries 50} (1994) 3637--3649},
  \href{http://arxiv.org/abs/hep-ph/0702143}{{\ttfamily arXiv:hep-ph/0702143}}.

\bibitem{Lee:1973iz}
T.~D. Lee, ``{A Theory of Spontaneous $T$ Violation},''
  \href{http://dx.doi.org/10.1103/PhysRevD.8.1226}{{\em Phys. Rev. D}
  {\bfseries 8} (1973) 1226--1239}.

\bibitem{Haber:1984rc}
H.~E. Haber and G.~L. Kane, ``{The Search for Supersymmetry: Probing Physics
  Beyond the Standard Model},''
  \href{http://dx.doi.org/10.1016/0370-1573(85)90051-1}{{\em Phys. Rept.}
  {\bfseries 117} (1985) 75--263}.

\bibitem{Kim:1986ax}
J.~E. Kim, ``{Light Pseudoscalars, Particle Physics and Cosmology},''
  \href{http://dx.doi.org/10.1016/0370-1573(87)90017-2}{{\em Phys. Rept.}
  {\bfseries 150} (1987) 1--177}.

\bibitem{Peccei:1977hh}
R.~D. Peccei and H.~R. Quinn, ``{$CP$ Conservation in the Presence of
  Instantons},'' \href{http://dx.doi.org/10.1103/PhysRevLett.38.1440}{{\em
  Phys. Rev. Lett.} {\bfseries 38} (1977) 1440--1443}.

\bibitem{Turok:1990zg}
N.~Turok and J.~Zadrozny, ``{Electroweak baryogenesis in the two doublet
  model},'' \href{http://dx.doi.org/10.1016/0550-3213(91)90356-3}{{\em Nucl.
  Phys. B} {\bfseries 358} (1991) 471--493}.

\bibitem{Konetschny:1977bn}
W.~Konetschny and W.~Kummer, ``{Nonconservation of Total Lepton Number with
  Scalar Bosons},'' \href{http://dx.doi.org/10.1016/0370-2693(77)90407-5}{{\em
  Phys. Lett. B} {\bfseries 70} (1977) 433--435}.

\bibitem{Cheng:1980qt}
T.~P. Cheng and L.-F. Li, ``{Neutrino Masses, Mixings and Oscillations in
  $SU(2) \times U(1)$ Models of Electroweak Interactions},''
  \href{http://dx.doi.org/10.1103/PhysRevD.22.2860}{{\em Phys. Rev. D}
  {\bfseries 22} (1980) 2860}.

\bibitem{Lazarides:1980nt}
G.~Lazarides, Q.~Shafi, and C.~Wetterich, ``{Proton Lifetime and Fermion Masses
  in an $SO(10)$ Model},''
  \href{http://dx.doi.org/10.1016/0550-3213(81)90354-0}{{\em Nucl. Phys. B}
  {\bfseries 181} (1981) 287--300}.

\bibitem{Schechter:1980gr}
J.~Schechter and J.~W.~F. Valle, ``{Neutrino Masses in $SU(2) \times U(1)$
  Theories},'' \href{http://dx.doi.org/10.1103/PhysRevD.22.2227}{{\em Phys.
  Rev. D} {\bfseries 22} (1980) 2227}.

\bibitem{Magg:1980ut}
M.~Magg and C.~Wetterich, ``{Neutrino Mass Problem and Gauge Hierarchy},''
  \href{http://dx.doi.org/10.1016/0370-2693(80)90825-4}{{\em Phys. Lett. B}
  {\bfseries 94} (1980) 61--64}.

\bibitem{Mohapatra:1980yp}
R.~N. Mohapatra and G.~Senjanovic, ``{Neutrino Masses and Mixings in Gauge
  Models with Spontaneous Parity Violation},''
  \href{http://dx.doi.org/10.1103/PhysRevD.23.165}{{\em Phys. Rev. D}
  {\bfseries 23} (1981) 165}.

\bibitem{CDF:2022hxs}
{\bfseries CDF} Collaboration, T.~Aaltonen {\em et~al.}, ``{High-precision
  measurement of the $W$ boson mass with the CDF II detector},''
  \href{http://dx.doi.org/10.1126/science.abk1781}{{\em Science} {\bfseries
  376} no.~6589, (2022) 170--176}.

\bibitem{Butterworth:2022dkt}
J.~Butterworth, J.~Heeck, S.~H. Jeon, O.~Mattelaer, and R.~Ruiz, ``{Testing the
  scalar triplet solution to CDF's heavy $W$ problem at the LHC},''
  \href{http://dx.doi.org/10.1103/PhysRevD.107.075020}{{\em Phys. Rev. D}
  {\bfseries 107} no.~7, (2023) 075020},
  \href{http://arxiv.org/abs/2210.13496}{{\ttfamily arXiv:2210.13496
  [hep-ph]}}.

\bibitem{Heeck:2022fvl}
J.~Heeck, ``{$W$-boson mass in the triplet seesaw model},''
  \href{http://dx.doi.org/10.1103/PhysRevD.106.015004}{{\em Phys. Rev. D}
  {\bfseries 106} no.~1, (2022) 015004},
  \href{http://arxiv.org/abs/2204.10274}{{\ttfamily arXiv:2204.10274
  [hep-ph]}}.

\bibitem{Strumia:2022qkt}
A.~Strumia, ``{Interpreting electroweak precision data including the $W$-mass
  CDF anomaly},'' \href{http://dx.doi.org/10.1007/JHEP08(2022)248}{{\em JHEP}
  {\bfseries 08} (2022) 248}, \href{http://arxiv.org/abs/2204.04191}{{\ttfamily
  arXiv:2204.04191 [hep-ph]}}.

\bibitem{Dorsner:2007fy}
I.~Dorsner and I.~Mocioiu, ``{Predictions from type II see-saw mechanism in
  $SU(5)$},'' \href{http://dx.doi.org/10.1016/j.nuclphysb.2007.12.004}{{\em
  Nucl. Phys. B} {\bfseries 796} (2008) 123--136},
  \href{http://arxiv.org/abs/0708.3332}{{\ttfamily arXiv:0708.3332 [hep-ph]}}.

\bibitem{FileviezPerez:2022lxp}
P.~Fileviez~Perez, H.~H. Patel, and A.~D. Plascencia, ``{On the $W$ mass and
  new Higgs bosons},''
  \href{http://dx.doi.org/10.1016/j.physletb.2022.137371}{{\em Phys. Lett. B}
  {\bfseries 833} (2022) 137371},
  \href{http://arxiv.org/abs/2204.07144}{{\ttfamily arXiv:2204.07144
  [hep-ph]}}.

\bibitem{Cheng:2022hbo}
Y.~Cheng, X.-G. He, F.~Huang, J.~Sun, and Z.-P. Xing, ``{Electroweak precision
  tests for triplet scalars},''
  \href{http://dx.doi.org/10.1016/j.nuclphysb.2023.116118}{{\em Nucl. Phys. B}
  {\bfseries 989} (2023) 116118},
  \href{http://arxiv.org/abs/2208.06760}{{\ttfamily arXiv:2208.06760
  [hep-ph]}}.

\bibitem{Rizzo:2022jti}
T.~G. Rizzo, ``{Kinetic mixing, dark Higgs triplets, and $M_W$},''
  \href{http://dx.doi.org/10.1103/PhysRevD.106.035024}{{\em Phys. Rev. D}
  {\bfseries 106} no.~3, (2022) 035024},
  \href{http://arxiv.org/abs/2206.09814}{{\ttfamily arXiv:2206.09814
  [hep-ph]}}.

\bibitem{Wang:2022dte}
J.-W. Wang, X.-J. Bi, P.-F. Yin, and Z.-H. Yu, ``{Electroweak dark matter model
  accounting for the CDF $W$-mass anomaly},''
  \href{http://dx.doi.org/10.1103/PhysRevD.106.055001}{{\em Phys. Rev. D}
  {\bfseries 106} no.~5, (2022) 055001},
  \href{http://arxiv.org/abs/2205.00783}{{\ttfamily arXiv:2205.00783
  [hep-ph]}}.

\bibitem{Chabab:2018ert}
M.~Chabab, M.~C. Peyran\`ere, and L.~Rahili, ``{Probing the Higgs sector of
  $Y=0$ Higgs Triplet Model at LHC},''
  \href{http://dx.doi.org/10.1140/epjc/s10052-018-6339-2}{{\em Eur. Phys. J. C}
  {\bfseries 78} no.~10, (2018) 873},
  \href{http://arxiv.org/abs/1805.00286}{{\ttfamily arXiv:1805.00286
  [hep-ph]}}.

\bibitem{Shimizu:2023rvi}
Y.~Shimizu and S.~Takeshita, ``{$W$ boson mass and grand unification via the
  type-II seesaw-like mechanism},''
  \href{http://dx.doi.org/10.1016/j.nuclphysb.2023.116290}{{\em Nucl. Phys. B}
  {\bfseries 994} (2023) 116290},
  \href{http://arxiv.org/abs/2303.11070}{{\ttfamily arXiv:2303.11070
  [hep-ph]}}.

\bibitem{Crivellin:2023xbu}
A.~Crivellin, M.~Kirk, and A.~Thapa, ``{Minimal model for the $W$-boson mass,
  $(g-2)_\mu$, $h\to\mu^+\mu^-$ and quark-mixing-matrix unitarity},''
  \href{http://dx.doi.org/10.1103/PhysRevD.108.L031702}{{\em Phys. Rev. D}
  {\bfseries 108} no.~3, (2023) L031702},
  \href{http://arxiv.org/abs/2305.03081}{{\ttfamily arXiv:2305.03081
  [hep-ph]}}.

\bibitem{Chowdhury:2022moc}
T.~A. Chowdhury, J.~Heeck, A.~Thapa, and S.~Saad, ``{$W$ boson mass shift and
  muon magnetic moment in the Zee model},''
  \href{http://dx.doi.org/10.1103/PhysRevD.106.035004}{{\em Phys. Rev. D}
  {\bfseries 106} no.~3, (2022) 035004},
  \href{http://arxiv.org/abs/2204.08390}{{\ttfamily arXiv:2204.08390
  [hep-ph]}}.

\bibitem{Dcruz:2022dao}
R.~Dcruz and A.~Thapa, ``{$W$ boson mass shift, dark matter, and $(g-2)_\ell$
  in a scotogenic-Zee model},''
  \href{http://dx.doi.org/10.1103/PhysRevD.107.015002}{{\em Phys. Rev. D}
  {\bfseries 107} no.~1, (2023) 015002},
  \href{http://arxiv.org/abs/2205.02217}{{\ttfamily arXiv:2205.02217
  [hep-ph]}}.

\bibitem{Babu:2022pdn}
K.~S. Babu, S.~Jana, and V.~P. K., ``{Correlating $W$-Boson Mass Shift with
  Muon $(g-2)$ in the Two Higgs Doublet Model},''
  \href{http://dx.doi.org/10.1103/PhysRevLett.129.121803}{{\em Phys. Rev.
  Lett.} {\bfseries 129} no.~12, (2022) 121803},
  \href{http://arxiv.org/abs/2204.05303}{{\ttfamily arXiv:2204.05303
  [hep-ph]}}.

\bibitem{Arcadi:2022dmt}
G.~Arcadi and A.~Djouadi, ``{2HD+a light pseudoscalar model for a combined
  explanation of the possible excesses in the CDF $M_W$ measurement and
  $(g-2)_\mu$ with dark matter},''
  \href{http://dx.doi.org/10.1103/PhysRevD.106.095008}{{\em Phys. Rev. D}
  {\bfseries 106} no.~9, (2022) 095008},
  \href{http://arxiv.org/abs/2204.08406}{{\ttfamily arXiv:2204.08406
  [hep-ph]}}.

\bibitem{Kim:2022hvh}
J.~Kim, S.~Lee, P.~Sanyal, and J.~Song, ``{CDF $W$-boson mass and muon $(g-2)$
  in a type-X two-Higgs-doublet model with a Higgs-phobic light
  pseudoscalar},'' \href{http://dx.doi.org/10.1103/PhysRevD.106.035002}{{\em
  Phys. Rev. D} {\bfseries 106} no.~3, (2022) 035002},
  \href{http://arxiv.org/abs/2205.01701}{{\ttfamily arXiv:2205.01701
  [hep-ph]}}.

\bibitem{Kim:2022xuo}
J.~Kim, ``{Compatibility of muon $g-2$, $W$ mass anomaly in type-X 2HDM},''
  \href{http://dx.doi.org/10.1016/j.physletb.2022.137220}{{\em Phys. Lett. B}
  {\bfseries 832} (2022) 137220},
  \href{http://arxiv.org/abs/2205.01437}{{\ttfamily arXiv:2205.01437
  [hep-ph]}}.

\bibitem{Chakrabarty:2022voz}
N.~Chakrabarty, ``{Muon $g-2$ and $W$-mass anomalies explained and the
  electroweak vacuum stabilized by extending the minimal type-II seesaw
  model},'' \href{http://dx.doi.org/10.1103/PhysRevD.108.075024}{{\em Phys.
  Rev. D} {\bfseries 108} no.~7, (2023) 075024},
  \href{http://arxiv.org/abs/2206.11771}{{\ttfamily arXiv:2206.11771
  [hep-ph]}}.

\bibitem{Chowdhury:2023uyd}
T.~A. Chowdhury, K.~Ezzat, S.~Khalil, E.~Ma, and D.~Nanda, ``{Higgs quadruplet
  impact on $W$ mass shift, dark matter, and LHC signatures},''
  \href{http://dx.doi.org/10.1103/PhysRevD.109.075039}{{\em Phys. Rev. D}
  {\bfseries 109} no.~7, (2024) 075039},
  \href{http://arxiv.org/abs/2312.11833}{{\ttfamily arXiv:2312.11833
  [hep-ph]}}.

\bibitem{Chen:2022ocr}
T.-K. Chen, C.-W. Chiang, and K.~Yagyu, ``{Explanation of the $W$ mass shift at
  CDF II in the extended Georgi-Machacek model},''
  \href{http://dx.doi.org/10.1103/PhysRevD.106.055035}{{\em Phys. Rev. D}
  {\bfseries 106} no.~5, (2022) 055035},
  \href{http://arxiv.org/abs/2204.12898}{{\ttfamily arXiv:2204.12898
  [hep-ph]}}.

\bibitem{Kanemura:2022ahw}
S.~Kanemura and K.~Yagyu, ``{Implication of the $W$ boson mass anomaly at CDF
  II in the Higgs triplet model with a mass difference},''
  \href{http://dx.doi.org/10.1016/j.physletb.2022.137217}{{\em Phys. Lett. B}
  {\bfseries 831} (2022) 137217},
  \href{http://arxiv.org/abs/2204.07511}{{\ttfamily arXiv:2204.07511
  [hep-ph]}}.

\bibitem{Ashanujjaman:2022ofg}
S.~Ashanujjaman, K.~Ghosh, and R.~Sahu, ``{Low-mass doubly charged Higgs bosons
  at the LHC},'' \href{http://dx.doi.org/10.1103/PhysRevD.107.015018}{{\em
  Phys. Rev. D} {\bfseries 107} no.~1, (2023) 015018},
  \href{http://arxiv.org/abs/2211.00632}{{\ttfamily arXiv:2211.00632
  [hep-ph]}}.

\bibitem{Ross:1975fq}
D.~A. Ross and M.~J.~G. Veltman, ``{Neutral Currents in Neutrino
  Experiments},'' \href{http://dx.doi.org/10.1016/0550-3213(75)90485-X}{{\em
  Nucl. Phys. B} {\bfseries 95} (1975) 135--147}.

\bibitem{Gunion:1989ci}
J.~F. Gunion, R.~Vega, and J.~Wudka, ``{Higgs triplets in the standard
  model},'' \href{http://dx.doi.org/10.1103/PhysRevD.42.1673}{{\em Phys. Rev.
  D} {\bfseries 42} (1990) 1673--1691}.

\bibitem{Blank:1997qa}
T.~Blank and W.~Hollik, ``{Precision observables in $SU(2) \times U(1)$ models
  with an additional Higgs triplet},''
  \href{http://dx.doi.org/10.1016/S0550-3213(97)00785-2}{{\em Nucl. Phys. B}
  {\bfseries 514} (1998) 113--134},
  \href{http://arxiv.org/abs/hep-ph/9703392}{{\ttfamily arXiv:hep-ph/9703392}}.

\bibitem{Forshaw:2003kh}
J.~R. Forshaw, A.~Sabio~Vera, and B.~E. White, ``{Mass bounds in a model with a
  triplet Higgs},'' \href{http://dx.doi.org/10.1088/1126-6708/2003/06/059}{{\em
  JHEP} {\bfseries 06} (2003) 059},
  \href{http://arxiv.org/abs/hep-ph/0302256}{{\ttfamily arXiv:hep-ph/0302256}}.

\bibitem{Chankowski:2006hs}
P.~H. Chankowski, S.~Pokorski, and J.~Wagner, ``{(Non)decoupling of the Higgs
  triplet effects},''
  \href{http://dx.doi.org/10.1140/epjc/s10052-007-0259-x}{{\em Eur. Phys. J. C}
  {\bfseries 50} (2007) 919--933},
  \href{http://arxiv.org/abs/hep-ph/0605302}{{\ttfamily arXiv:hep-ph/0605302}}.

\bibitem{Chen:2006pb}
M.-C. Chen, S.~Dawson, and T.~Krupovnickas, ``{Higgs triplets and limits from
  precision measurements},''
  \href{http://dx.doi.org/10.1103/PhysRevD.74.035001}{{\em Phys. Rev. D}
  {\bfseries 74} (2006) 035001},
  \href{http://arxiv.org/abs/hep-ph/0604102}{{\ttfamily arXiv:hep-ph/0604102}}.

\bibitem{Chivukula:2007koj}
R.~S. Chivukula, N.~D. Christensen, and E.~H. Simmons, ``{Low-energy effective
  theory, unitarity, and non-decoupling behavior in a model with heavy
  Higgs-triplet fields},''
  \href{http://dx.doi.org/10.1103/PhysRevD.77.035001}{{\em Phys. Rev. D}
  {\bfseries 77} (2008) 035001},
  \href{http://arxiv.org/abs/0712.0546}{{\ttfamily arXiv:0712.0546 [hep-ph]}}.

\bibitem{Bandyopadhyay:2020otm}
P.~Bandyopadhyay and A.~Costantini, ``{Obscure Higgs boson at Colliders},''
  \href{http://dx.doi.org/10.1103/PhysRevD.103.015025}{{\em Phys. Rev. D}
  {\bfseries 103} no.~1, (2021) 015025},
  \href{http://arxiv.org/abs/2010.02597}{{\ttfamily arXiv:2010.02597
  [hep-ph]}}.

\bibitem{Lazarides:2022spe}
G.~Lazarides, R.~Maji, R.~Roshan, and Q.~Shafi, ``{Heavier $W$ boson, dark
  matter, and gravitational waves from strings in an $SO(10)$ axion model},''
  \href{http://dx.doi.org/10.1103/PhysRevD.106.055009}{{\em Phys. Rev. D}
  {\bfseries 106} no.~5, (2022) 055009},
  \href{http://arxiv.org/abs/2205.04824}{{\ttfamily arXiv:2205.04824
  [hep-ph]}}.

\bibitem{Butterworth:2023rnw}
J.~Butterworth, H.~Debnath, P.~Fileviez~Perez, and F.~Mitchell, ``{Custodial
  symmetry breaking and Higgs boson signatures at the LHC},''
  \href{http://dx.doi.org/10.1103/PhysRevD.109.095014}{{\em Phys. Rev. D}
  {\bfseries 109} no.~9, (2024) 095014},
  \href{http://arxiv.org/abs/2309.10027}{{\ttfamily arXiv:2309.10027
  [hep-ph]}}.

\bibitem{Senjanovic:2022zwy}
G.~Senjanovi\'c and M.~Zantedeschi, ``{$SU(5)$ grand unification and $W$-boson
  mass},'' \href{http://dx.doi.org/10.1016/j.physletb.2022.137653}{{\em Phys.
  Lett. B} {\bfseries 837} (2023) 137653},
  \href{http://arxiv.org/abs/2205.05022}{{\ttfamily arXiv:2205.05022
  [hep-ph]}}.

\bibitem{Crivellin:2023gtf}
A.~Crivellin, M.~Kirk, and A.~Thapa, ``{Minimal model for the $W$-boson mass,
  $(g-2)_\mu$, $h\to\mu^+\mu^-$ and quark-mixing-matrix unitarity},''
  \href{http://dx.doi.org/10.1103/PhysRevD.108.L031702}{{\em Phys. Rev. D}
  {\bfseries 108} no.~3, (2023) L031702},
  \href{http://arxiv.org/abs/2305.03081}{{\ttfamily arXiv:2305.03081
  [hep-ph]}}.

\bibitem{Chen:2023ins}
T.-K. Chen, C.-W. Chiang, and K.~Yagyu, ``{$CP$ violation in a model with Higgs
  triplets},'' \href{http://dx.doi.org/10.1007/JHEP06(2023)069}{{\em JHEP}
  {\bfseries 06} (2023) 069}, \href{http://arxiv.org/abs/2303.09294}{{\ttfamily
  arXiv:2303.09294 [hep-ph]}}. [Erratum: JHEP 07, 169 (2023)].

\bibitem{Ashanujjaman:2023etj}
S.~Ashanujjaman, S.~Banik, G.~Coloretti, A.~Crivellin, B.~Mellado, and A.-T.
  Mulaudzi, ``{$SU(2)_L$ triplet scalar as the origin of the 95~GeV excess?},''
  \href{http://dx.doi.org/10.1103/PhysRevD.108.L091704}{{\em Phys. Rev. D}
  {\bfseries 108} no.~9, (2023) L091704},
  \href{http://arxiv.org/abs/2306.15722}{{\ttfamily arXiv:2306.15722
  [hep-ph]}}.

\bibitem{Degrassi:2024qsf}
G.~Degrassi and P.~Slavich, ``{On the two-loop BSM corrections to
  $h\to\gamma\gamma$ in a triplet extension of the SM},''
  \href{http://arxiv.org/abs/2407.18185}{{\ttfamily arXiv:2407.18185
  [hep-ph]}}.

\bibitem{vonBuddenbrock:2016rmr}
S.~von Buddenbrock, N.~Chakrabarty, A.~S. Cornell, D.~Kar, M.~Kumar, T.~Mandal,
  B.~Mellado, B.~Mukhopadhyaya, R.~G. Reed, and X.~Ruan, ``{Phenomenological
  signatures of additional scalar bosons at the LHC},''
  \href{http://dx.doi.org/10.1140/epjc/s10052-016-4435-8}{{\em Eur. Phys. J. C}
  {\bfseries 76} no.~10, (2016) 580},
  \href{http://arxiv.org/abs/1606.01674}{{\ttfamily arXiv:1606.01674
  [hep-ph]}}.

\bibitem{vonBuddenbrock:2017gvy}
S.~von Buddenbrock, A.~S. Cornell, A.~Fadol, M.~Kumar, B.~Mellado, and X.~Ruan,
  ``{Multi-lepton signatures of additional scalar bosons beyond the Standard
  Model at the LHC},'' \href{http://dx.doi.org/10.1088/1361-6471/aae3d6}{{\em
  J. Phys. G} {\bfseries 45} no.~11, (2018) 115003},
  \href{http://arxiv.org/abs/1711.07874}{{\ttfamily arXiv:1711.07874
  [hep-ph]}}.

\bibitem{vonBuddenbrock:2019ajh}
S.~Buddenbrock, A.~S. Cornell, Y.~Fang, A.~Fadol~Mohammed, M.~Kumar,
  B.~Mellado, and K.~G. Tomiwa, ``{The emergence of multi-lepton anomalies at
  the LHC and their compatibility with new physics at the EW scale},''
  \href{http://dx.doi.org/10.1007/JHEP10(2019)157}{{\em JHEP} {\bfseries 10}
  (2019) 157}, \href{http://arxiv.org/abs/1901.05300}{{\ttfamily
  arXiv:1901.05300 [hep-ph]}}.

\bibitem{vonBuddenbrock:2020ter}
S.~von Buddenbrock, R.~Ruiz, and B.~Mellado, ``{Anatomy of inclusive $t\bar tW$
  production at hadron colliders},''
  \href{http://dx.doi.org/10.1016/j.physletb.2020.135964}{{\em Phys. Lett. B}
  {\bfseries 811} (2020) 135964},
  \href{http://arxiv.org/abs/2009.00032}{{\ttfamily arXiv:2009.00032
  [hep-ph]}}.

\bibitem{Hernandez:2019geu}
Y.~Hernandez, M.~Kumar, A.~S. Cornell, S.-E. Dahbi, Y.~Fang, B.~Lieberman,
  B.~Mellado, K.~Monnakgotla, X.~Ruan, and S.~Xin, ``{The anomalous production
  of multi-lepton and its impact on the measurement of $Wh$ production at the
  LHC},'' \href{http://dx.doi.org/10.1140/epjc/s10052-021-09137-1}{{\em Eur.
  Phys. J. C} {\bfseries 81} no.~4, (2021) 365},
  \href{http://arxiv.org/abs/1912.00699}{{\ttfamily arXiv:1912.00699
  [hep-ph]}}.

\bibitem{Coloretti:2023wng}
G.~Coloretti, A.~Crivellin, S.~Bhattacharya, and B.~Mellado, ``{Searching for
  low-mass resonances decaying into $W$ bosons},''
  \href{http://dx.doi.org/10.1103/PhysRevD.108.035026}{{\em Phys. Rev. D}
  {\bfseries 108} no.~3, (2023) 035026},
  \href{http://arxiv.org/abs/2302.07276}{{\ttfamily arXiv:2302.07276
  [hep-ph]}}.

\bibitem{Banik:2023vxa}
S.~Banik, G.~Coloretti, A.~Crivellin, and B.~Mellado, ``{Uncovering New Higgses
  in the LHC Analyses of Differential $t\bar t$ Cross Sections},''
  \href{http://arxiv.org/abs/2308.07953}{{\ttfamily arXiv:2308.07953
  [hep-ph]}}.

\bibitem{Coloretti:2023yyq}
G.~Coloretti, A.~Crivellin, and B.~Mellado, ``{Combined explanation of LHC
  multilepton, diphoton, and top-quark excesses},''
  \href{http://dx.doi.org/10.1103/PhysRevD.110.073001}{{\em Phys. Rev. D}
  {\bfseries 110} no.~7, (2024) 073001},
  \href{http://arxiv.org/abs/2312.17314}{{\ttfamily arXiv:2312.17314
  [hep-ph]}}.

\bibitem{Fischer:2021sqw}
O.~Fischer {\em et~al.}, ``{Unveiling hidden physics at the LHC},''
  \href{http://dx.doi.org/10.1140/epjc/s10052-022-10541-4}{{\em Eur. Phys. J.
  C} {\bfseries 82} no.~8, (2022) 665},
  \href{http://arxiv.org/abs/2109.06065}{{\ttfamily arXiv:2109.06065
  [hep-ph]}}.

\bibitem{Crivellin:2021ubm}
A.~Crivellin, Y.~Fang, O.~Fischer, S.~Bhattacharya, M.~Kumar, E.~Malwa,
  B.~Mellado, N.~Rapheeha, X.~Ruan, and Q.~Sha, ``{Accumulating evidence for
  the associated production of a new Higgs boson at the LHC},''
  \href{http://dx.doi.org/10.1103/PhysRevD.108.115031}{{\em Phys. Rev. D}
  {\bfseries 108} no.~11, (2023) 115031},
  \href{http://arxiv.org/abs/2109.02650}{{\ttfamily arXiv:2109.02650
  [hep-ph]}}.

\bibitem{Bhattacharya:2023lmu}
S.~Bhattacharya, G.~Coloretti, A.~Crivellin, S.-E. Dahbi, Y.~Fang, M.~Kumar,
  and B.~Mellado, ``{Growing Excesses of New Scalars at the Electroweak
  Scale},'' \href{http://arxiv.org/abs/2306.17209}{{\ttfamily arXiv:2306.17209
  [hep-ph]}}.

\bibitem{Sirunyan:2021ybb}
{\bfseries CMS} Collaboration, A.~M. Sirunyan {\em et~al.}, ``{Measurements of
  Higgs boson production cross sections and couplings in the diphoton decay
  channel at $ \sqrt{\mathrm{s}} $ = 13 TeV},''
  \href{http://dx.doi.org/10.1007/JHEP07(2021)027}{{\em JHEP} {\bfseries 07}
  (2021) 027}, \href{http://arxiv.org/abs/2103.06956}{{\ttfamily
  arXiv:2103.06956 [hep-ex]}}.

\bibitem{ATLAS:2020pvn}
{\bfseries ATLAS} Collaboration, ``{Measurement of the properties of Higgs
  boson production at $\sqrt{s}$ = 13 TeV in the $H\to \gamma\gamma$ channel
  using 139 fb$^{-1}$ of $pp$ collision data with the ATLAS experiment},''.

\bibitem{Aad:2020ivc}
{\bfseries ATLAS} Collaboration, G.~Aad {\em et~al.}, ``{$CP$ Properties of
  Higgs Boson Interactions with Top Quarks in the $t\bar{t}H$ and $tH$
  Processes Using $H \to \gamma\gamma$ with the ATLAS Detector},''
  \href{http://dx.doi.org/10.1103/PhysRevLett.125.061802}{{\em Phys. Rev.
  Lett.} {\bfseries 125} no.~6, (2020) 061802},
  \href{http://arxiv.org/abs/2004.04545}{{\ttfamily arXiv:2004.04545
  [hep-ex]}}.

\bibitem{Sirunyan:2020sum}
{\bfseries CMS} Collaboration, A.~M. Sirunyan {\em et~al.}, ``{Measurements of
  $t\bar{t}H$ Production and the $CP$ Structure of the Yukawa Interaction
  between the Higgs Boson and Top Quark in the Diphoton Decay Channel},''
  \href{http://dx.doi.org/10.1103/PhysRevLett.125.061801}{{\em Phys. Rev.
  Lett.} {\bfseries 125} no.~6, (2020) 061801},
  \href{http://arxiv.org/abs/2003.10866}{{\ttfamily arXiv:2003.10866
  [hep-ex]}}.

\bibitem{Aad:2021qks}
{\bfseries ATLAS} Collaboration, G.~Aad {\em et~al.}, ``{Search for dark matter
  in events with missing transverse momentum and a Higgs boson decaying into
  two photons in $pp$ collisions at $\sqrt{s}$ = 13 TeV with the ATLAS
  detector},'' \href{http://dx.doi.org/10.1007/JHEP10(2021)013}{{\em JHEP}
  {\bfseries 10} (2021) 013}, \href{http://arxiv.org/abs/2104.13240}{{\ttfamily
  arXiv:2104.13240 [hep-ex]}}.

\bibitem{CMS:2018nlv}
{\bfseries CMS} Collaboration, A.~M. Sirunyan {\em et~al.}, ``{Search for dark
  matter produced in association with a Higgs boson decaying to $\gamma\gamma$
  or $\tau^+\tau^-$ at $\sqrt{s} =$ 13 TeV},''
  \href{http://dx.doi.org/10.1007/JHEP09(2018)046}{{\em JHEP} {\bfseries 09}
  (2018) 046}, \href{http://arxiv.org/abs/1806.04771}{{\ttfamily
  arXiv:1806.04771 [hep-ex]}}.

\bibitem{Sirunyan:2018tbk}
{\bfseries CMS} Collaboration, A.~M. Sirunyan {\em et~al.}, ``{Search for the
  decay of a Higgs boson in the $\ell\ell\gamma$ channel in proton-proton
  collisions at $\sqrt{s}=$ 13 TeV},''
  \href{http://dx.doi.org/10.1007/JHEP11(2018)152}{{\em JHEP} {\bfseries 11}
  (2018) 152}, \href{http://arxiv.org/abs/1806.05996}{{\ttfamily
  arXiv:1806.05996 [hep-ex]}}.

\bibitem{ATLAS:2020fcp}
{\bfseries ATLAS} Collaboration, G.~Aad {\em et~al.}, ``{Measurements of $WH$
  and $ZH$ production in the $H \to b\bar{b}$ decay channel in $pp$ collisions
  at 13 TeV with the ATLAS detector},''
  \href{http://dx.doi.org/10.1140/epjc/s10052-020-08677-2}{{\em Eur. Phys. J.
  C} {\bfseries 81} no.~2, (2021) 178},
  \href{http://arxiv.org/abs/2007.02873}{{\ttfamily arXiv:2007.02873
  [hep-ex]}}.

\bibitem{Ashanujjaman:2024pky}
S.~Ashanujjaman, S.~Banik, G.~Coloretti, A.~Crivellin, S.~P. Maharathy, and
  B.~Mellado, ``{Explaining the $\gamma\gamma+X$ Excesses at $\approx$151.5 GeV
  via the Drell-Yan Production of a Higgs Triplet},''
  \href{http://arxiv.org/abs/2402.00101}{{\ttfamily arXiv:2402.00101
  [hep-ph]}}.

\bibitem{Crivellin:2024uhc}
A.~Crivellin, S.~Ashanujjaman, S.~Banik, G.~Coloretti, S.~P. Maharathy, and
  B.~Mellado, ``{Growing Evidence for a Higgs Triplet},''
  \href{http://arxiv.org/abs/2404.14492}{{\ttfamily arXiv:2404.14492
  [hep-ph]}}.

\bibitem{Banik:2024ftv}
S.~Banik and A.~Crivellin, ``{Explanation of the excesses in associated
  di-photon production at 152 GeV in 2HDM},''
  \href{http://dx.doi.org/10.1007/JHEP10(2024)203}{{\em JHEP} {\bfseries 10}
  (2024) 203}, \href{http://arxiv.org/abs/2407.06267}{{\ttfamily
  arXiv:2407.06267 [hep-ph]}}.

\bibitem{ATLAS:2023omk}
{\bfseries ATLAS} Collaboration, G.~Aad {\em et~al.}, ``{Model-independent
  search for the presence of new physics in events including
  $H\rightarrow\gamma\gamma$ with $\sqrt{s}$ = 13 TeV $pp$ data recorded by the
  ATLAS detector at the LHC},''
  \href{http://dx.doi.org/10.1007/JHEP07(2023)176}{{\em JHEP} {\bfseries 07}
  (2023) 176}, \href{http://arxiv.org/abs/2301.10486}{{\ttfamily
  arXiv:2301.10486 [hep-ex]}}.

\bibitem{ATLAS:2024lhu}
{\bfseries ATLAS} Collaboration, G.~Aad {\em et~al.}, ``{Search for
  non-resonant Higgs boson pair production in final states with leptons, taus,
  and photons in $pp$ collisions at $\sqrt{s}$ = 13 TeV with the ATLAS
  detector},'' \href{http://dx.doi.org/10.1007/JHEP08(2024)164}{{\em JHEP}
  {\bfseries 08} (2024) 164}, \href{http://arxiv.org/abs/2405.20040}{{\ttfamily
  arXiv:2405.20040 [hep-ex]}}.

\bibitem{CMS:2018cyk}
{\bfseries CMS} Collaboration, A.~M. Sirunyan {\em et~al.}, ``{Search for a
  standard model-like Higgs boson in the mass range between 70 and 110 GeV in
  the diphoton final state in proton-proton collisions at $\sqrt{s}=$ 8 and 13
  TeV},'' \href{http://dx.doi.org/10.1016/j.physletb.2019.03.064}{{\em Phys.
  Lett. B} {\bfseries 793} (2019) 320--347},
  \href{http://arxiv.org/abs/1811.08459}{{\ttfamily arXiv:1811.08459
  [hep-ex]}}.

\bibitem{CMS:2023yay}
{\bfseries CMS} Collaboration, ``{Search for a standard model-like Higgs boson
  in the mass range between 70 and 110~GeV in the diphoton final state in
  proton-proton collisions at $\sqrt{s}=13$~TeV},''.

\bibitem{CMS:2022goy}
{\bfseries CMS} Collaboration, A.~Tumasyan {\em et~al.}, ``{Searches for
  additional Higgs bosons and for vector leptoquarks in $\tau\tau$ final states
  in proton-proton collisions at $\sqrt{s}$ = 13 TeV},''
  \href{http://dx.doi.org/10.1007/JHEP07(2023)073}{{\em JHEP} {\bfseries 07}
  (2023) 073}, \href{http://arxiv.org/abs/2208.02717}{{\ttfamily
  arXiv:2208.02717 [hep-ex]}}.

\bibitem{ATLAS:2018xad}
{\bfseries ATLAS} Collaboration, ``{Search for resonances in the 65 to 110 GeV
  diphoton invariant mass range using 80 fb$^{-1}$ of $pp$ collisions collected
  at $\sqrt{s}=13$ TeV with the ATLAS detector},''.

\bibitem{ATLAS:2022yrq}
{\bfseries ATLAS} Collaboration, G.~Aad {\em et~al.}, ``{Measurements of Higgs
  boson production cross-sections in the~$H\to\tau^{+}\tau^{-}$ decay channel
  in pp collisions at $ \sqrt{s} $ = 13 TeV with the ATLAS detector},''
  \href{http://dx.doi.org/10.1007/JHEP08(2022)175}{{\em JHEP} {\bfseries 08}
  (2022) 175}, \href{http://arxiv.org/abs/2201.08269}{{\ttfamily
  arXiv:2201.08269 [hep-ex]}}.

\bibitem{LEPWorkingGroupforHiggsbosonsearches:2003ing}
{\bfseries LEP Working Group for Higgs boson searches, ALEPH, DELPHI, L3, OPAL}
  Collaboration, R.~Barate {\em et~al.}, ``{Search for the standard model Higgs
  boson at LEP},'' \href{http://dx.doi.org/10.1016/S0370-2693(03)00614-2}{{\em
  Phys. Lett. B} {\bfseries 565} (2003) 61--75},
  \href{http://arxiv.org/abs/hep-ex/0306033}{{\ttfamily arXiv:hep-ex/0306033}}.

\bibitem{Chen:2023bqr}
T.-K. Chen, C.-W. Chiang, S.~Heinemeyer, and G.~Weiglein, ``{95~GeV Higgs boson
  in the Georgi-Machacek model},''
  \href{http://dx.doi.org/10.1103/PhysRevD.109.075043}{{\em Phys. Rev. D}
  {\bfseries 109} no.~7, (2024) 075043},
  \href{http://arxiv.org/abs/2312.13239}{{\ttfamily arXiv:2312.13239
  [hep-ph]}}.

\bibitem{Lopez-Val:2014jva}
D.~L\'opez-Val and T.~Robens, ``{$\Delta r$ and the $W$-boson mass in the
  singlet extension of the standard model},''
  \href{http://dx.doi.org/10.1103/PhysRevD.90.114018}{{\em Phys. Rev. D}
  {\bfseries 90} (2014) 114018},
  \href{http://arxiv.org/abs/1406.1043}{{\ttfamily arXiv:1406.1043 [hep-ph]}}.

\bibitem{tHooft:1979rat}
G.~'t~Hooft, ``{Naturalness, chiral symmetry, and spontaneous chiral symmetry
  breaking},'' \href{http://dx.doi.org/10.1007/978-1-4684-7571-5_9}{{\em NATO
  Sci. Ser. B} {\bfseries 59} (1980) 135--157}.

\bibitem{Cirelli:2005uq}
M.~Cirelli, N.~Fornengo, and A.~Strumia, ``{Minimal dark matter},''
  \href{http://dx.doi.org/10.1016/j.nuclphysb.2006.07.012}{{\em Nucl. Phys. B}
  {\bfseries 753} (2006) 178--194},
  \href{http://arxiv.org/abs/hep-ph/0512090}{{\ttfamily arXiv:hep-ph/0512090}}.

\bibitem{FileviezPerez:2008bj}
P.~Fileviez~Perez, H.~H. Patel, M.~J. Ramsey-Musolf, and K.~Wang, ``{Triplet
  Scalars and Dark Matter at the LHC},''
  \href{http://dx.doi.org/10.1103/PhysRevD.79.055024}{{\em Phys. Rev. D}
  {\bfseries 79} (2009) 055024},
  \href{http://arxiv.org/abs/0811.3957}{{\ttfamily arXiv:0811.3957 [hep-ph]}}.

\bibitem{Kanemura:2012rs}
S.~Kanemura and K.~Yagyu, ``{Radiative corrections to electroweak parameters in
  the Higgs triplet model and implication with the recent Higgs boson
  searches},'' \href{http://dx.doi.org/10.1103/PhysRevD.85.115009}{{\em Phys.
  Rev. D} {\bfseries 85} (2012) 115009},
  \href{http://arxiv.org/abs/1201.6287}{{\ttfamily arXiv:1201.6287 [hep-ph]}}.

\bibitem{Pal:1994jk}
P.~B. Pal, ``{What is the equivalence theorem really?},''
  \href{http://arxiv.org/abs/hep-ph/9405362}{{\ttfamily arXiv:hep-ph/9405362}}.

\bibitem{Horejsi:1995jj}
J.~Horejsi, ``{Electroweak interactions and high-energy limit: An Introduction
  to equivalence theorem},''
  \href{http://dx.doi.org/10.1023/A:1021177005216}{{\em Czech. J. Phys.}
  {\bfseries 47} (1997) 951--977},
  \href{http://arxiv.org/abs/hep-ph/9603321}{{\ttfamily arXiv:hep-ph/9603321}}.

\bibitem{Camargo-Molina:2014pwa}
J.~E. Camargo-Molina, B.~Garbrecht, B.~O'Leary, W.~Porod, and F.~Staub,
  ``{Constraining the Natural MSSM through tunneling to color-breaking vacua at
  zero and non-zero temperature},''
  \href{http://dx.doi.org/10.1016/j.physletb.2014.08.036}{{\em Phys. Lett. B}
  {\bfseries 737} (2014) 156--161},
  \href{http://arxiv.org/abs/1405.7376}{{\ttfamily arXiv:1405.7376 [hep-ph]}}.

\bibitem{Camargo-Molina:2013qva}
J.~E. Camargo-Molina, B.~O'Leary, W.~Porod, and F.~Staub,
  ``{$\mathbf{Vevacious}$: A Tool For Finding The Global Minima Of One-Loop
  Effective Potentials With Many Scalars},''
  \href{http://dx.doi.org/10.1140/epjc/s10052-013-2588-2}{{\em Eur. Phys. J. C}
  {\bfseries 73} no.~10, (2013) 2588},
  \href{http://arxiv.org/abs/1307.1477}{{\ttfamily arXiv:1307.1477 [hep-ph]}}.

\bibitem{Porod:2003um}
W.~Porod, ``{SPheno, a program for calculating supersymmetric spectra, SUSY
  particle decays and SUSY particle production at $e^+ e^-$ colliders},''
  \href{http://dx.doi.org/10.1016/S0010-4655(03)00222-4}{{\em Comput. Phys.
  Commun.} {\bfseries 153} (2003) 275--315},
  \href{http://arxiv.org/abs/hep-ph/0301101}{{\ttfamily arXiv:hep-ph/0301101}}.

\bibitem{Porod:2011nf}
W.~Porod and F.~Staub, ``{SPheno 3.1: Extensions including flavour, $CP$-phases
  and models beyond the MSSM},''
  \href{http://dx.doi.org/10.1016/j.cpc.2012.05.021}{{\em Comput. Phys.
  Commun.} {\bfseries 183} (2012) 2458--2469},
  \href{http://arxiv.org/abs/1104.1573}{{\ttfamily arXiv:1104.1573 [hep-ph]}}.

\bibitem{Goodsell:2023iac}
M.~D. Goodsell and A.~Joury, ``{BSMArt: Simple and fast parameter space
  scans},'' \href{http://dx.doi.org/10.1016/j.cpc.2023.109057}{{\em Comput.
  Phys. Commun.} {\bfseries 297} (2024) 109057},
  \href{http://arxiv.org/abs/2301.01154}{{\ttfamily arXiv:2301.01154
  [hep-ph]}}.

\bibitem{Logan:2022uus}
H.~E. Logan, ``{Lectures on perturbative unitarity and decoupling in Higgs
  physics},'' \href{http://arxiv.org/abs/2207.01064}{{\ttfamily
  arXiv:2207.01064 [hep-ph]}}.

\bibitem{Anastasiou:2016cez}
C.~Anastasiou, C.~Duhr, F.~Dulat, E.~Furlan, T.~Gehrmann, F.~Herzog,
  A.~Lazopoulos, and B.~Mistlberger, ``{High precision determination of the
  gluon fusion Higgs boson cross-section at the LHC},''
  \href{http://dx.doi.org/10.1007/JHEP05(2016)058}{{\em JHEP} {\bfseries 05}
  (2016) 058}, \href{http://arxiv.org/abs/1602.00695}{{\ttfamily
  arXiv:1602.00695 [hep-ph]}}.

\bibitem{LHCHiggsCrossSectionWorkingGroup:2013rie}
{\bfseries LHC Higgs Cross Section Working Group} Collaboration, J.~R. Andersen
  {\em et~al.}, ``{Handbook of LHC Higgs Cross Sections: 3. Higgs
  Properties},'' \href{http://arxiv.org/abs/1307.1347}{{\ttfamily
  arXiv:1307.1347 [hep-ph]}}.

\bibitem{Staub:2013tta}
F.~Staub, ``{SARAH 4 : A tool for (not only SUSY) model builders},''
  \href{http://dx.doi.org/10.1016/j.cpc.2014.02.018}{{\em Comput. Phys.
  Commun.} {\bfseries 185} (2014) 1773--1790},
  \href{http://arxiv.org/abs/1309.7223}{{\ttfamily arXiv:1309.7223 [hep-ph]}}.

\bibitem{Staub:2015kfa}
F.~Staub, ``{Exploring new models in all detail with SARAH},''
  \href{http://dx.doi.org/10.1155/2015/840780}{{\em Adv. High Energy Phys.}
  {\bfseries 2015} (2015) 840780},
  \href{http://arxiv.org/abs/1503.04200}{{\ttfamily arXiv:1503.04200
  [hep-ph]}}.

\bibitem{Degrande:2011ua}
C.~Degrande, C.~Duhr, B.~Fuks, D.~Grellscheid, O.~Mattelaer, and T.~Reiter,
  ``{UFO - The Universal FeynRules Output},''
  \href{http://dx.doi.org/10.1016/j.cpc.2012.01.022}{{\em Comput. Phys.
  Commun.} {\bfseries 183} (2012) 1201--1214},
  \href{http://arxiv.org/abs/1108.2040}{{\ttfamily arXiv:1108.2040 [hep-ph]}}.

\bibitem{Alloul:2013bka}
A.~Alloul, N.~D. Christensen, C.~Degrande, C.~Duhr, and B.~Fuks, ``{FeynRules
  2.0 - A complete toolbox for tree-level phenomenology},''
  \href{http://dx.doi.org/10.1016/j.cpc.2014.04.012}{{\em Comput. Phys.
  Commun.} {\bfseries 185} (2014) 2250--2300},
  \href{http://arxiv.org/abs/1310.1921}{{\ttfamily arXiv:1310.1921 [hep-ph]}}.

\bibitem{Degrande:2014vpa}
C.~Degrande, ``{Automatic evaluation of UV and R2 terms for beyond the Standard
  Model Lagrangians: a proof-of-principle},''
  \href{http://dx.doi.org/10.1016/j.cpc.2015.08.015}{{\em Comput. Phys.
  Commun.} {\bfseries 197} (2015) 239--262},
  \href{http://arxiv.org/abs/1406.3030}{{\ttfamily arXiv:1406.3030 [hep-ph]}}.

\bibitem{Alwall:2011uj}
J.~Alwall, M.~Herquet, F.~Maltoni, O.~Mattelaer, and T.~Stelzer, ``{MadGraph 5
  : Going Beyond},'' \href{http://dx.doi.org/10.1007/JHEP06(2011)128}{{\em
  JHEP} {\bfseries 06} (2011) 128},
  \href{http://arxiv.org/abs/1106.0522}{{\ttfamily arXiv:1106.0522 [hep-ph]}}.

\bibitem{Alwall:2014hca}
J.~Alwall, R.~Frederix, S.~Frixione, V.~Hirschi, F.~Maltoni, O.~Mattelaer,
  H.~S. Shao, T.~Stelzer, P.~Torrielli, and M.~Zaro, ``{The automated
  computation of tree-level and next-to-leading order differential cross
  sections, and their matching to parton shower simulations},''
  \href{http://dx.doi.org/10.1007/JHEP07(2014)079}{{\em JHEP} {\bfseries 07}
  (2014) 079}, \href{http://arxiv.org/abs/1405.0301}{{\ttfamily arXiv:1405.0301
  [hep-ph]}}.

\bibitem{Ball:2013hta}
{\bfseries NNPDF} Collaboration, R.~D. Ball, V.~Bertone, S.~Carrazza,
  L.~Del~Debbio, S.~Forte, A.~Guffanti, N.~P. Hartland, and J.~Rojo, ``{Parton
  distributions with QED corrections},''
  \href{http://dx.doi.org/10.1016/j.nuclphysb.2013.10.010}{{\em Nucl. Phys. B}
  {\bfseries 877} (2013) 290--320},
  \href{http://arxiv.org/abs/1308.0598}{{\ttfamily arXiv:1308.0598 [hep-ph]}}.

\bibitem{Ruiz:2015zca}
R.~Ruiz, ``{QCD Corrections to Pair Production of Type III Seesaw Leptons at
  Hadron Colliders},'' \href{http://dx.doi.org/10.1007/JHEP12(2015)165}{{\em
  JHEP} {\bfseries 12} (2015) 165},
  \href{http://arxiv.org/abs/1509.05416}{{\ttfamily arXiv:1509.05416
  [hep-ph]}}.

\bibitem{Ajjath:2023ugn}
A.~A~H, B.~Fuks, H.-S. Shao, and Y.~Simon, ``{Precision predictions for exotic
  lepton production at the Large Hadron Collider},''
  \href{http://dx.doi.org/10.1103/PhysRevD.107.075011}{{\em Phys. Rev. D}
  {\bfseries 107} no.~7, (2023) 075011},
  \href{http://arxiv.org/abs/2301.03640}{{\ttfamily arXiv:2301.03640
  [hep-ph]}}.

\bibitem{ATLAS:2022vkf}
{\bfseries ATLAS} Collaboration, G.~Aad {\em et~al.}, ``{A detailed map of
  Higgs boson interactions by the ATLAS experiment ten years after the
  discovery},'' \href{http://dx.doi.org/10.1038/s41586-022-04893-w}{{\em
  Nature} {\bfseries 607} no.~7917, (2022) 52--59},
  \href{http://arxiv.org/abs/2207.00092}{{\ttfamily arXiv:2207.00092
  [hep-ex]}}. [Erratum: Nature 612, E24 (2022)].

\bibitem{CMS:2022dwd}
{\bfseries CMS} Collaboration, A.~Tumasyan {\em et~al.}, ``{A portrait of the
  Higgs boson by the CMS experiment ten years after the discovery.},''
  \href{http://dx.doi.org/10.1038/s41586-022-04892-x}{{\em Nature} {\bfseries
  607} no.~7917, (2022) 60--68},
  \href{http://arxiv.org/abs/2207.00043}{{\ttfamily arXiv:2207.00043
  [hep-ex]}}. [Erratum: Nature 623, (2023)].

\bibitem{Chen:2013vi}
C.-S. Chen, C.-Q. Geng, D.~Huang, and L.-H. Tsai, ``{New Scalar Contributions
  to $h\to Z\gamma$},''
  \href{http://dx.doi.org/10.1103/PhysRevD.87.075019}{{\em Phys. Rev. D}
  {\bfseries 87} (2013) 075019},
  \href{http://arxiv.org/abs/1301.4694}{{\ttfamily arXiv:1301.4694 [hep-ph]}}.

\bibitem{CMS:2021kom}
{\bfseries CMS} Collaboration, A.~M. Sirunyan {\em et~al.}, ``{Measurements of
  Higgs boson production cross sections and couplings in the diphoton decay
  channel at $\sqrt{s}$ = 13 TeV},''
  \href{http://dx.doi.org/10.1007/JHEP07(2021)027}{{\em JHEP} {\bfseries 07}
  (2021) 027}, \href{http://arxiv.org/abs/2103.06956}{{\ttfamily
  arXiv:2103.06956 [hep-ex]}}.

\bibitem{ATLAS:2022tnm}
{\bfseries ATLAS} Collaboration, G.~Aad {\em et~al.}, ``{Measurement of the
  properties of Higgs boson production at $\sqrt{s} = 13$ TeV in the
  $H\to\gamma\gamma$ channel using $139$ fb$^{-1}$ of $pp$ collision data with
  the ATLAS experiment},''
  \href{http://dx.doi.org/10.1007/JHEP07(2023)088}{{\em JHEP} {\bfseries 07}
  (2023) 088}, \href{http://arxiv.org/abs/2207.00348}{{\ttfamily
  arXiv:2207.00348 [hep-ex]}}.

\bibitem{ATLAS:2020qcv}
{\bfseries ATLAS} Collaboration, G.~Aad {\em et~al.}, ``{A search for the
  $Z\gamma$ decay mode of the Higgs boson in $pp$ collisions at $\sqrt{s}$ = 13
  TeV with the ATLAS detector},''
  \href{http://dx.doi.org/10.1016/j.physletb.2020.135754}{{\em Phys. Lett. B}
  {\bfseries 809} (2020) 135754},
  \href{http://arxiv.org/abs/2005.05382}{{\ttfamily arXiv:2005.05382
  [hep-ex]}}.

\bibitem{CMS:2022ahq}
{\bfseries CMS} Collaboration, A.~Tumasyan {\em et~al.}, ``{Search for Higgs
  boson decays to a $Z$ boson and a photon in proton-proton collisions at
  $\sqrt{s}$ = 13 TeV},'' \href{http://dx.doi.org/10.1007/JHEP05(2023)233}{{\em
  JHEP} {\bfseries 05} (2023) 233},
  \href{http://arxiv.org/abs/2204.12945}{{\ttfamily arXiv:2204.12945
  [hep-ex]}}.

\bibitem{ATLAS:2023yqk}
{\bfseries ATLAS, CMS} Collaboration, G.~Aad {\em et~al.}, ``{Evidence for the
  Higgs Boson Decay to a $Z$ Boson and a Photon at the LHC},''
  \href{http://dx.doi.org/10.1103/PhysRevLett.132.021803}{{\em Phys. Rev.
  Lett.} {\bfseries 132} no.~2, (2024) 021803},
  \href{http://arxiv.org/abs/2309.03501}{{\ttfamily arXiv:2309.03501
  [hep-ex]}}.

\bibitem{ATLAS:2020rej}
{\bfseries ATLAS} Collaboration, G.~Aad {\em et~al.}, ``{Higgs boson production
  cross-section measurements and their EFT interpretation in the $4\ell $ decay
  channel at $\sqrt{s}=$13 TeV with the ATLAS detector},''
  \href{http://dx.doi.org/10.1140/epjc/s10052-020-8227-9}{{\em Eur. Phys. J. C}
  {\bfseries 80} no.~10, (2020) 957},
  \href{http://arxiv.org/abs/2004.03447}{{\ttfamily arXiv:2004.03447
  [hep-ex]}}. [Erratum: Eur.Phys.J.C 81, 29 (2021), Erratum: Eur.Phys.J.C 81,
  398 (2021)].

\bibitem{Rizzo:1980gz}
T.~G. Rizzo, ``{Decays of Heavy Higgs Bosons},''
  \href{http://dx.doi.org/10.1103/PhysRevD.22.722}{{\em Phys. Rev. D}
  {\bfseries 22} (1980) 722}.

\bibitem{Keung:1984hn}
W.-Y. Keung and W.~J. Marciano, ``{HIGGS SCALAR DECAYS: $H \to W^\pm X$},''
  \href{http://dx.doi.org/10.1103/PhysRevD.30.248}{{\em Phys. Rev. D}
  {\bfseries 30} (1984) 248}.

\bibitem{Djouadi:1997rp}
A.~Djouadi, ``{Decays of the Higgs bosons},'' in {\em {International Workshop
  on Quantum Effects in the Minimal Supersymmetric Standard Model}},
  pp.~197--222.
\newblock 12, 1997.
\newblock \href{http://arxiv.org/abs/hep-ph/9712334}{{\ttfamily
  arXiv:hep-ph/9712334}}.

\bibitem{Djouadi:2005gi}
A.~Djouadi, ``{The Anatomy of electro-weak symmetry breaking. I: The Higgs
  boson in the standard model},''
  \href{http://dx.doi.org/10.1016/j.physrep.2007.10.004}{{\em Phys. Rept.}
  {\bfseries 457} (2008) 1--216},
  \href{http://arxiv.org/abs/hep-ph/0503172}{{\ttfamily arXiv:hep-ph/0503172}}.

\bibitem{Djouadi:2005gj}
A.~Djouadi, ``{The Anatomy of electro-weak symmetry breaking. II. The Higgs
  bosons in the minimal supersymmetric model},''
  \href{http://dx.doi.org/10.1016/j.physrep.2007.10.005}{{\em Phys. Rept.}
  {\bfseries 459} (2008) 1--241},
  \href{http://arxiv.org/abs/hep-ph/0503173}{{\ttfamily arXiv:hep-ph/0503173}}.

\bibitem{ATLAS:2023fsi}
{\bfseries ATLAS} Collaboration, ``{Improved $W$ boson Mass Measurement using 7
  TeV Proton-Proton Collisions with the ATLAS Detector},''.

\bibitem{LHCb:2021bjt}
{\bfseries LHCb} Collaboration, R.~Aaij {\em et~al.}, ``{Measurement of the $W$
  boson mass},'' \href{http://dx.doi.org/10.1007/JHEP01(2022)036}{{\em JHEP}
  {\bfseries 01} (2022) 036}, \href{http://arxiv.org/abs/2109.01113}{{\ttfamily
  arXiv:2109.01113 [hep-ex]}}.

\bibitem{Abazov:2009cp}
{\bfseries D0} Collaboration, V.~M. Abazov {\em et~al.}, ``{Measurement of the
  $W$ boson mass},''
  \href{http://dx.doi.org/10.1103/PhysRevLett.103.141801}{{\em Phys. Rev.
  Lett.} {\bfseries 103} (2009) 141801},
  \href{http://arxiv.org/abs/0908.0766}{{\ttfamily arXiv:0908.0766 [hep-ex]}}.

\bibitem{D0:2013jba}
{\bfseries D0} Collaboration, V.~M. Abazov {\em et~al.}, ``{Measurement of the
  $W$ boson mass with the D0 detector},''
  \href{http://dx.doi.org/10.1103/PhysRevD.89.012005}{{\em Phys. Rev. D}
  {\bfseries 89} no.~1, (2014) 012005},
  \href{http://arxiv.org/abs/1310.8628}{{\ttfamily arXiv:1310.8628 [hep-ex]}}.

\bibitem{Aaboud:2017svj}
{\bfseries ATLAS} Collaboration, M.~Aaboud {\em et~al.}, ``{Measurement of the
  $W$-boson mass in $pp$ collisions at $\sqrt{s}=7$ TeV with the ATLAS
  detector},'' \href{http://dx.doi.org/10.1140/epjc/s10052-017-5475-4}{{\em
  Eur. Phys. J. C} {\bfseries 78} no.~2, (2018) 110},
  \href{http://arxiv.org/abs/1701.07240}{{\ttfamily arXiv:1701.07240
  [hep-ex]}}. [Erratum: Eur.Phys.J.C 78, 898 (2018)].

\bibitem{ALEPH:2013dgf}
{\bfseries ALEPH, DELPHI, L3, OPAL, LEP Electroweak} Collaboration, S.~Schael
  {\em et~al.}, ``{Electroweak Measurements in Electron-Positron Collisions at
  $W$-Boson-Pair Energies at LEP},''
  \href{http://dx.doi.org/10.1016/j.physrep.2013.07.004}{{\em Phys. Rept.}
  {\bfseries 532} (2013) 119--244},
  \href{http://arxiv.org/abs/1302.3415}{{\ttfamily arXiv:1302.3415 [hep-ex]}}.

\bibitem{LHC-TeVMWWorkingGroup:2023zkn}
{\bfseries LHC-TeV~MW~Working~Group} Collaboration, S.~Amoroso {\em et~al.},
  ``{Compatibility and combination of world $W$-boson mass measurements},''
  \href{http://dx.doi.org/10.1140/epjc/s10052-024-12532-z}{{\em Eur. Phys. J.
  C} {\bfseries 84} no.~5, (2024) 451},
  \href{http://arxiv.org/abs/2308.09417}{{\ttfamily arXiv:2308.09417
  [hep-ex]}}.

\bibitem{deBlas:2022hdk}
J.~de~Blas, M.~Pierini, L.~Reina, and L.~Silvestrini, ``{Impact of the Recent
  Measurements of the Top-Quark and $W$-Boson Masses on Electroweak Precision
  Fits},'' \href{http://dx.doi.org/10.1103/PhysRevLett.129.271801}{{\em Phys.
  Rev. Lett.} {\bfseries 129} no.~27, (2022) 271801},
  \href{http://arxiv.org/abs/2204.04204}{{\ttfamily arXiv:2204.04204
  [hep-ph]}}.

\bibitem{ATLAS:2024erm}
{\bfseries ATLAS} Collaboration, G.~Aad {\em et~al.}, ``{Measurement of the
  $W$-boson mass and width with the ATLAS detector using proton-proton
  collisions at $\sqrt{s}$ = 7 TeV},''
  \href{http://arxiv.org/abs/2403.15085}{{\ttfamily arXiv:2403.15085
  [hep-ex]}}.

\bibitem{CMS:2024nau}
{\bfseries CMS} Collaboration, ``{Measurement of the $W$ boson mass in
  proton-proton collisions at $\sqrt{s}$ = 13 TeV},''.

\bibitem{ATLAS:2019gti}
{\bfseries ATLAS} Collaboration, G.~Aad {\em et~al.}, ``{Search for direct stau
  production in events with two hadronic $\tau$-leptons in $\sqrt{s} = 13$ TeV
  $pp$ collisions with the ATLAS detector},''
  \href{http://dx.doi.org/10.1103/PhysRevD.101.032009}{{\em Phys. Rev. D}
  {\bfseries 101} no.~3, (2020) 032009},
  \href{http://arxiv.org/abs/1911.06660}{{\ttfamily arXiv:1911.06660
  [hep-ex]}}.

\bibitem{CMS:2019eln}
{\bfseries CMS} Collaboration, A.~M. Sirunyan {\em et~al.}, ``{Search for
  direct pair production of supersymmetric partners to the $\tau$ lepton in
  proton-proton collisions at $\sqrt{s}=$ 13 TeV},''
  \href{http://dx.doi.org/10.1140/epjc/s10052-020-7739-7}{{\em Eur. Phys. J. C}
  {\bfseries 80} no.~3, (2020) 189},
  \href{http://arxiv.org/abs/1907.13179}{{\ttfamily arXiv:1907.13179
  [hep-ex]}}.

\bibitem{CMS:2022rqk}
{\bfseries CMS} Collaboration, A.~Tumasyan {\em et~al.}, ``{Search for direct
  pair production of supersymmetric partners of $\tau$ leptons in the final
  state with two hadronically decaying $\tau$ leptons and missing transverse
  momentum in proton-proton collisions at $\sqrt{s}$ = 13 TeV},''
  \href{http://dx.doi.org/10.1103/PhysRevD.108.012011}{{\em Phys. Rev. D}
  {\bfseries 108} no.~1, (2023) 012011},
  \href{http://arxiv.org/abs/2207.02254}{{\ttfamily arXiv:2207.02254
  [hep-ex]}}.

\bibitem{ATLAS:2024fub}
{\bfseries ATLAS} Collaboration, G.~Aad {\em et~al.}, ``{Search for electroweak
  production of supersymmetric particles in final states with two
  \ensuremath{\tau}-leptons in $\sqrt{s}$ = 13 TeV $pp$ collisions with the
  ATLAS detector},'' \href{http://dx.doi.org/10.1007/JHEP05(2024)150}{{\em
  JHEP} {\bfseries 05} (2024) 150},
  \href{http://arxiv.org/abs/2402.00603}{{\ttfamily arXiv:2402.00603
  [hep-ex]}}.

\bibitem{CMS:2019lwf}
{\bfseries CMS} Collaboration, A.~M. Sirunyan {\em et~al.}, ``{Search for
  physics beyond the standard model in multilepton final states in
  proton-proton collisions at $\sqrt{s}=$ 13 TeV},''
  \href{http://dx.doi.org/10.1007/JHEP03(2020)051}{{\em JHEP} {\bfseries 03}
  (2020) 051}, \href{http://arxiv.org/abs/1911.04968}{{\ttfamily
  arXiv:1911.04968 [hep-ex]}}.

\bibitem{CMS:2021zkl}
{\bfseries CMS} Collaboration, ``{Inclusive nonresonant multilepton probes of
  new phenomena at $\sqrt{s}=13~\mathrm{TeV}$},''.

\bibitem{ATLAS:2021yyr}
{\bfseries ATLAS} Collaboration, G.~Aad {\em et~al.}, ``{Search for
  supersymmetry in events with four or more charged leptons in 139 fb$^{-1}$ of
  $ \sqrt{s} $ = 13 TeV $pp$ collisions with the ATLAS detector},''
  \href{http://dx.doi.org/10.1007/JHEP07(2021)167}{{\em JHEP} {\bfseries 07}
  (2021) 167}, \href{http://arxiv.org/abs/2103.11684}{{\ttfamily
  arXiv:2103.11684 [hep-ex]}}.

\bibitem{ATLAS:2021wob}
{\bfseries ATLAS} Collaboration, G.~Aad {\em et~al.}, ``{Search for new
  phenomena in three- or four-lepton events in $pp$ collisions at $\sqrt{s}$ =
  13 TeV with the ATLAS detector},''
  \href{http://dx.doi.org/10.1016/j.physletb.2021.136832}{{\em Phys. Lett. B}
  {\bfseries 824} (2022) 136832},
  \href{http://arxiv.org/abs/2107.00404}{{\ttfamily arXiv:2107.00404
  [hep-ex]}}.

\bibitem{ATLAS:2022nmi}
{\bfseries ATLAS} Collaboration, G.~Aad {\em et~al.}, ``{A search for new
  resonances in multiple final states with a high transverse momentum $Z$ boson
  in $\sqrt{s}=13$ TeV $pp$ collisions with the ATLAS detector},''
  \href{http://dx.doi.org/10.1007/JHEP06(2023)036}{{\em JHEP} {\bfseries 06}
  (2023) 036}, \href{http://arxiv.org/abs/2209.15345}{{\ttfamily
  arXiv:2209.15345 [hep-ex]}}.

\bibitem{ATLAS:2022yhd}
{\bfseries ATLAS} Collaboration, G.~Aad {\em et~al.}, ``{Search for type-III
  seesaw heavy leptons in leptonic final states in $pp$ collisions at $\sqrt{s}
  = 13$~TeV with the ATLAS detector},''
  \href{http://dx.doi.org/10.1140/epjc/s10052-022-10785-0}{{\em Eur. Phys. J.
  C} {\bfseries 82} no.~11, (2022) 988},
  \href{http://arxiv.org/abs/2202.02039}{{\ttfamily arXiv:2202.02039
  [hep-ex]}}.

\bibitem{ATLAS:2022pbd}
{\bfseries ATLAS} Collaboration, G.~Aad {\em et~al.}, ``{Search for doubly
  charged Higgs boson production in multi-lepton final states using
  139~fb$^{-1}$ of proton-proton collisions at $\sqrt{s}$ = 13~TeV with the
  ATLAS detector},''
  \href{http://dx.doi.org/10.1140/epjc/s10052-023-11578-9}{{\em Eur. Phys. J.
  C} {\bfseries 83} no.~7, (2023) 605},
  \href{http://arxiv.org/abs/2211.07505}{{\ttfamily arXiv:2211.07505
  [hep-ex]}}.

\bibitem{ATLAS:2021kog}
{\bfseries ATLAS} Collaboration, G.~Aad {\em et~al.}, ``{Measurements of
  differential cross-sections in four-lepton events in 13 TeV proton-proton
  collisions with the ATLAS detector},''
  \href{http://dx.doi.org/10.1007/JHEP07(2021)005}{{\em JHEP} {\bfseries 07}
  (2021) 005}, \href{http://arxiv.org/abs/2103.01918}{{\ttfamily
  arXiv:2103.01918 [hep-ex]}}.

\bibitem{Frederix:2018nkq}
R.~Frederix, S.~Frixione, V.~Hirschi, D.~Pagani, H.~S. Shao, and M.~Zaro,
  ``{The automation of next-to-leading order electroweak calculations},''
  \href{http://dx.doi.org/10.1007/JHEP11(2021)085}{{\em JHEP} {\bfseries 07}
  (2018) 185}, \href{http://arxiv.org/abs/1804.10017}{{\ttfamily
  arXiv:1804.10017 [hep-ph]}}. [Erratum: JHEP 11, 085 (2021)].

\bibitem{Sjostrand:2014zea}
T.~Sj\"ostrand, S.~Ask, J.~R. Christiansen, R.~Corke, N.~Desai, P.~Ilten,
  S.~Mrenna, S.~Prestel, C.~O. Rasmussen, and P.~Z. Skands, ``{An introduction
  to PYTHIA 8.2}'' \href{http://dx.doi.org/10.1016/j.cpc.2015.01.024}{{\em
  Comput. Phys. Commun.} {\bfseries 191} (2015) 159--177},
  \href{http://arxiv.org/abs/1410.3012}{{\ttfamily arXiv:1410.3012 [hep-ph]}}.

\bibitem{deFavereau:2013fsa}
{\bfseries DELPHES 3} Collaboration, J.~de~Favereau, C.~Delaere, P.~Demin,
  A.~Giammanco, V.~Lema\^\i{}tre, A.~Mertens, and M.~Selvaggi, ``{DELPHES 3, A
  modular framework for fast simulation of a generic collider experiment},''
  \href{http://dx.doi.org/10.1007/JHEP02(2014)057}{{\em JHEP} {\bfseries 02}
  (2014) 057}, \href{http://arxiv.org/abs/1307.6346}{{\ttfamily arXiv:1307.6346
  [hep-ex]}}.

\bibitem{Cacciari:2008gp}
M.~Cacciari, G.~P. Salam, and G.~Soyez, ``{The anti-$k_t$ jet clustering
  algorithm},'' \href{http://dx.doi.org/10.1088/1126-6708/2008/04/063}{{\em
  JHEP} {\bfseries 04} (2008) 063},
  \href{http://arxiv.org/abs/0802.1189}{{\ttfamily arXiv:0802.1189 [hep-ph]}}.

\bibitem{Cacciari:2011ma}
M.~Cacciari, G.~P. Salam, and G.~Soyez, ``{FastJet User Manual},''
  \href{http://dx.doi.org/10.1140/epjc/s10052-012-1896-2}{{\em Eur. Phys. J. C}
  {\bfseries 72} (2012) 1896}, \href{http://arxiv.org/abs/1111.6097}{{\ttfamily
  arXiv:1111.6097 [hep-ph]}}.

\bibitem{Cepeda:2019klc}
M.~Cepeda {\em et~al.}, ``{Report from Working Group 2: Higgs Physics at the
  HL-LHC and HE-LHC},''
  \href{http://dx.doi.org/10.23731/CYRM-2019-007.221}{{\em CERN Yellow Rep.
  Monogr.} {\bfseries 7} (2019) 221--584},
  \href{http://arxiv.org/abs/1902.00134}{{\ttfamily arXiv:1902.00134
  [hep-ph]}}.

\bibitem{ATLAS:2018hxb}
{\bfseries ATLAS} Collaboration, M.~Aaboud {\em et~al.}, ``{Measurements of
  Higgs boson properties in the diphoton decay channel with 36 fb$^{-1}$ of
  $pp$ collision data at $\sqrt{s} = 13$ TeV with the ATLAS detector},''
  \href{http://dx.doi.org/10.1103/PhysRevD.98.052005}{{\em Phys. Rev. D}
  {\bfseries 98} (2018) 052005},
  \href{http://arxiv.org/abs/1802.04146}{{\ttfamily arXiv:1802.04146
  [hep-ex]}}.

\bibitem{Cowan:2010js}
G.~Cowan, K.~Cranmer, E.~Gross, and O.~Vitells, ``{Asymptotic formulae for
  likelihood-based tests of new physics},''
  \href{http://dx.doi.org/10.1140/epjc/s10052-011-1554-0}{{\em Eur. Phys. J. C}
  {\bfseries 71} (2011) 1554}, \href{http://arxiv.org/abs/1007.1727}{{\ttfamily
  arXiv:1007.1727 [physics.data-an]}}. [Erratum: Eur.Phys.J.C 73, 2501 (2013)].

\bibitem{Georgi:1985nv}
H.~Georgi and M.~Machacek, ``{DOUBLY CHARGED HIGGS BOSONS},''
  \href{http://dx.doi.org/10.1016/0550-3213(85)90325-6}{{\em Nucl. Phys. B}
  {\bfseries 262} (1985) 463--477}.

\bibitem{ATLAS:2023gsl}
{\bfseries ATLAS} Collaboration, G.~Aad {\em et~al.}, ``{Inclusive and
  differential cross-sections for dilepton $ t\bar t $ production measured in $
  \sqrt{s} $ = 13 TeV pp collisions with the ATLAS detector},''
  \href{http://dx.doi.org/10.1007/JHEP07(2023)141}{{\em JHEP} {\bfseries 07}
  (2023) 141}, \href{http://arxiv.org/abs/2303.15340}{{\ttfamily
  arXiv:2303.15340 [hep-ex]}}.

\bibitem{Inoue:2015pza}
S.~Inoue, G.~Ovanesyan, and M.~J. Ramsey-Musolf, ``{Two-Step Electroweak
  Baryogenesis},'' \href{http://dx.doi.org/10.1103/PhysRevD.93.015013}{{\em
  Phys. Rev. D} {\bfseries 93} (2016) 015013},
  \href{http://arxiv.org/abs/1508.05404}{{\ttfamily arXiv:1508.05404
  [hep-ph]}}.

\bibitem{Ferreira:2004yd}
P.~M. Ferreira, R.~Santos, and A.~Barroso, ``{Stability of the tree-level
  vacuum in two Higgs doublet models against charge or $CP$ spontaneous
  violation},'' \href{http://dx.doi.org/10.1016/j.physletb.2004.10.022}{{\em
  Phys. Lett. B} {\bfseries 603} (2004) 219--229},
  \href{http://arxiv.org/abs/hep-ph/0406231}{{\ttfamily arXiv:hep-ph/0406231}}.
  [Erratum: Phys.Lett.B 629, 114--114 (2005)].

\bibitem{Barroso:2005sm}
A.~Barroso, P.~M. Ferreira, and R.~Santos, ``{Charge and $CP$ symmetry breaking
  in two Higgs doublet models},''
  \href{http://dx.doi.org/10.1016/j.physletb.2005.11.031}{{\em Phys. Lett. B}
  {\bfseries 632} (2006) 684--687},
  \href{http://arxiv.org/abs/hep-ph/0507224}{{\ttfamily arXiv:hep-ph/0507224}}.

\bibitem{Ferreira:2019hfk}
P.~M. Ferreira and B.~L. Gon\c{c}alves, ``{Stability of neutral minima against
  charge breaking in the Higgs triplet model},''
  \href{http://dx.doi.org/10.1007/JHEP02(2020)182}{{\em JHEP} {\bfseries 02}
  (2020) 182}, \href{http://arxiv.org/abs/1911.09746}{{\ttfamily
  arXiv:1911.09746 [hep-ph]}}.

\bibitem{Azevedo:2020mjg}
D.~Azevedo, P.~Ferreira, H.~E. Logan, and R.~Santos, ``{Vacuum structure of the
  $\mathbb{Z}_2$ symmetric Georgi-Machacek model},''
  \href{http://dx.doi.org/10.1007/JHEP03(2021)221}{{\em JHEP} {\bfseries 03}
  (2021) 221}, \href{http://arxiv.org/abs/2012.07758}{{\ttfamily
  arXiv:2012.07758 [hep-ph]}}.

\bibitem{Hundi:2023tdq}
R.~S. Hundi, ``{Study on the global minimum and $H\to\gamma\gamma$ in the Dirac
  scotogenic model},''
  \href{http://dx.doi.org/10.1103/PhysRevD.108.015006}{{\em Phys. Rev. D}
  {\bfseries 108} no.~1, (2023) 015006},
  \href{http://arxiv.org/abs/2303.04655}{{\ttfamily arXiv:2303.04655
  [hep-ph]}}.

\end{thebibliography}\endgroup

\end{document}